\documentclass[%
 reprint,
 superscriptaddress,preprintnumbers,
 nofootinbib,
 amsmath,
 amssymb,
 aps,
 pra
]{revtex4-2}

\allowdisplaybreaks %Allows page breaks in multi-line math environments
\usepackage{dsfont}
\usepackage[utf8]{inputenc}
\usepackage{hyperref}
\usepackage{graphicx}   
\usepackage[table]{xcolor}
\usepackage{colortbl}
\usepackage{url}
\usepackage[normalem]{ulem}
\usepackage{slashed,centernot}
\usepackage{bbold}
\usepackage[capitalise, english]{cleveref}
\usepackage[noindentafter]{titlesec} %Removes the indentation after new section
\usepackage{epigraph}
\usepackage{graphicx}
\usepackage{booktabs} 
\usepackage{upgreek}

\usepackage[shortlabels]{enumitem}
\setlist{itemsep=.1em,topsep=.5em}
\SetEnumerateShortLabel{i}{\textit{\roman*}}

\usepackage{siunitx}
\usepackage{xspace}
\usepackage{enumitem}

    \makeatletter
    \def\CT@@do@color{%
      \global\let\CT@do@color\relax
            \@tempdima\wd\z@
            \advance\@tempdima\@tempdimb
            \advance\@tempdima\@tempdimc
    \advance\@tempdimb\tabcolsep
    \advance\@tempdimc\tabcolsep
    \advance\@tempdima2\tabcolsep
            \kern-\@tempdimb
            \leaders\vrule
                    \hskip\@tempdima\@plus  1fill
            \kern-\@tempdimc
            \hskip-\wd\z@ \@plus -1fill }
    \makeatother

\usepackage{accents}
\DeclareMathSymbol{\widetildesym}{\mathord}{largesymbols}{"65}

\def\thesubsection{\arabic{section}.\arabic{subsection}}

\def\thesection{\arabic{section}}
\titleformat*{\subsubsection}{\normalfont \small \bfseries \boldmath}
\renewcommand{\paragraph}[1]{\vspace{.3em} \indent {\itshape #1 ---}\xspace }
\makeatletter
    \renewcommand{\p@subsection}{}
    \renewcommand{\p@subsubsection}{}
\makeatother

\definecolor{red}{rgb}{0.6,.0706,.1373}
\definecolor{blue}{rgb}{0,0.396,0.741}
\definecolor{green}{rgb}{0.25,0.6,0.2}
\newcommand\myshade{80}
\colorlet{mylinkcolor}{violet}
\colorlet{mycitecolor}{violet}
\colorlet{myurlcolor}{violet}

\hypersetup{
  linkcolor  = mylinkcolor!\myshade!black,
  citecolor  = mycitecolor!\myshade!black,
  urlcolor   = myurlcolor!\myshade!black,
  colorlinks = true
}
\newcommand{\U}{\mathrm{U}}

\newcommand{\X}{\mathcal{X}}

\newcommand{\diag}{\mathop{\mathrm{diag}}}

\newcommand{\eminus}{\vcenter{\hbox{\scalebox{0.6}[1]{$ - $}}}}	%Narrow minus signed (for e.g. negative exponents)

\newcommand{\transpose}{^\intercal}
\newcommand{\andeq}{\quad \mathrm{and} \quad}
\newcommand{\tr}{\mathop{\mathrm{tr}} }
\newcommand{\hc}{\; + \; \mathrm{H.c.} \;}
\newcommand{\Z}{\mathbb{Z}}

\newcommand{\sscript}[1]{{\scriptscriptstyle \mathrm{#1}}}

\newcommand{\vew}{v_\sscript{EW}}

\def\L{\mathcal{L}}
\newcommand{\gx}{$ \U(1)_{X_p}$\xspace}
\newcommand{\BL}{$ \U(1)_{B-L}$\xspace}

\usepackage{physics}

\begin{document}

\title{ \boldmath Insights on the Cosmic Origin of Matter from Proton Stability}

\author{Admir Greljo}
\email{admir.greljo@unibas.ch}
\affiliation{Department of Physics, University of Basel, Klingelbergstrasse 82,  CH-4056 Basel, 
Switzerland}

\author{Xavier Ponce D\'iaz}
\email{xavier.poncediaz@unibas.ch}
\affiliation{Department of Physics, University of Basel, Klingelbergstrasse 82,  CH-4056 Basel, 
Switzerland}

\author{Anders Eller Thomsen}
\email{anders.thomsen@unibe.ch}
\affiliation{Albert Einstein Center for Fundamental Physics, Institute for Theoretical Physics, University of Bern, Sidlerstrasse 5, CH-3012 Bern, Switzerland}

% \date{\today}

%\preprint{XXX}
%=============================================================================

\begin{abstract}

We investigate the phenomenology of a model~\cite{Davighi:2022qgb} in which the proton is rendered \textit{absolutely} stable by an IR mechanism that remains robust against unknown quantum gravity effects. A linear combination of baryon number and lepton flavors is gauged and spontaneously broken to a residual $\mathbb{Z}_9$ discrete gauge symmetry enforcing a strict selection rule: $\Delta B = 0\,(\mathrm{mod}\,3)$. Despite its minimal field content, the model successfully accounts for established empirical evidence of physics beyond the SM.
High-scale symmetry breaking simultaneously provides a seesaw mechanism explaining the smallness of neutrino masses, minimal thermal leptogenesis, and a viable phenomenology of the majoron as dark matter.
Any cosmic string--wall network remaining after inflation is unstable for numerous charge assignments.
Lepton flavor non-universality, central to the construction, leads to predictive neutrino textures testable via oscillation experiments, neutrinoless double beta decay, and cosmology. The model motivates searches in $X$- and $\gamma$-ray lines, neutrino telescopes, and predicts CMB imprints. 

\end{abstract}

\maketitle

\tableofcontents

%-----------------------------------------------------------------------------
\section{Introduction} 
\label{sec:intro}
%-----------------------------------------------------------------------------

What if the proton is absolutely stable? Within the Standard Model (SM), proton stability is not a consequence of a fundamental symmetry but rather an accidental feature arising from the gauge structure and particle content. While many extensions of the SM predict proton decay through baryon number-violating interactions~\cite{Ohlsson:2023ddi}, decades of experimental searches have produced no evidence for such processes~\cite{Super-Kamiokande:2020wjk}. Upcoming experiments will continue to push the experimental limits further~\cite{Hyper-Kamiokande:2018ofw, DUNE:2020lwj, JUNO:2015sjr}. Given a persistent null result, it is worth entertaining a provocative possibility: could the proton be \textit{absolutely} stable, protected by a deeper principle?

While academic in nature, this question presents a nontrivial challenge for model building, as it is widely believed that global symmetries cannot survive in a theory consistently coupled to quantum gravity: they must either be gauged or they will be explicitly broken~\cite{Banks:1988yz, Giddings:1988cx, Banks:2010zn}. Given that the SM is understood as a low-energy effective field theory (SMEFT) emerging from an unknown ultraviolet (UV) completion, one expects that higher-dimensional operators violating baryon number are induced (at least) by quantum gravity effects. Such operators would, in turn, lead to proton decay. 

In this paper, we focus on an infrared (IR) mechanism that ensures \textit{exact} proton stability, independent of the details of the UV completion. A particularly elegant proposal introduced in Ref.~\cite{Davighi:2022qgb} (see also~\cite{Babu:2003qh} for a related approach) extends the SM by a $\U(1)_X$ gauge symmetry, spontaneously broken above the electroweak (EW) scale. The key idea is that when $X$ is an appropriate anomaly-free linear combination of baryon number and lepton flavor numbers,\footnote{The baryon number symmetry, $\U(1)_{B}$, is anomalous within the SM and cannot be consistently gauged without introducing additional exotic chiral fermions~\cite{FileviezPerez:2010gw, Ma:2020quj, Boos:2022jvc}.} the breaking of $\U(1)_X$ gives rise to a residual discrete $\Z_9$ gauge symmetry. The resulting residual symmetry, which remains unbroken in the deep IR and assigns unit charge to all quarks while leaving other SM fields neutral, forbids all proton decay operators ($\Delta B = 1$) of any dimension. In this framework, baryon number violation is restricted to occur only in multiples of three,
\begin{equation}\label{eq:selectrule}
\Delta B = 0\ (\mathrm{mod}\ 3)\,, 
\end{equation}
since the $\mathbb{Z}_9$-invariant $B$-violating operators involve (a multiple of) nine quark fields. This allows processes such as sphaleron transitions, while forbidding both proton decay and neutron–antineutron oscillations.

This mechanism links proton stability to the non-universality of lepton flavor. Remarkably, it accomplishes this with a minimal set of ingredients, nearly at the level of a toy model. Aside from the SM fields and two SM-singlet scalars needed to appropriately break the $\U(1)_X$ symmetry, anomaly cancellation necessitates only three right-handed neutrinos to complete the field content. In this paper, we perform a detailed phenomenological study of the model. Despite simplicity and minimality, we show that the model, in its original form, can simultaneously account for established \textit{empirical} evidence for physics beyond the SM: neutrino masses and mixings, the presence of dark matter, and the matter-antimatter asymmetry of the universe. Each of these phenomena finds a natural origin within this framework, converging on a common region of parameter space characterized by high-scale $\U(1)_X$ breaking.

The spontaneous breaking of the $\U(1)_X$ gauge symmetry is achieved via two scalar fields—the minimal set required to generate a realistic neutrino mass matrix, via the type-I seesaw mechanism~\cite{Minkowski:1977sc, Yanagida:1980xy, Gell-Mann:1979vob, Mohapatra:1979ia, Schechter:1980gr, CentellesChulia:2024uzv}, consistent with the observed PMNS mixing matrix~\cite{Pontecorvo:1967fh, Maki:1962mu}. An especially attractive feature of this scalar content is the prediction of a (pseudo-)Nambu–Goldstone Boson (pNGB). While one of the Goldstone bosons is eaten by the $Z'$ gauge boson, the other remains in the spectrum as a physical state: a majoron \cite{Chikashige:1980ui, Chikashige:1980qk, Schechter:1981cv, Gelmini:1980re}, which is identified as the dark matter. 

This construction offers a conceptual advantage over traditional majoron dark matter models that rely on global symmetries, which are expected to be violated by quantum gravity as discussed above. Here, the lepton number arises as an accidental symmetry, protected at the renormalizable level by the underlying gauge structure~\cite{Rothstein:1992rh}, and is broken explicitly only by Planck-suppressed higher-dimensional operators. The $\U(1)_X$ charges of the proton stability mechanism naturally generate a sufficiently light majoron for it to be a viable dark matter candidate.

The high-scale breaking of $\U(1)_X$ proves advantageous for several interconnected reasons. It enables the high-scale type-I seesaw mechanism, explaining the smallness of neutrino masses relative to those of charged fermions; it supports minimal thermal leptogenesis~\cite{Fukugita:1986hr}, consistent with the Davidson–Ibarra bound~\cite{Davidson:2002qv}; and it ensures the correct relic abundance and viable phenomenology of the majoron as a dark matter candidate. Thus, the model economically unifies the cosmic origins of visible and dark matter with the neutrino sector within a common parameter space. 

Despite its high-scale realization, the model makes concrete predictions that can be tested in the future. One of the most striking predictions of lepton non-universality lies in the neutrino textures, which can be tested through the neutrino program, including long-baseline experiments~\cite{DUNE:2015lol, Hyper-Kamiokande:2018ofw} and searches for neutrinoless double beta decay \cite{nEXO:2021ujk, LEGEND:2021bnm, CUPID:2022jlk, Adams:2022jwx}. Additionally, there are broader implications for cosmology and astroparticle physics, motivating searches for $X-$ and $\gamma-$ray lines, signals in neutrino telescopes, imprints in the CMB, and stochastic gravitational wave backgrounds. And, of course, no proton decay.

The paper is organized as follows. Section~\ref{sec:model} introduces the model and its field content. Predictions for the neutrino sector are discussed in Section~\ref{sec:PMNS}, followed by thermal leptogenesis in Section~\ref{sec:leptogenesis}. Majoron dark matter phenomenology is presented in Section~\ref{sec:DM}. Topological defects are discussed in Section~\ref{sec:topological}. Section~\ref{sec:summary} summarizes the key constraints and prospects, while conclusions are given in Section~\ref{sec:conc}. Technical details are relegated to the Appendices.

%%%%%%%%%%%
\section{A model for proton stability}
\label{sec:model}
%%%%%%%%%%%

\begin{table*}[t] 
    \begin{center} \renewcommand{\arraystretch}{1.1}
    \begin{tabular}{|c|c|c|c|c|}
    \hline  \rowcolor{black!15}
    & Fields & \gx & $ \mathds{Z}_9\subseteq \Gamma $& $ \U(1)_{B-L}$\\
    \hline
    Quarks & $q_i$, $u_i$,  $d_i$ & $m$ & $ 1 $& $ 1/3 $\\
    \hline
    Specific leptons & $\ell_p$, $e_p$, $N_p$ & $-2n -3m$ & $ 0 $ & $ -1 $\\ 
    \hline
    Common leptons ($q\neq p$) & $\ell_{q}$, $e_{q}$, $N_{q}$ & $n-3m$ & $ 0 $ & $ -1 $\\
    \hline
    Higgs & $H$ & $0$ & $ 0 $ & $ 0 $\\
    \hline
    New scalars & $\phi_1$ & $6m + n $ & $ 0 $ & $ 2 $\\ 
    & $\phi_{2}$ & $6m -2n$ & $ 0 $ & $ 2 $\\
    \hline
    \end{tabular}
\caption{The field content of the proton-stability model. In addition to the SM fields plus three RHNs, there is a \gx gauge field with flavor non-universal couplings to SM leptons and a pair of SM singlets $\phi_{1}$ and $\phi_2$ whose VEVs break \gx to discrete gauge symmetry $\Gamma$. By contrast, $\U(1)_{B-L}$ emerges as an accidental symmetry of the renormalizable Lagrangian, and its associated Nambu-Goldstone boson serves as the dark matter. }
    \label{tab:fields}
    \end{center}
\end{table*}

Our starting point is the \gx group from Ref.~\cite{Davighi:2022qgb}, which guarantees exact proton stability:
	\begin{equation} \label{eq:Xsymm}
	X_p = 3m(B - L) - n\left(3 L_p-  L \right)\, , \quad \gcd(m, n)= 1,
	\end{equation}
where $ B $ is the baryon number, $ L $ the total lepton number, and $ L_ p $ the lepton number of a particular flavor $p$.
An overview of the fields is found in Tab.~\ref{tab:fields}. Cancellation of chiral anomalies is achieved by adding three right-handed neutrinos (RHNs), $ N_i$, to the SM field content. The lepton flavor $ p \in \{e,\, \mu,\, \tau\} $ is charged differently from the other two generations; we will refer to $ X_p $ as the $ p $-specific scenario. Each scenario gives distinct predictions for the low-energy neutrino phenomenology.

Two new complex scalar singlets $ \phi_a$ are responsible for breaking \gx at high scales. While SM singlets, the fields carry charges 
\begin{equation}
\label{eq:scalar_charges}
    [\phi_{1}]_{X_p} = 6m+ n \,,\quad \textrm{and} \quad [\phi_{2}]_{X_p} = 6m- 2n \, .
\end{equation}
Their VEVs break the gauge symmetry:
    \begin{equation} \label{eq:SSB_pattern}
    \U(1)_{X_p} \xrightarrow{\langle \phi \rangle\neq 0} \Gamma  \cong 
    \begin{cases}
    \Z_9^B,  &\text{for~}    n  \in 2\Z +1 \\
    \Z_9^B \times \Z_2^f,  &\text{for~}  n \in 2\Z  \\
    \end{cases},
    \end{equation}
where the remnant group $ \Z_9^B $ guarantees exact proton stability.\footnote{The $ \Z_2^f $ remnant group that is sometimes embedded in \gx is nothing but fermion parity. It is generated by the element $ e^{i\pi} \in \U(1)_{X_p} $ when $ n \equiv 0 \pmod{2} $.} This scenario is not completely generic, but is realized if and only if $ (m,\,n)$ satisfies~\cite{Davighi:2022qgb}
\begin{equation} \label{eq:m_and_n}
    (m,\, n) = \big(3a +1,\, 9b + 3\big), \quad \text{for} \quad (a,\, b) \in \mathbb{Z}^2 
\end{equation}
subject to $\gcd\!\big(3a + 1,\, b - a\big)=1$ implying $\gcd(m,n)=1$.\footnote{We restrict our analysis to one half of the solutions identified in~\cite{Davighi:2022qgb}; the other half is physically equivalent and can be obtained by switching the sign of all charges.}
The symmetry-breaking pattern~\eqref{eq:SSB_pattern} is a consequence of both $\phi_a$ scalars acquiring a VEV. As explained in the next section, this is the minimal choice of scalar fields reproducing the observed neutrino oscillation data for generic $(m,n)$ satisfying~\eqref{eq:m_and_n}.

While Eq.~\eqref{eq:m_and_n} suggests large charge ratios, which might appear baroque, this feature is in fact advantageous for generating a sufficiently light majoron via gravitational effects, in a way that remains compatible with experimental constraints on dark matter (Section~\ref{sec:DM}).

The exact $\mathds{Z}_9 \subseteq \Gamma$ symmetry in the deep IR assigns unit charge to quarks while leaving all other SM fields neutral (see Tab.~\ref{tab:fields}). This enforces a strict selection rule, $\Delta B \equiv 0 \pmod{3}$, since forming a $\mathds{Z}_9$ singlet requires nine quark fields. As anticipated, proton decay and neutron--antineutron oscillations are forbidden, while sphaleron processes remain allowed as mandatory in leptogenesis scenarios. Triple nucleon decays ($\Delta B = 3$) arise from dimension-15 operators and are therefore highly suppressed~\cite{Babu:2003qh}.

When the $ \phi_a $ fields develop their VEVs, they provide a mass to the gauge boson $ Z' $ associated with the \gx symmetry. The two radial modes $ \rho_{1,2} $ of the $ \phi_a $ fields are, on general grounds, also expected to get masses of a size similar to the VEVs. Of the two angular fields, one is eaten by the heavy $ Z' $, leaving a single pNGB, $ a $, in the spectrum. The majoron $ a $ is the dark matter of the model and obtains a small mass through higher-dimensional operators, as discussed in Section~\ref{sec:DM}. The symmetry breaking also provides masses to the neutrinos.

%%%%%%%%%%%%%%%%%%%%%%%%%%%%%%%%%%%%%
\section{Neutrino masses and mixings}
\label{sec:PMNS}
%%%%%%%%%%%%%%%%%%%%%%%%%%%%%%%%%%%%%

The Higgs interactions in the quark sector remain unchanged from the SM, while those in the lepton sector take the form
    \begin{equation} \label{eq:L_UV}
    \L \supset 
    -Y_N^{ij} \overline{\ell}_i \tilde{H} N_j - Y_e^{ij} \overline{\ell}_i H e_j   - \tfrac{1}{2} \overline{N}^c_i Y^{ij}_{\phi,a} \phi_{a} N_j \hc
    \end{equation}
The \gx gauge symmetry leads to distinct textures in the coupling matrices, determined by the specific charge assignments.

\begin{figure*}[t]
    \centering
    \includegraphics[width=\linewidth]{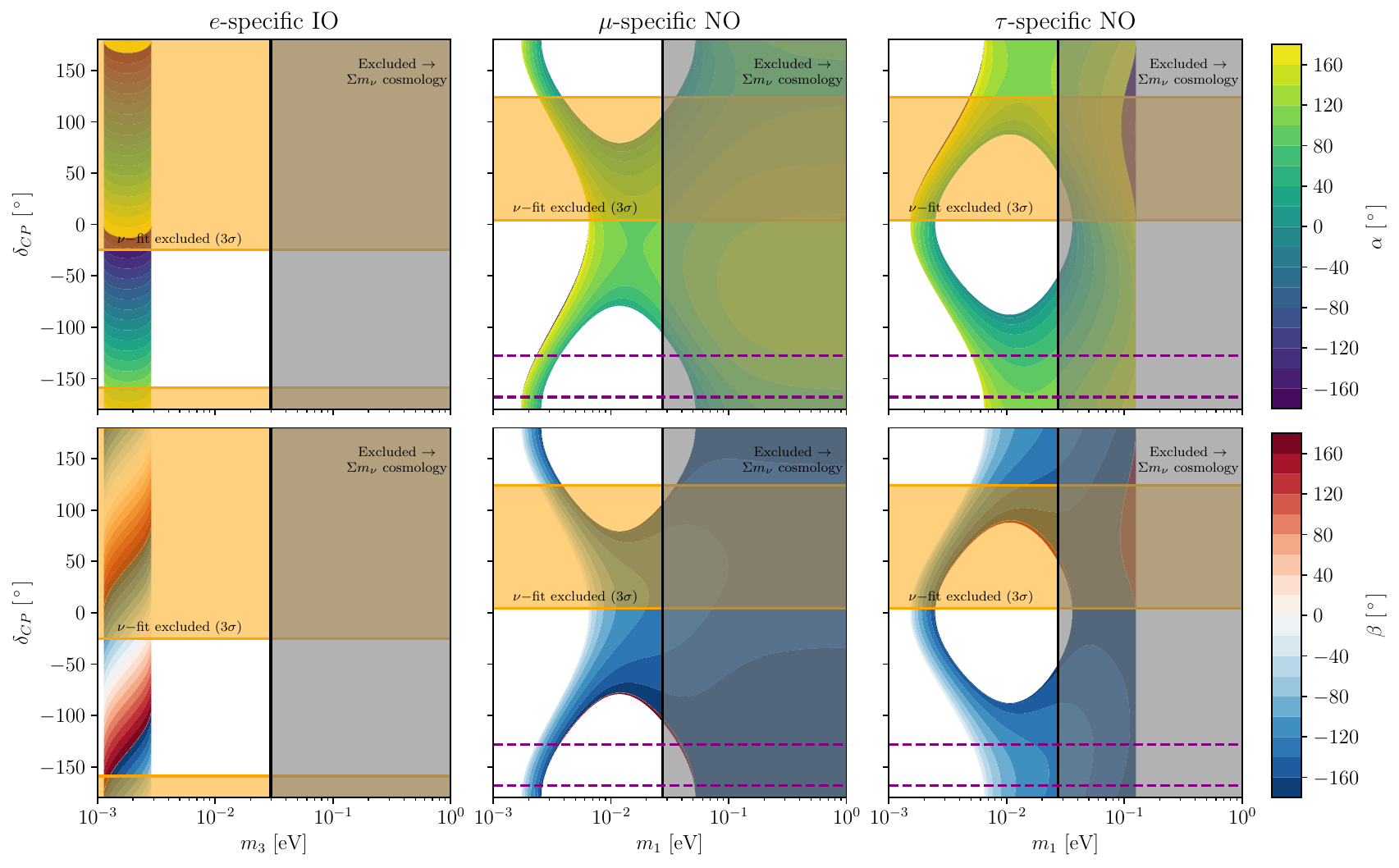}
    \caption{One of the two solutions for the Majorana phases as a function of the lightest neutrino mass and Dirac phase $\delta_\sscript{CP}$. Upper and lower rows are for $\alpha$ and $\beta$, respectively. The columns correspond to the three models allowed by the current data: $e$-IO, $\mu$-NO, and $\tau$-NO. The remaining neutrino parameters are fixed to their central values from Ref.~\cite{Esteban:2024eli}. The future $2\sigma$-sensitivity of DUNE on $\delta_\sscript{CP}$-NO is denoted in dashed purple lines, assuming $\delta_{\sscript{CP}}$ is the current central value preferred by the global fits. See Section~\ref{sec:PMNS} for details.}
    \label{fig:majorana_phases}
\end{figure*}

Populating the $3 \times 3$ Majorana mass matrix using scalar VEVs requires distinguishing a 1-dimensional $p$ block, a 2-dimensional $q$ block, and a mixed block connecting them.\footnote{For $(m,n) = (\eminus 2, 3)$, $ (1,3) $, and $(1, \eminus 6)$, a renormalisable Majorana mass term can be written for the $p$, $q$, and mixed blocks, respectively. We do not pursue these isolated scenarios further.} A single scalar VEV can populate only one of these blocks and, therefore, cannot generate masses for all heavy neutrinos. Among the three possible combinations of two scalar fields, populating the $p$ block and the mixed block leaves one right-handed neutrino massless, while populating the $p$ and $q$ blocks forbids mixing between the $p$ and $q$ flavors. Thus, only one viable choice remains: populating the $q$ block and the mixed block.\footnote{For $(m,n) = (1,12)$, the two scalars have opposite charges, rendering one of them redundant. In this scenario, no majoron is present.}

We introduce permutation matrices $ \mathcal{P}_p $ given by 
    \begin{equation}
    \label{eq:permutations}
    \mathcal{P}_e = \mathds{1},\quad 
    \mathcal{P}_\mu = \begin{pmatrix}
        \phantom{m} & 1 & \phantom{m} \\
        1 & \phantom{m} &  \\
          &  & 1 
        \end{pmatrix},\quad
    \mathcal{P}_\tau = \begin{pmatrix}
        \phantom{m} & \phantom{m} & 1\\
        & 1 & \phantom{m}   \\
        1 & &
        \end{pmatrix},
    \end{equation}
with which the coupling textures are
    \begin{align}
    Y_{\phi,1} &\sim 
    \mathcal{P}_p \begin{pmatrix}
    0 & \times & \times \\
    \times & 0 & 0 \\
    \times  & 0 & 0 
    \end{pmatrix}\mathcal{P}_p\,, &
    Y_{\phi,2} &\sim 
    \mathcal{P}_p \begin{pmatrix}
    0& 0 &0 \\
    0 & \times & \times \\
    0 & \times & \times \\
    \end{pmatrix}\mathcal{P}_p\,, \nonumber \\
    Y_{e,N} &\sim 
    \mathcal{P}_p \begin{pmatrix}
    \times & 0 & 0 \\
    0 & \times & \times \\
    0  & \times & \times 
    \end{pmatrix} \mathcal{P}_p\,, \label{eq:e_yuk}
    \end{align}
where $ \times $ indicates the entries not forbidden by the gauge symmetry, which are generically populated.
Observe that the gauge and mass eigenstates for the $ p $-specific charged lepton coincide in all cases; hence, the mass states are appropriate for labeling the scenarios. 
With the high-scale spontaneous symmetry breaking, the new scalars develop vacuum expectation values (VEVs) $ \langle \phi_a \rangle = \tfrac{1}{\sqrt{2}} v_a > 0 $ and the right-handed neutrinos receive a Majorana mass matrix 
    \begin{equation}
    M_N  =  \frac{v_a}{\sqrt{2}} Y_{\phi,a}\,.
    \end{equation}
Both VEVs are necessary to reproduce the observed pattern of light-neutrino oscillations, as discussed earlier.

Exploiting the flavor symmetry of the kinetic terms, we define a flavor basis of Eq.~\eqref{eq:L_UV}. The choice of basis and the corresponding parameter counting are detailed in Appendix~\ref{app:flavorcounting}. Altogether, Eq.~\eqref{eq:UVparams} introduces nine real couplings and three complex phases in the neutrino sector, which serve as input (UV) parameters for the numerical scan presented later.

The type-I seesaw mechanism~\cite{Minkowski:1977sc, Yanagida:1980xy, Gell-Mann:1979vob, Mohapatra:1979ia, Schechter:1980gr, CentellesChulia:2024uzv} predicts the mass matrix of the light (active) neutrinos to be
\begin{equation}
    \label{eq:seesaw_formula}
    m^*_\nu = - \vew^2 Y_N M_N^{\eminus 1} Y^T_N = U \hat m_\nu U^T~.
\end{equation}
where $ \vew = \SI{174}{GeV} $ is the Higgs VEV. 
Here, $\hat{m}_\nu = \mathrm{diag}(m_1, m_2, m_3)$ denotes the light neutrino mass eigenvalues, which can follow either a normal ($m_1 < m_2 < m_3$) or inverted ordering ($m_3 < m_1 < m_2$). The matrix $U$ is the PMNS mixing matrix, defined up to a $\mathrm{U}(1)^3$ phase redefinition of the form $m^{ij}_\nu \to e^{i(\theta_i + \theta_j)} m^{ij}_\nu$.

\begin{figure*}[t]
    \centering
    \includegraphics[width=0.7\linewidth]{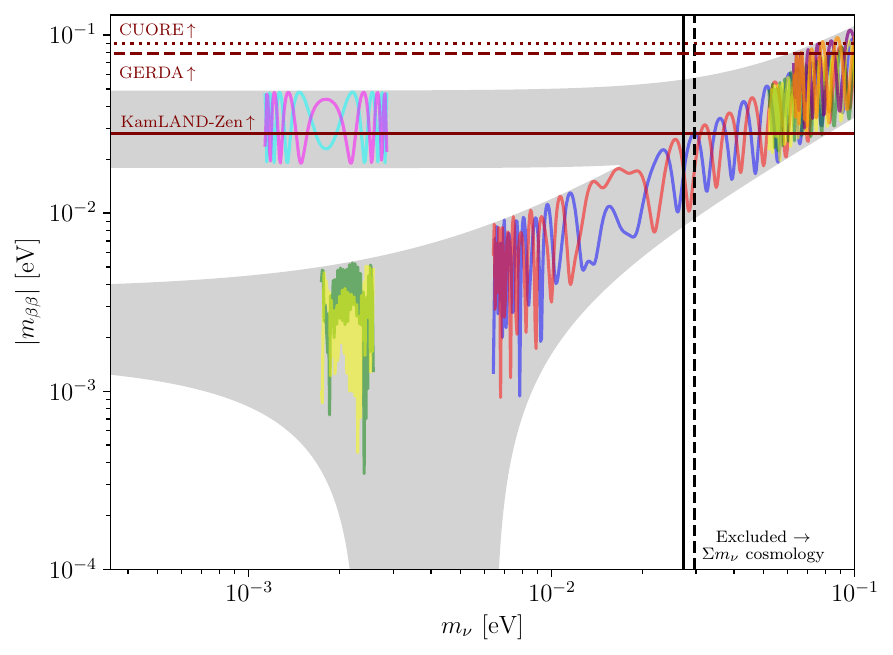}
    \caption{Predictions for the $0\nu\beta\beta$ effective Majorana mass $m_{\beta\beta}$ as a function of the lightest neutrino mass for different scenarios. The color bands represent the two possible solutions for the Majorana phase, with $\delta_\sscript{CP}$ fixed to its current best-fit value in each scenario: $e$-specific IO (cyan–purple), $\mu$-specific NO (yellow–green), $\tau$-specific NO (red–blue), and the excluded $\tau$-specific IO model (orange–purple).   See Section~\ref{sec:PMNS} for details.}
    \label{fig:neutrinolesss}
\end{figure*}

Interestingly, due to the lepton flavor non-universality ($n\neq 0$), the neutrino mass matrix $m_\nu$ is not completely free, and there are important predictions for future data. As shown in Appendix~\ref{app:neutrino_mass_matrix}, the texture of the individual factors in Eq.~\eqref{eq:seesaw_formula} imply a tree-level condition on the light-neutrino Majorana mass matrix, namely that the $p$'th minor of the neutrino matrix must vanish:\footnote{Radiative corrections generically invalidate this condition, but are expected to be small barring tuned cancellations. See Appendix~\ref{app:radiative} for details.}
\begin{align}
    [m_\nu]_{pp} \equiv (m_\nu)_{ii}  &(m_\nu)_{jj} -  (m_\nu)_{ij}^2 = 0, \\ &\text{for}\; i<j \; \text{and}\; i,j\neq p\, , \nonumber
\end{align}
This condition can be cast as the closure of a triangle in the complex plane as detailed in Appendix~\ref{app:neutrino_mass_matrix}. The length of each side of the triangle is governed by a separate neutrino mass (cf. Eq.~\eqref{eq:Delta_m_generic_form}). This has several important consequences:
\begin{enumerate}[i)]
    \item Only three out of six scenarios can fit current data: $\tau$- and $\mu$-specific normal ordering (NO) and $e$-specific inverted ordering (IO).
    \item All scenarios predict a nonzero lower (and some, an upper) limit on the lightest neutrino mass. 
    \item There exists a one to two map between the lightest neutrino mass and the Dirac phase, and two possible solutions for the Majorana phases: $(m_\nu,\,\delta_\sscript{CP}) \to (\alpha_{1,2},\, \beta_{1,2})$. 
\end{enumerate}
The final point can be visualized in Fig.~\ref{fig:majorana_phases}, where we plot one of the two possible solutions for the Majorana phases $(\alpha,\, \beta)$ as a function of the lightest neutrino mass and $\delta_\sscript{CP}$ for the three valid models. The other solution for the Majorana phases is obtained by conjugating the triangle associated with $ [m_\nu]_{pp} $ in the complex plane (Eq.~\eqref{eq:conjugate_solutions}).

As a cross-check, these relations also emerge from a ``top-down'' approach: by scanning over the UV parameters and fitting to low-energy neutrino oscillation data, Fig.~\ref{fig:scan_majoranaphases} demonstrates how the same patterns are reproduced, giving numerical predictions consistent with those obtained from Eq.~\eqref{eq:Delta_m_generic_form}. In the numerical scan, both solutions for the Majorana phases coexist in the same parameter space, confirming the two-fold ambiguity.

\begin{figure*}[t]
    \centering
    \includegraphics[width=0.5\linewidth]{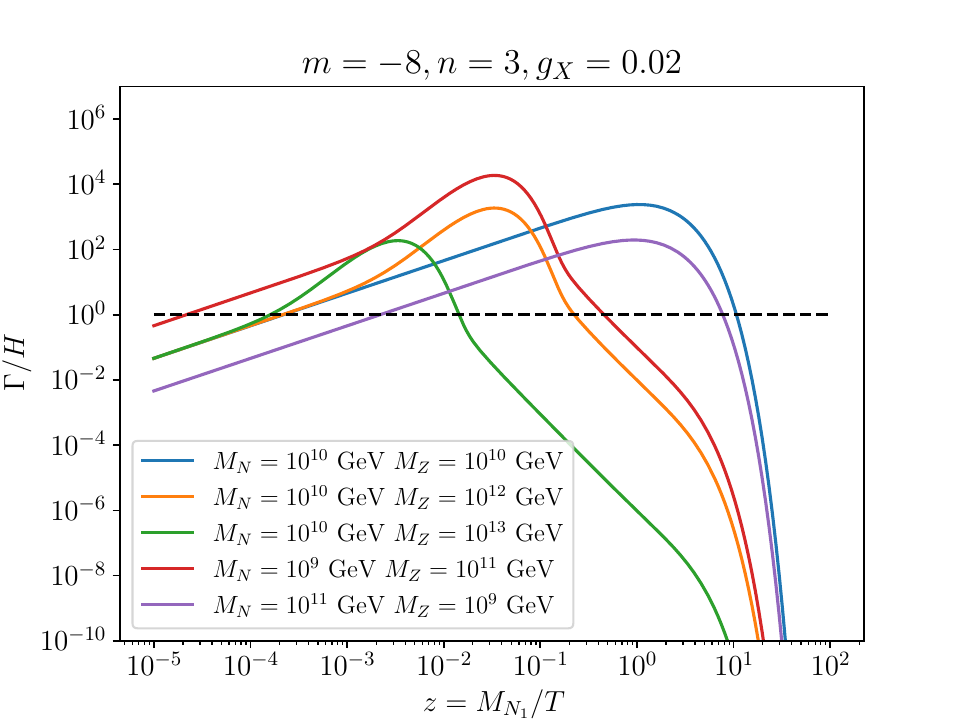}\!\! \includegraphics[width=0.5\linewidth]{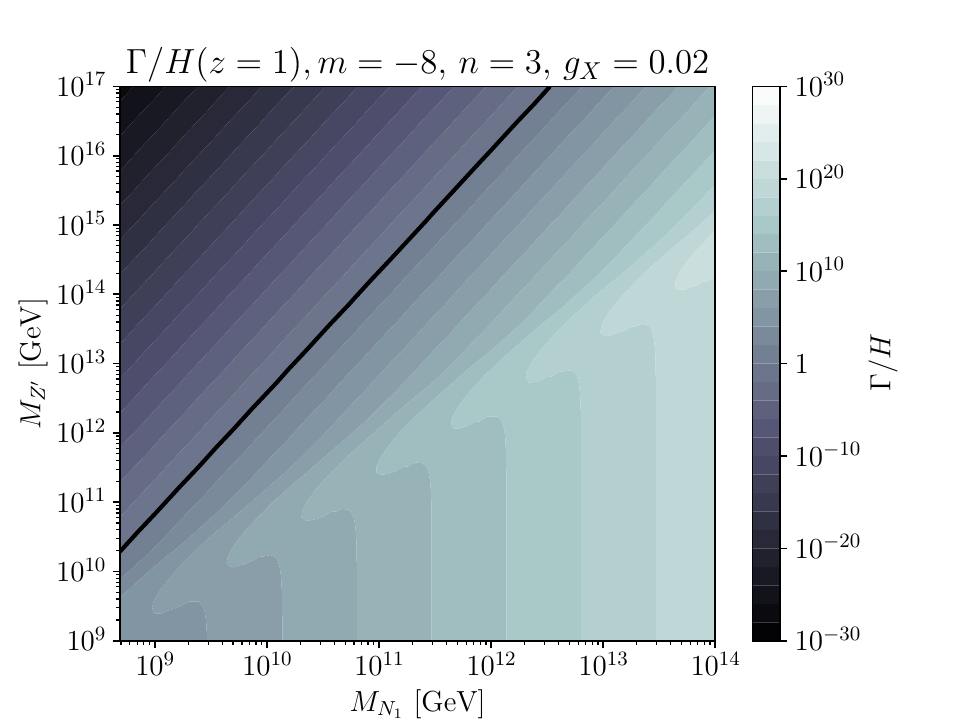}
    \caption{Thermalization rate $ \Gamma/H$ of the process $NN\to \textrm{SM}\, \textrm{SM}$ via $Z'$ interactions as a function of $z=M_{N_1}/T$ (\textit{left}) and as a function of $M_{N_1}$ and $M_Z$ at $z=1$ (\textit{right}). The black lines mark where $\Gamma/H =1$. We have plotted a benchmark scenario with the gauge symmetry charges $ X_p $ determined by $ (m,n) = (\eminus 8, 3) $ and a coupling $ g_X = 0.02 $. See Section~\ref{sec:leptogenesis} for details.}
    \label{fig:GammaH_ratio_Zprime}
\end{figure*}

The future neutrino program is crucial to test this model. DUNE \cite{DUNE:2015lol} and Hyper-Kamiokande (HK) \cite{Hyper-Kamiokande:2018ofw} are expected to measure the Dirac CP-violating phase and determine the neutrino mass ordering \cite{DUNE:2022aul}, while the sum of neutrino masses will be constrained indirectly through cosmological observations~\cite{DiValentino:2024xsv}. 
Once the Dirac phase $\delta_\sscript{CP}$ is fixed, the size of the effective Majorana mass $m_{\beta\beta}$ for neutrinoless double-$ \beta$ decay depends solely on the lightest neutrino mass $m_\nu$. This prediction is illustrated in Fig.~\ref{fig:neutrinolesss}, where we plot the two possible solutions for the Majorana phases across the three allowed models: $e-$IO (cyan-purple), $\mu-$NO (yellow-green), and $\tau-$NO (red-blue) for the current best fit value of $\delta_\sscript{CP}$. The complementary colors represent the two conjugate solutions arising from the triangle inequalities (see Eq.~\ref{eq:conjugate_solutions}). An additional model, $\tau-$IO (orange-purple), is also included; however, this scenario is excluded not only by cosmological bounds on the sum of neutrino masses but also by results from the KamLAND-Zen experiment~\cite{KamLAND-Zen:2024eml}. Other bounds from CUORE~\cite{CUORE:2024ikf} and GERDA~\cite{GERDA:2020xhi} are also found in Fig.~\ref{fig:neutrinolesss} in dotted and dashed lines, respectively.

Future experiments such as nEXO~\cite{nEXO:2021ujk}, LEGEND~\cite{LEGEND:2021bnm}, and CUPID~\cite{CUPID:2022jlk} are expected to place stringent constraints, potentially ruling out the inverted ordering (IO) scenario---see Ref.~\cite{Granelli:2025cho}---provided that nuclear matrix elements are determined with sufficient precision. While precise predictions for the Majorana phases require highly accurate measurements of $\delta_\sscript{CP}$, robust limits on the lightest neutrino mass can still be achieved with a $\delta_\sscript{CP}$ uncertainty of about $20^\circ-30^\circ$. Such precision appears attainable in upcoming experiments \cite{DUNE:2022aul, Hyper-Kamiokande:2018ofw}.

%%%%%%%%%%%%%%%%
\section{Minimal thermal leptogenesis}
\label{sec:leptogenesis}
%%%%%%%%%%%%%%%%

The type-I seesaw naturally accounts for the smallness of neutrino masses relative to the electroweak scale when the RHN masses are large, thereby motivating high-scale breaking of the \gx gauge symmetry. Another compelling argument favoring a high-scale symmetry breaking arises from cosmology: high-scale breaking readily accommodates the observed baryon asymmetry of the universe via minimal thermal leptogenesis. In this context, the Davidson–Ibarra bound requires the mass of the lightest right-handed neutrino to satisfy $M_{N_1} \gtrsim 10^9\,$GeV~\cite{Davidson:2002qv}.

\begin{figure*}[t]
    \centering
    \includegraphics[width=0.8\linewidth]{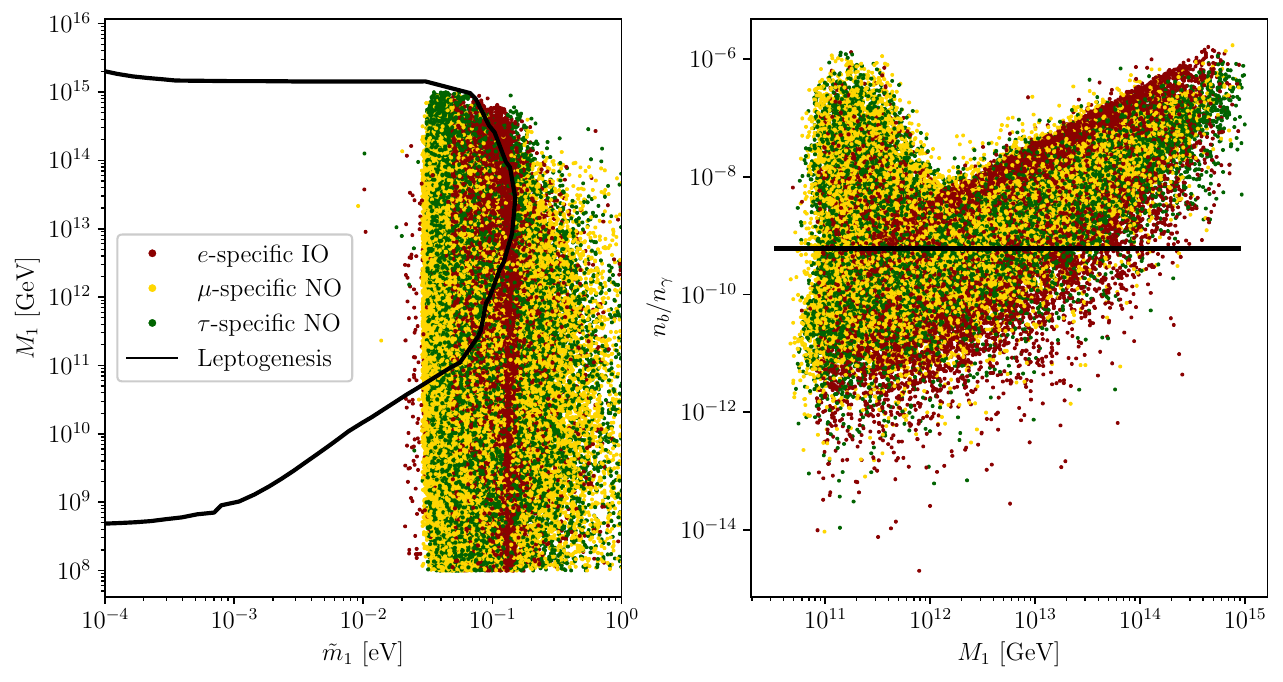}
    \caption{Numerical scan of the UV parameters for different scenarios, where all points reproduce the low-energy neutrino data within $3\sigma$. We show the relevant quantities for leptogenesis: the mass of the lightest RHN versus $\tilde{m}_1$ (\textit{left}) and the baryon-to-photon ratio (\textit{right}). The black line in the \textit{left} panel encloses the parameter region identified in Ref.~\cite{Giudice:2003jh}, where successful leptogenesis is possible. On the \textit{right}, the observed value of the baryon-to-photon ratio is also shown. See Section~\ref{sec:leptogenesis} for details.}
    \label{fig:leptogenesis_scan}
\end{figure*}

The minimal thermal leptogenesis scenario assumes a hierarchical mass spectrum among the RHNs: $M_{N_1} \ll M_{N_{2,3}}$. Our \gx gauge model can dynamically generate this through a mild hierarchy in the VEVs, $v_1/v_2 \sim 10^{\eminus 3}$--$ 10^{\eminus 1}$.\footnote{The VEV hierarchy can be lifted in the seesaw formula~\eqref{eq:seesaw_formula}, e.g., by letting a single coupling, namely $y^N_1$, in Eq.~\eqref{eq:UVparams} be a factor of $v_1/v_2$ smaller than the remaining entries in $ Y_N $ (which are expected to be of a similar size). This naturally leads to large PMNS mixing angles.} 
For comparable Yukawa couplings to $ \phi_a $, this leads to a mass relation of the form $M_{N_1} \sim M_{N_{2,3}} (v_1/v_2)^2$.\footnote{The option of inverting the hierarchy $v_1\gg v_2$ is viable, however the lighter neutrino mass is proportional to $M_{N_1}\sim v_2+\mathcal{O}(v_2^2/v_1)$.} For simplicity, we focus on the mass spectrum where $M_{N_{2,3}}, M_{Z'}$, and $M_{\rho_{2}}$ are all of the order $v_2$ while $M_{\rho_1}\sim v_1$ and are collectively referred to as the heavy states. By contrast, the lightest RHN acquires a suppressed mass, $M_{N_1} \sim v_1^2 / v_2$. This mass hierarchy allows $N_1$ to depart from equilibrium with the thermal bath before decaying. 

Successful leptogenesis requires a reheating temperature $T_\sscript{RH} \gtrsim M_{N_1}$ to produce $N_1$ efficiently. Thermal production of $N_1$ can proceed through the $\ell HN$ Yukawa interactions as well as four-fermion interactions mediated by $Z'$ exchange~\cite{Plumacher:1996kc}. Fig.~\ref{fig:GammaH_ratio_Zprime} shows the thermal rate $\Gamma/H$ for $NN\to \textrm{SM} \,\textrm{SM}$ processes via $Z'$, obtained by adapting equations from~\cite{Heeck:2016oda}. As expected, these processes typically lead to a nearly thermal initial abundance for $N_1$ even for moderate $T_\sscript{RH} < v_2$. To satisfy one of the Sakharov conditions, this interaction must decouple before $M_{N_1}/T \lesssim 1$, ensuring that $N_1$ departs from the thermal equilibrium before decaying~\cite{Sakharov:1967dj}. As illustrated in Fig.~\ref{fig:GammaH_ratio_Zprime}, the bound is easily satisfied for a mass hierarchy $M_{N_1} / M_{Z'} \lesssim 10^{-2}$. When the reheating temperature $T_\sscript{RH} > v_2$, the heavy states ($N_{2,3}$, $Z'$, and $\rho_{1,2}$) are temporarily populated in the thermal bath. However, strong washout effects prevent the heavier RHNs $N_{2,3}$ from contributing significantly to the matter asymmetry.

In summary, our scenario naturally yields a thermal initial abundance of the lightest right-handed neutrino $N_1$, which departs from equilibrium around $T \sim M_{N_1}$. Subsequently, the lepton asymmetry is generated through CP-violating decays of $N_1$ occurring out of equilibrium at late times. Lastly, sphaleron transitions, essential for transferring the asymmetry from the lepton sector to quarks, are $\Delta B = 3$ processes, consistent with the exact selection rule in Eq.~\eqref{eq:selectrule}. 

After this general discussion, we now demonstrate that our model can achieve successful leptogenesis. Fig.~\ref{fig:leptogenesis_scan} shows a numerical scan of the UV parameters for the three scenarios, consistent with low-energy neutrino data from Ref.~\cite{Esteban:2024eli} at the $3\sigma$ level, except for $\delta_\sscript{CP}$, which is left unconstrained. This scan uses the basis defined in Appendix~\ref{app:flavorcounting} with the UV parameters of Eq.~\eqref{eq:UVparams}. In addition, we impose $M_{N_2}\geq 10\times M_{N_1}$ and $\sum_i m_i \leq 0.113 (0.145)\,$eV for the NO (IO) ordering~\cite{DESI:2024mwx}. 

In the left panel of Fig.~\ref{fig:leptogenesis_scan}, we also display in black the result from Ref.~\cite{Giudice:2003jh}, indicating the region in the $\tilde{m}_1$–$M_1$ plane compatible with successful thermal leptogenesis. Here $\tilde{m}_1\equiv( Y_N^\dagger Y_N)_{11} \vew^2/M_{N_1} $ denotes the contribution to the neutrino masses mediated by $N_1$~\cite{Giudice:2003jh}. All three models lie in the strong washout regime (see Fig.~8 of~\cite{Giudice:2003jh}), resulting from the lower bound on the neutrino mass implied by the triangle equations (see Appendix~\ref{app:triangle_equations}). In the right panel of Fig.~\ref{fig:leptogenesis_scan}, we show the predicted baryon-to-photon ratio for the scanned points, using the results of Ref.~\cite{Giudice:2003jh}, demonstrating that our model successfully accounts for the observed matter-antimatter asymmetry while remaining consistent with low-energy neutrino data.

%%%%%%%%%%%%%%%%
\section{Majoron Dark Mater}
\label{sec:DM}
%%%%%%%%%%%%%%%%

When the RHN masses are generated via spontaneous symmetry breaking (SSB) of a scalar field, there are two distinct possibilities: If the broken symmetry is local, the angular mode of the complex field is eaten by the gauge boson, leaving no light field at low energies. If instead the symmetry is global, the massless angular mode becomes a physical Nambu-Goldstone boson. The latter case corresponds to the original formulation of the majoron~\cite{Chikashige:1980ui, Chikashige:1980qk, Schechter:1981cv}.\footnote{See also Ref.~\cite{Gelmini:1980re} for the original majoron in the Type-II seesaw mechanism.}

In the model presented here, both mechanisms are realized. The \gx charges of the two scalar fields are coprimes, resulting in a renormalizable potential with an accidental global Abelian symmetry, which can be identified with \BL (cf. Tab.~\ref{tab:fields}). 
In this scenario, a linear combination of the two angular fields is eaten by the gauge boson, while the orthogonal component remains as a physical Nambu-Goldstone boson. However, we expect the accidental symmetry to be broken by gravitational effects. This breaking can generically be parametrized as\footnote{We estimate the couplings using naive dimensional analysis for strongly coupled theories, see~\cite{Cohen:1997rt} and Appendix~A of~\cite{Martucci:2024trp}, as well as references therein. As a word of caution, quantum gravity may provide mechanisms that enhance the quality of global symmetries beyond naive expectations. See, for example,~\cite{Petrossian-Byrne:2025mto}.}
\begin{align} \label{eq:explicit_breaking_term}
     V_\textrm{grav.}&= (4\pi)^2\frac{\eta}{M_\textrm{Pl}^{|s|+|t|-4}}\phi_1^{[s]} \phi_2^{[t]} + \textrm{H.c.} \, , \\
    \text{where}& \quad \phi_i^{[s]} = \begin{cases}
        \phi_i^{|s|} & \text{for } s\geq 0, \nonumber \\
        (\phi_i^\ast)^{|s|} & \text{for } s< 0,
    \end{cases}
\end{align}
with $M_\textrm{Pl}=2.435\times10^{18}\,$GeV and $ \eta $ expected to be an $ \mathcal{O}(1) $ coupling constant.
Without any protection mechanism, the leading breaking would typically appear at dimension 5 (or even at the renormalizable level), as in some early studies of the majoron~\cite{Akhmedov:1992hi,Cline:1993ht}. In our case, the gauge symmetry forbids gravitationally induced operators up to a certain order. This type of protection against gravitational corrections is well-known in axion models~\cite{Barr:1992qq}, and has previously been employed for majorons in the context of gauge $B-L$ symmetries~\cite{Rothstein:1992rh}.

\begin{table}[t]
    \centering
    \begin{tabular}{|*{2}{l}|*{11}{c}|}
    \hline
    &  &  &  &  &  & & $b$  &  &  &  & &   \\
    && -5 & -4 & -3 & -2 & -1 & 0 & 1 & 2 & 3 & 4 & 5 \\
    \hline
    &-5 & \cellcolor{gray!50} & \cellcolor{blue!25}15 & \cellcolor{gray!50} & \cellcolor{blue!25}17 & \cellcolor{gray!50} & \cellcolor{blue!25}19 & \cellcolor{gray!50} & \cellcolor{gray!50} & \cellcolor{gray!50} & 23 & \cellcolor{gray!50} \\
    &-4 & \cellcolor{green!25}7 & \cellcolor{gray!50} & \cellcolor{green!25}6 & \cellcolor{blue!25}13 & \cellcolor{blue!25}7 & \cellcolor{blue!25}15 & 8 & 17 & 9 & 19 & \cellcolor{blue!25}10 \\
    &-3 & \cellcolor{gray!50} & \cellcolor{blue!25}11 & \cellcolor{gray!50} & \cellcolor{blue!25}9 & \cellcolor{gray!50} & \cellcolor{blue!25}11 & \cellcolor{gray!50} & 13 & \cellcolor{gray!50} & \cellcolor{blue!25}15 & \cellcolor{gray!50} \\
    &-2 & \cellcolor{blue!25}7 & \cellcolor{blue!25}11 & \cellcolor{brown!60}4 & \cellcolor{gray!50} & \cellcolor{brown!60}3 & \cellcolor{blue!25}7 & \cellcolor{brown!60}4 & \cellcolor{blue!25}9 & \cellcolor{gray!50} & \cellcolor{blue!25}13 & \cellcolor{blue!25}8 \\
    &-1 & \cellcolor{gray!50} & \cellcolor{blue!25}11 & \cellcolor{gray!50} & \cellcolor{blue!25}5 & \cellcolor{gray!50} &  \cellcolor{black!70} & \cellcolor{gray!50} & \cellcolor{blue!25}7 & \cellcolor{gray!50}\phantom{13} & 13 & \cellcolor{gray!50} \\
    $a$&0 & 7 & 11 & \cellcolor{brown!60}4 & \cellcolor{blue!25}5 &  \cellcolor{black!70}\phantom{13}& \cellcolor{black!70} & \cellcolor{black!70}\phantom{13} & \cellcolor{blue!25}7 & \cellcolor{blue!25}5 & 13 & 8 \\
    &1 & \cellcolor{gray!50} & \cellcolor{blue!25}11 & \cellcolor{gray!50} & \cellcolor{blue!25}7 & \cellcolor{gray!50} & \cellcolor{blue!25}5 & \cellcolor{gray!50} & \cellcolor{blue!25}7 & \cellcolor{gray!50} & \cellcolor{blue!25}13 & \cellcolor{gray!50} \\
    &2 & \cellcolor{gray!50} & \cellcolor{blue!25}13 & \cellcolor{blue!25}6 & 11 & \cellcolor{blue!25}5 & \cellcolor{blue!25}9 & \cellcolor{brown!60}4 & \cellcolor{gray!50} & \cellcolor{green!25}5 & \cellcolor{blue!25}13 & \cellcolor{blue!25}8 \\
    &3 & \cellcolor{gray!50} & 17 & \cellcolor{gray!50} & \cellcolor{gray!50} & \cellcolor{gray!50} & \cellcolor{blue!25}13 & \cellcolor{gray!50} & \cellcolor{blue!25}11 & \cellcolor{gray!50} & \cellcolor{blue!25}13 & \cellcolor{gray!50} \\
    &4 & 11 & 21 & 10 & 19 & \cellcolor{blue!25}9 & \cellcolor{blue!25}17 & \cellcolor{blue!25}8 & \cellcolor{blue!25}15 & \cellcolor{green!25}7 & \cellcolor{gray!50} & \cellcolor{green!25}8 \\
    &5 & \cellcolor{gray!50} & 25 & \cellcolor{gray!50} & 23 & \cellcolor{gray!50} & \cellcolor{blue!25}21 & \cellcolor{gray!50} & \cellcolor{blue!25}19 & \cellcolor{gray!50} & \cellcolor{blue!25}17 & \cellcolor{gray!50} \\
    \hline
    \end{tabular}
    \caption{
    Values for $|s|+|t|$, which determine the dimension of the $ V_\mathrm{grav.} $ term in Eq.~\eqref{eq:explicit_breaking_term}, for various \gx charge assignments specified by Eq.~\eqref{eq:m_and_n}; see also Eq.~\eqref{eq:combined} and Section~\ref{sec:majoron}.
    The blue (green) entries indicate that local (global) strings have $ N_W = 1$ and can dissipate domain walls, preventing overclosure of the universe, as discussed in Section~\ref{sec:topological}.  
    The gray entries without numbers do not ensure exact proton stability, while the black entries are special cases discussed below Eq.~\eqref{eq:L_UV}. Finally, the brown entries with $|s|+|t|\leq4$ generate renormalizable breaking terms such that the majoron would not be a pNGB.}
    \label{tab:models}
\end{table}

%%%%%
\subsection{Majoron properties}
\label{sec:majoron}
%%%%%

In models with two scalars, one linear combination of angular modes is eaten by the gauge boson, while the orthogonal one remains massless. Following SSB, the mixing term $ \partial^\mu a Z'_\mu$ must vanish for the mass eigenstate. For details, we refer to Appendix~\ref{app:Majoron}, summarizing here only the key results. In particular, the scale associated with the Nambu--Goldstone boson $a$ in Eq.~\eqref{eq:a_in_phi} is given by
\begin{equation}
    f_a = \frac{v_1 v_2}{v_X } \mathcal{X}_1 \mathcal{X}_2 \equiv \tfrac{1}{2} v_X s_{2\varphi} \, ,
\end{equation}
where the rotation angle $\varphi $ determines the orthogonal matrix diagonalizing the angular modes~(\ref{eq:rotation}--\ref{eq:angles}), and $v_X \equiv \sqrt{\mathcal{X}_1^2 v_1^2 + v_2^2 \mathcal{X}^2_2}$, with $\mathcal{X}_{1,2}$ denoting the charges of $\phi_1$ and $\phi_2$ under the \gx symmetry.\footnote{This expression is analogous to that of the original Weinberg-Wilczek model \cite{Weinberg:1977ma, Wilczek:1977pj}, where two $\mathrm{SU}(2)_L$ doublets are replaced by two singlets.}

In the $v_2 \gg v_1$ limit explored in this work, the two bosons are approximately diagonal: $a_1 \simeq a$ corresponds to the physical majoron, while $a_2 \simeq \varphi_X$ is the Nambu-Goldstone boson absorbed by the $Z'$. Moreover, the majoron scale simplifies to $f_a \simeq \mathcal{X}_1 v_1$, with the dependence on the charges $\mathcal{X}_{1,2}$ dropping out of its couplings with neutrinos. 

\paragraph{Majoron mass} The gravitational term~\eqref{eq:explicit_breaking_term} gives the majoron a mass through the explicit breaking of \BL while respecting the \gx gauge symmetry. The leading breaking term minimizes $|s|+|t| > 0$. The constraint from $\mathrm{U}(1)_{X_p}$–invariance is 
    \begin{equation}
    0 = \tfrac{1}{3} [\phi_1^{[s]}\, \phi_2^{[t]}]_{X_p}
    = 2(3a+1)(s+t) + (3b+1)(s- 2t) \, ,
    \end{equation}
noting that $ 3a +1 $ and $ 3b+1 $ are coprime. 
The smallest non-zero solution to this homogeneous linear Diophantine equation is\footnote{The solution with opposite sign produces a term related by complex conjugation.}
    \begin{align}
    (s,\, t) &= k_{b}(2b-2a,\; b+1+2a) =\frac{k_{b}}{9}(-\X_2, \,\X_1),\nonumber
    \\ &\textrm{with} \quad
    k_{b} =
    \begin{cases}
    1 & b \in 2\mathbb{Z}\\
    \tfrac{1}{2} & b \in 2\mathbb{Z}+1
    \end{cases} . 
    \label{eq:combined}
    \end{align}
This gives the exponents for the leading $ V_\mathrm{grav.} $ term. Any other \BL--breaking term will have an integer multiple of these field exponents (times \BL--preserving factors). 
For values of $a,\, b \in [-5,\,5]$ the dimension of non-renormalizable operators ranges from $|s|+|t|\in [5, \,25]$ with the full set of possible combinations shown in Tab.~\ref{tab:models}. 

The mass of the majoron can now be computed by writing the complex fields in the polar form $\phi_i \sim \tfrac{1}{\sqrt{2}} v_i e^{i a_i/v_i} $ and diagonalizing the components. This leads to a potential for the majoron of the form
\begin{align}
    &V_{\textrm{grav}}= \abs{\eta}(4\pi M_{\textrm{Pl}}^2)^2 \times\nonumber\\ & \left( \frac{v_1}{\sqrt{2}M_{\textrm{Pl}}} \right)^{\!\! |s|} \left( \frac{v_2}{\sqrt{2}M_{\textrm{Pl}}} \right)^{\!\!|t|} \cos \! \left(\alpha_\eta -9\frac{ st  }{k_{b} }\frac{a}{f_a} \right) \, ,
\end{align}
where the would-be nambu-Goldstone boson $\varphi_X$ drops out, and $\alpha_\eta = \arg \eta $ is a generic complex phase which can be set to $ \pi $ by using the remaining shift symmetry of the majoron. The mass term of the majoron is obtained by expanding the potential to second order around the minimum:
\begin{equation}
    m_a^2= 8\pi^2 \abs{\eta} M_\textrm{Pl}^2 \frac{v_1^{|s|-2} v_2^{|t|}}{ (\sqrt{2}M_\textrm{Pl})^{|s|+|t|-2}} \! \left(s^2+  \dfrac{v_1^2}{v_2^2} t^2\right).
\end{equation}
Thus, for each model with $(a,\, b)$ fixed, the mass can be obtained as a function of the VEVs and Eq.~\eqref{eq:combined}. As discussed later, a phenomenologically viable majoron must be sufficiently light, which necessitates a large enough $|s|+|t|$.

\paragraph{Majoron couplings}
In the traditional majoron model~\cite{Chikashige:1980ui, Chikashige:1980qk, Schechter:1981cv}, the majoron couples to right-handed neutrinos with interactions aligned to their mass matrix, yielding couplings proportional to $ M_N$, that is, diagonal in the mass basis. However, in models where the majoron arises from multiple fields or the scalar fields carry non-universal charges~\cite{Heeck:2017xbu,deGiorgi:2023tvn}, this alignment is generally no longer true. In Appendix~\ref{app:Majoron} we derive the most minimal expressions for the tree-level couplings to the light and heavy neutrinos. 
As a simplified, order-of-magnitude approximation of the full expressions in Eqs.~\eqref{eq:RHNLag2} and \eqref{eq:alightnu}, the flavor-conserving majoron couplings scale as $\mathcal{O}(m_\nu / v_1)$ for light neutrinos, and as $\mathcal{O}(M_{N_1} / v_1)$ for heavy ones.

Finally, the majoron also couples to SM particles via one- and two-loop radiative corrections~\cite{Heeck:2019guh}. These couplings can be matched onto the generic ALP Lagrangian, where we focus only on the phenomenologically relevant interactions, with the electron and the photon, defined as:
\begin{equation}
    \mathcal{L} = a\left(ig_{ae} \bar{e} \gamma_5 e + g_{a\gamma\gamma}F\tilde{F}\right)\,,
\end{equation}
with 
    \begin{align}
    g_{ae}& \simeq\frac{m_e}{16\pi^2 v_1} \left(\Tr K-2K_{ee}\right) \, , \\
     g_{a\gamma\gamma}&\simeq-\dfrac{\alpha_\text{em}}{8\pi^3 v_1} \left[\Tr K\sum\limits_f N_c^f Q_f^2 T_3^f\, h\! 
    \left( \dfrac{m_a^2}{4m_f^2} \right) +\right.\nonumber \\ &\qquad\qquad\qquad\qquad\left. \sum \limits_{\ell=e, \mu, \tau}K_{\ell\ell}\, h\! \left( \dfrac{m_a^2}{4m_\ell^2} \right) \right]\,,
    \label{eq:gagaga1}   
    \end{align}
where $ Q_f $ is the electric charge of fermion $ f $. 
The matrix $K$ is defined as the product of the Dirac Yukawas
\begin{equation}
    K\equiv Y_N Y^\dagger_N\, ,
\end{equation}
while the loop function $h(x)$ is defined as 
\begin{align}
    \label{eq:2loop}
    h(x)\equiv -\dfrac{1}{4x}&\left[\log\left(1-2x+2\sqrt{x(x-1)}\right)\right]^2-1 \nonumber \\ &\simeq \begin{cases}
        \, \dfrac{x}{3} & \text{for } x \to 0 \\
        \, -1 & \text{for } x \to \infty
    \end{cases}  \, .
\end{align}
Although the combination $v^2 Y_N^\dagger Y_N / v_1 \sim m_\nu$ is related to the light neutrino masses, the mismatch $v_1\neq M_N$ precludes a straightforward analytical matching. Consequently, these couplings will be computed numerically in the subsequent analysis.

%%%%%
\subsection{Thermal production}
\label{sec:therm}
%%%%%

The majoron interacts very weakly with the SM fields, with all its couplings suppressed by neutrino masses and loop effects. However, if the reheating temperature is sufficiently high to bring the RHNs into thermal equilibrium with the SM bath, the majoron can also be thermalized. This thermalization can occur via interactions with the heavy RHNs or through couplings to the radial modes of the complex scalar fields, provided the temperature is high enough.

\begin{figure*}[t]
    \centering
    \includegraphics[width=0.49\linewidth]{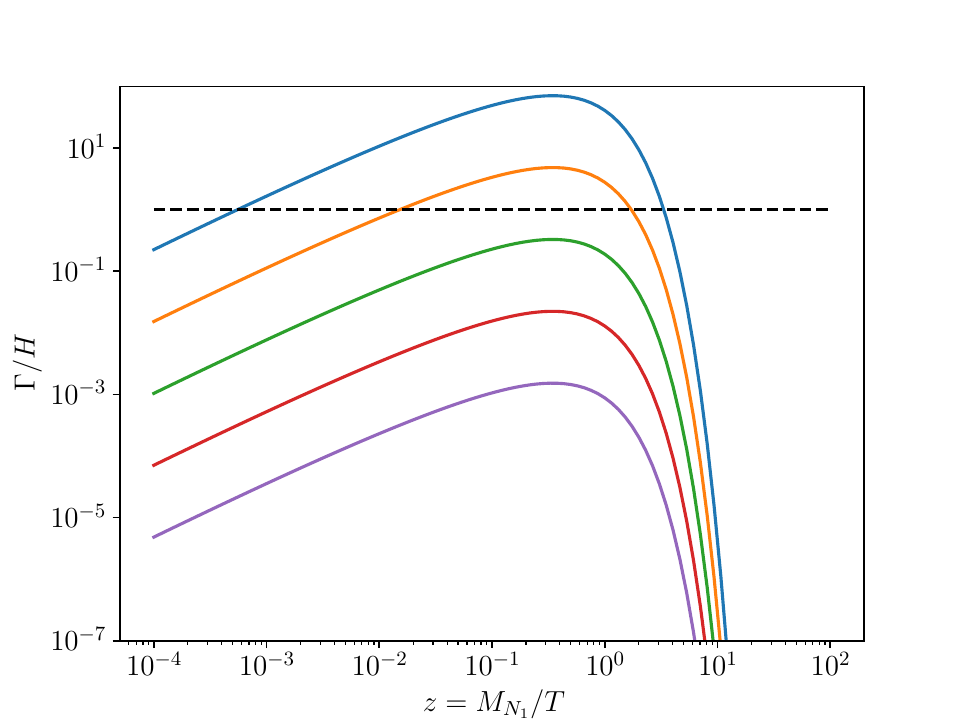} \includegraphics[width=0.5\linewidth]{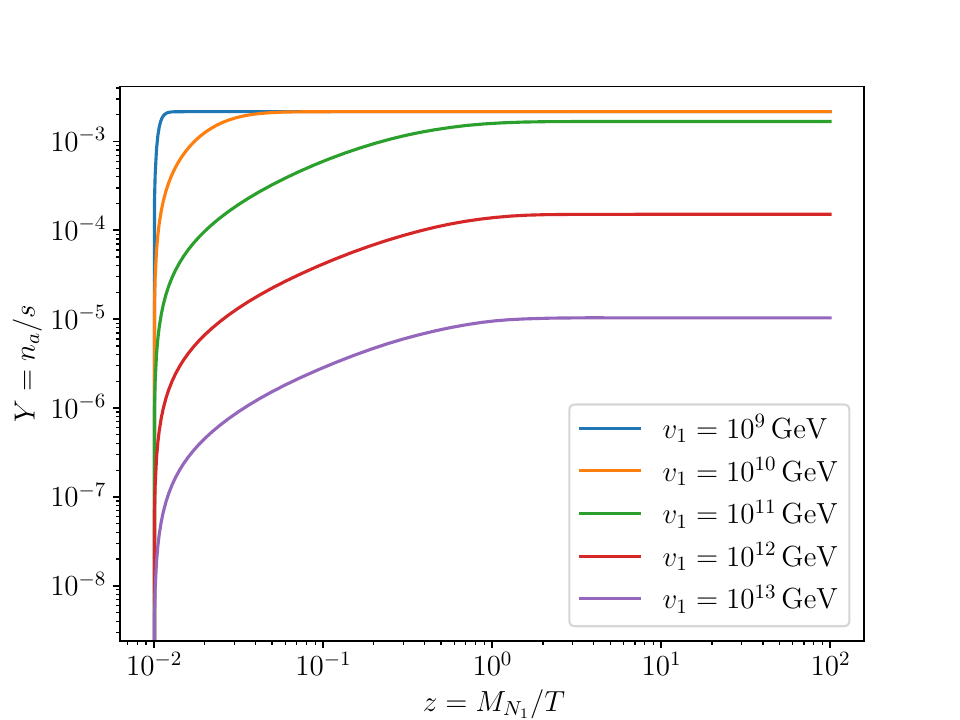}
    \caption{$\Gamma/H$ of the $t$-channel $N_1\, N_1\to a \, a$ (\textit{left}) and the freeze-in yield as a function of $z=M_{N_1}/T$ (\textit{right}), where we assume that only the lightest RHN is in thermal equilibrium. In both figures, we assume $v_1/v_2=0.1$.}
    \label{fig:majoron_freezein}
\end{figure*}

In such a scenario, the majoron decouples relativistically once the temperature drops below the mass of the heavy mediators. The resulting relic abundance is given by its mass~\cite{Akhmedov:1992hi,Reig:2019sok}:
\begin{align}
    \Omega_\textrm{fo} h^2 = \frac{\zeta(3) g_s(T_0) m_a T_0^3}{\pi^2 g_s(T_\textrm{dec}) \rho_c} h^2
    = 0.12\frac{m_a}{166.62 \, \textrm{eV}} \frac{106.75}{g_s(T_\textrm{dec})} ,
\end{align}
where $g_s(T)$ is the effective number of entropy degrees of freedom at temperature $T$, while $ \rho_c$ and $T_0$ are the critical energy density and temperature today, respectively. 

Since particles with such low masses remain relativistic during structure formation, they would erase small-scale structures. This effect imposes a lower bound on the majoron mass of $m_a \gtrsim 5\,\textrm{keV}$, as inferred from analyses of the Lyman-$\alpha$ forest and other small-scale structure observations~\cite{Baur:2015jsy, Yeche:2017upn, Baur:2017stq, Irsic:2017ixq}.\footnote{This bound can vary depending on the specific production mechanism~\cite{Heeck:2017xbu, Boulebnane:2017fxw}.} A consistent scenario requires the thermal contribution to the dark matter relic abundance, $\Omega_\textrm{fo} h^2$, to be subdominant, at most a few percent of the total abundance, thereby placing an upper bound on $m_a$ (cf. Fig.~\ref{fig:light_majoron}). For $m_a \lesssim 1\,\textrm{eV}$, the thermal relic behaves
like dark radiation and contributes to $N_{\rm eff}$.
In conclusion, if majorons thermalize in the early universe, the observed dark matter density must be explained by a non-thermal production mechanism.

The minimal leptogenesis scenario assumes a mass hierarchy among the RHNs with $M_{N_{2,3}} \gg M_{N_1}$. A consistent thermal history may involve a reheating temperature sufficient to populate only the lightest right-handed neutrino, $N_1$, while neither restoring the symmetry nor bringing the heavier states into equilibrium. Under these conditions, the majoron may never reach thermal equilibrium with the SM plasma, and heavier majoron masses become viable without violating structure formation constraints. This is illustrated in the left panel of Fig.~\ref{fig:majoron_freezein}, where it is shown that for VEVs above $\sim 10^{11}\,\textrm{GeV}$, the process $N_1 N_1 \to a a$ does not suffice to thermalize the majoron~\cite{Gu:2009hn}.

As discussed earlier, the lightest RHN tends to thermalize with the SM plasma, either due to interactions with a $Z'$ gauge boson (Fig.~\ref{fig:GammaH_ratio_Zprime}) or due to the strong washout regime (Fig.~\ref{fig:leptogenesis_scan}). Even if its interactions are too weak to thermalize the majoron, they can still lead to its production via the \textit{freeze-in} mechanism~\cite{Frigerio:2011in,Boulebnane:2017fxw}. The corresponding yield $Y \equiv n_a / s$ (the ratio of majoron number density to entropy density) can be computed by solving the Boltzmann equations and is shown in the right panel of Fig.~\ref{fig:majoron_freezein}, for an effective Yukawa coupling of $M_{N_1}/v_1 \simeq v_1/v_2 = 0.1$. The present-day dark matter abundance is given by:
\begin{equation}
    \Omega_\textrm{DM} \simeq \frac{2\pi^2 g_\ast(T_0) T_0^3 m_a}{45 \rho_c} Y(T_\textrm{dec}) \,.
\end{equation}
This provides a consistent thermal production mechanism within the parameter space where the majoron mass lies above the structure formation bound ($m_a \gtrsim 5\,\textrm{keV}$) and extends up to $m_a \lesssim 1\,\text{MeV}$, beyond which additional constraints kick in (cf. Figs.~\ref{fig:heavier_majoron} and~\ref{fig:photon_electron}).

As a final remark, Majorons can also be produced via the decays of the heavier right-handed neutrinos, $N_{2,3} \to N_1 a$ \cite{Heeck:2017xbu, Boulebnane:2017fxw}. In this model, the coupling responsible for off-diagonal transitions is relatively large; however, for light majorons with $m_a \lesssim 1\,\text{eV}$, the resulting contribution is subleading. For heavier majorons, as discussed earlier, the $N_{2,3}$ states must not be in thermal equilibrium with the SM bath, and thus this production channel is inactive.

%%%%%
\subsection{Non-thermal production}
\label{sec:nontherm}
%%%%%

As discussed in the previous section, for sub-keV majorons, thermal production can not account for the observed dark matter abundance, necessitating alternative non-thermal production mechanisms. For light pNGBs, the misalignment mechanism, originally proposed in the context of the QCD axion~\cite{Abbott:1982af, Dine:1982ah, Preskill:1982cy}, offers a compelling way to generate a cold population of majorons. This mechanism has since been generalized to arbitrary axion-like particles (ALPs)~\cite{Arias:2012az, Blinov:2019rhb} and applied in the context of majorons as well~\cite{Rothstein:1992rh, Reig:2019sok, Biggio:2023gtm, Chun:2023eqc,Liang:2024vnd}.

The misalignment mechanism involves solving the equation of motion for a massive angular field, $\theta \equiv a/f_a$, in an expanding universe:
\begin{equation}
    \label{eq:eom_misalignment}
    \ddot\theta + 3H\dot\theta + m_a^2(t) \theta \simeq 0\, ,
\end{equation}
where the potential has been expanded to leading order. At early times, when the Hubble parameter $H$ dominates over the mass term, the field is overdamped and remains frozen at an initial value $\theta_i$. As the universe cools, the mass becomes comparable to the Hubble rate, $m_a \simeq H(T_\text{osc})$, and the field begins to oscillate coherently around the minimum of its potential. These oscillations redshift as non-relativistic matter and contribute to the present-day dark matter abundance. 

Assuming the majoron mass does not receive significant thermal corrections, i.e., $m_a(t) = m_a$, and that oscillations commence during radiation domination, the relic abundance from misalignment is approximately:
\begin{align}
\Omega_\textrm{DM} h^2 &\simeq \nonumber\\ &0.12 \left( \frac{\theta_i f_a}{1.9\times 10^{13}} \right)^{\!\! 2} \left(\frac{m_a}{1 \upmu   \textrm{eV}}\right)^{\!\! 1/2} \left( \frac{90}{g_\ast(T_\textrm{osc})} \right)^{\!\! 1/4} \, .
\end{align}
The initial misalignment angle $\theta_i$ depends on the cosmological history. If the global symmetry is broken before inflation and never restored afterward, inflation selects a homogeneous patch of the universe, allowing $\theta_i$ to take any value in the interval $[-\pi, \pi]$. In this pre-inflationary breaking scenario, the presence of isocurvature perturbations in the cosmic microwave background \cite{Crotty:2003rz, Beltran:2005xd, Beltran:2006sq} imposes stringent constraints on the inflationary scale, which must be sufficiently low to evade current bounds.

Conversely, if the symmetry is broken after inflation, the universe is populated with uncorrelated patches of different $\theta$, leading to an ``average'' value $\theta_i = \sqrt{\langle \theta^2 \rangle} \simeq \pi/\sqrt{3}$. See Refs.~\cite{GrillidiCortona:2015jxo, DiLuzio:2020wdo} for a detailed discussion in the context of axions. In this post-inflationary scenario, topological defects, particularly cosmic strings and domain walls, form. Dark matter production from topological defects serves as an efficient mechanism, yielding a contribution comparable to that from coherent oscillations~\cite{Reig:2019sok}. In addition, domain walls can come to dominate the universe’s energy density~\cite{Barr:1992qq, Rothstein:1992rh} and must decay to ensure a consistent cosmological history, thereby imposing additional constraints. In the next section, we explore the formation of topological defects in this class of models.

%%%%%%
\section{Topological defects}
\label{sec:topological}
%%%%%%
Topological defects may form during the stages of symmetry-breaking involving the $ \phi_a $ fields. The breaking of \gx will give rise to cosmic strings, in which the phases of the scalars form a non-trivial vortex string through the universe (see, e.g, \cite{Hindmarsh:1994re}). After that, the slight bias introduced to the majoron vacuum by $ V_\mathrm{grav.}$ produces multiple degenerate discrete vacua between which domain walls may form. The domain walls may come to catastrophically dominate the energy density of the universe, which is known from axion physics as the domain wall problem~\cite{DiLuzio:2020wdo, Sikivie:2006ni}.   
If the topological defects form after inflation, they may persist in the universe and dramatically change the cosmological history unless the network dissipates efficiently. Our concrete scenario involving two scalar fields closely resembles the models~\cite{Barr:1992qq, Rothstein:1992rh}.

\subsection{Cosmic strings}
The formation of cosmic strings depends on the exact symmetry-breaking history.
The case $ v_2 \gg v_1 $ seems particularly promising for leptogenesis scenarios. Here we assume that the hierarchy is large enough (and the dynamics right) that we can consider the phase transition from the development of the $ \phi_2 $ VEV as isolated from the subsequent phase transition from the development of the $ \phi_1 $ VEV. In this event, the symmetry breaking proceeds as 
    \begin{equation}
    \U(1)_{X_p} \xrightarrow[\langle \phi_2 \rangle]{} \Gamma' \xrightarrow[\langle \phi_1 \rangle]{} \Gamma
    \end{equation}
where the intermediate stability group $ \Gamma' \simeq \mathbb{Z}_{\mathcal{X}_2} $ is determined by the charge of $ \phi_2 $.
With the $ \phi_2 $ phase transition, we expect the formation of \emph{local} strings involving the $ X_\mu $ and $ \phi_2 $ fields. Following the arguments of Appendix~\ref{sec:senario_simultaneous_breaking}, the long distance scalar field behaves as $ \bar{\phi}_2 = \rho_2[g(\theta)] v_2 $, $ \rho_a(g) $ being the representation of a \gx element acting on $\phi_a$,  in cylindrical coordinates $ (r,\theta)$ with boundary condition 
	\begin{equation}
	g(2\pi) \in \Gamma' \implies g(2\pi)= e^{2\pi i\, w_2/\mathcal{X}_2}.
	\end{equation}
More precisely, we let $ w_2 $ be the $ \phi_2 $ winding number of the string if and only if we can smoothly deform $ g(\theta) \to e^{i\,\theta w_2/ \mathcal{X}_2} $ for $ w_2\neq 0 $.

So far, $ \phi_1 $ has been a spectator, but as the temperature continues to drop, we expect a second phase transition with the formation of the $ \langle \phi_1 \rangle $ VEV. This has the effect of both modifying the solution to the \emph{local} strings already present after $ \phi_2 $ breaking and giving rise to new \emph{global} $ \phi_1 $ strings. Let us begin by introducing $ \phi_1 $ defects to the local strings. The long distance solution $ \bar{\phi}_1(\theta) $ of $ \phi_1 $ should coincide with its vacuum value, i.e., $ |\bar{\phi}_1| = v_1 $. The gauge field wrapped around the string now ends up dragging the phase of $ \bar{\phi}_1 $ in such a way that it may be impossible to eliminate its kinetic energy. We discriminate the solutions by their topology, i.e., by the winding number $ w_1 $ of the $ \phi_1 $ field around the local string: $ \bar{\phi}_1(\theta) \sim \tfrac{1}{\sqrt{2}} v_1 e^{i \theta w_1}$. The kinetic energy of the long-range field is 
\begin{equation}
\begin{split}
	\mathcal{E}_\mathrm{kin} \sim  \int \!\! \dd^2 x \, |D_i \bar{\phi}_1|^2 &=   \int \!\! \dd^2 x \, |\partial_i \bar{\phi}'_1|^2,  \\ \bar{\phi}'_1(\theta) &= \rho_1[g^{\eminus 1}(\theta)] \bar{\phi}_1(\theta).
\end{split}
\end{equation}
The phase of the $ \bar{\phi}_1' $ field is equivalent to (can be continuously transformed into) 
    \begin{equation}
    \bar{\phi}'_1(\theta) \sim \tfrac{1}{\sqrt{2}} v_1 e^{i\theta(w_1 - w_2 \mathcal{X}_1 / \mathcal{X}_2)} , 
    \end{equation}
and it is readily apparent that the minimal-energy configuration is the configuration such that 
    \begin{equation}
    \abs{w_1 \mathcal{X}_2 - w_2 \mathcal{X}_1} = \min_{w\in \mathbb{Z}} \abs{w \mathcal{X}_2 - w_2 \mathcal{X}_1}. 
    \end{equation}
The minimal configuration will try to match the phase shift from the gauge field configuration as closely as possible; however, the resulting kinetic energy is non-zero---in fact, logarithmically divergent---when $ g(2\pi) \in \Gamma' \setminus \Gamma $. With formula~\eqref{eq:combined} for the scalar charges, the minimization condition can alternatively be cast as 
    \begin{equation}
    \abs{s w_1 + t w_2} = \min_{w\in \mathbb{Z}} \abs{s w + t w_2 }
    \end{equation}
in terms of the powers of the fields in the $\U(1)_{B-L}$--breaking potential term~\eqref{eq:explicit_breaking_term}. 

In addition to the modification of the local strings, \emph{global} $ \phi_1 $ strings may also form. Assuming that the $ X_\mu $ field is too heavy (from the $ v_2 $ VEV) to be part of the string solution, the $ \phi_1 $ field may produce string solutions only involving itself ($w_2 = 0 $). The configuration will have 
    \begin{equation}
    \bar{\phi}_1(\theta) = \tfrac{1}{\sqrt{2}} v_1 e^{i\theta w_1}, \qquad  w_1 \in \mathbb{Z} , 
    \end{equation}
with a minimal string energy for winding number $ w_1 = \pm 1 $.

\subsection{Domain walls}
Eventually, as the temperature continues to decrease after the formation of the strings, the explicit breaking of $ \U(1)_{B-L} $ from the term~\eqref{eq:explicit_breaking_term} becomes relevant: it breaks the phase degeneracy of the vacuum solutions. The long-distance fields from the string solutions receive an additional potential contribution to their energy density proportional to
    \begin{equation}
    V_\mathrm{grav.}\!(\bar{\phi}_{1},\, \bar{\phi}_{2}) \to V_\mathrm{grav.}\!(v_1, v_2) \cos\!\left(s w_1\, \theta + t w_2\, \theta  \right).
    \end{equation} 
The potential term $ V_\mathrm{grav.} $ gives rise to the formation of $ N_W = \big|s w_1 + t w_2 \big|  $ degenerate minima around strings with the corresponding winding numbers separated by the same number of walls (the topology of the interpolating field ensures that there is a wall even when there is a single, unique minimum: $ N_W = 1$). 
For the minimal strings we have~\cite{Barr:1992qq,Rothstein:1992rh} 
    \begin{itemize}
    \item global strings: $ N_W = |s| $,
    \item local strings: $ N_W = \min_{w\in \mathbb{Z}} \abs{s w + t } $.
    \end{itemize}

Cosmic string--wall networks with $ N_W> 1 $ will generally not decay and would dramatically impact the cosmological history. By contrast, strings with $ N_W =1 $ can effectively cut up string-wall networks as they collide with other walls and strings. They experience a force towards the one wall attached to them, causing them to travel in that direction while unzipping the wall~\cite{Barr:1992qq, Rothstein:1992rh, Vilenkin:1982ks}.

\subsection{Cosmological history} \label{sec:cosmic_history}
Depending on the details of inflation, there are various avenues to obtain a viable cosmological history. This is primarily dictated by $ T_\sscript{RH} $ and the inflation temperature $ T_\sscript{I} =\tfrac{1}{2\pi} H_\sscript{I} $, which is related to the Hubble rate during inflation. If these are both lower than a symmetry-breaking scale, the symmetry is never restored after or during inflation and must have been broken pre-inflation~\cite{Marsh:2015xka,Cirelli:2024ssz}. This leaves the following viable options: 
\begin{enumerate}[i)]
    \item $ \max(T_\sscript{RH}, T_\sscript{I}) \gtrsim v_2 $: Both scalar field VEVs are formed after inflation, and $ \theta_i $ may take different values in patches of the universe that are initially out of causal contact. This produces a network with both global and local strings, which can dissipate if either of the strings has $ N_W = 1 $. As shown in Tab.~\ref{tab:models}, it is common for models to allow for $ N_W = 1 $ strings.
    \item $ v_2 \gtrsim \max(T_\sscript{RH}, T_\sscript{I}) \gtrsim v_1 $: The $ \phi_2 $ VEV forms pre-inflation and the local strings are inflated away. By contrast, the global strings, formed alongside the $ \phi_1 $ VEV, populate the universe along with the domain walls. As shown in Tab.~\ref{tab:models}, only a minority of the model charges allow for global strings with $ N_W = 1 $ to dissipate the network.
    \item $ v_1 \gtrsim \max(T_\sscript{RH}, T_\sscript{I}) \gtrsim M_{N_1} $: All strings are inflated away and will be absent in our universe. Inflation selects a single patch of the pre-inflation universe, leading to a homogeneous value for $ \theta_i $ inside the Horizon. As a consequence, the axion field will roll into the same minimum everywhere per the misalignment mechanism, and no domain walls will form. This scenario does not impose any constraints on the charges of the model; however, quantum fluctuations of the fields during inflation generate isocurvature perturbation, which are strongly bounded by CMB, see for instance \cite{Crotty:2003rz,DiValentino:2014eea}. These constraints translate into a bound on the scale of inflation $H_I \lesssim 10^7\,\textrm{GeV} (f_a/10^{12}\,\textrm{GeV})$ \cite{DiLuzio:2020wdo}.
\end{enumerate}
If the reheating temperature is any lower than $ M_{N_1} $, one would not obtain a sizable population of RHNs as required in thermal leptogenesis.

An alternative solution to the domain wall problem, sometimes pursued in axion models~\cite{Sikivie:1982qv}, is to have a higher-order potential term introduce a bias to the degenerate vacua around the strings, introducing a pressure to the walls to collapse them. Without introducing new fields, the \gx symmetry generically limits higher-order $ \U(1)_{B-L}$ breaking terms in the potential to be powers of the lowest order term $ V_\mathrm{grav.} $ (or its conjugate). None of these terms can break the degeneracy, and the minimal model must thus rely on strings with $ N_W = 1$ to remove the topological defects.     

Topological defects discussed in this section can generate a stochastic gravitational wave background, potentially detectable by present and future interferometers~\cite{NANOGrav:2023hvm, Auclair:2019wcv}. A detailed analysis of this signal is left for future work.

\begin{figure*}[t]
    \centering
    \includegraphics[width=\linewidth]{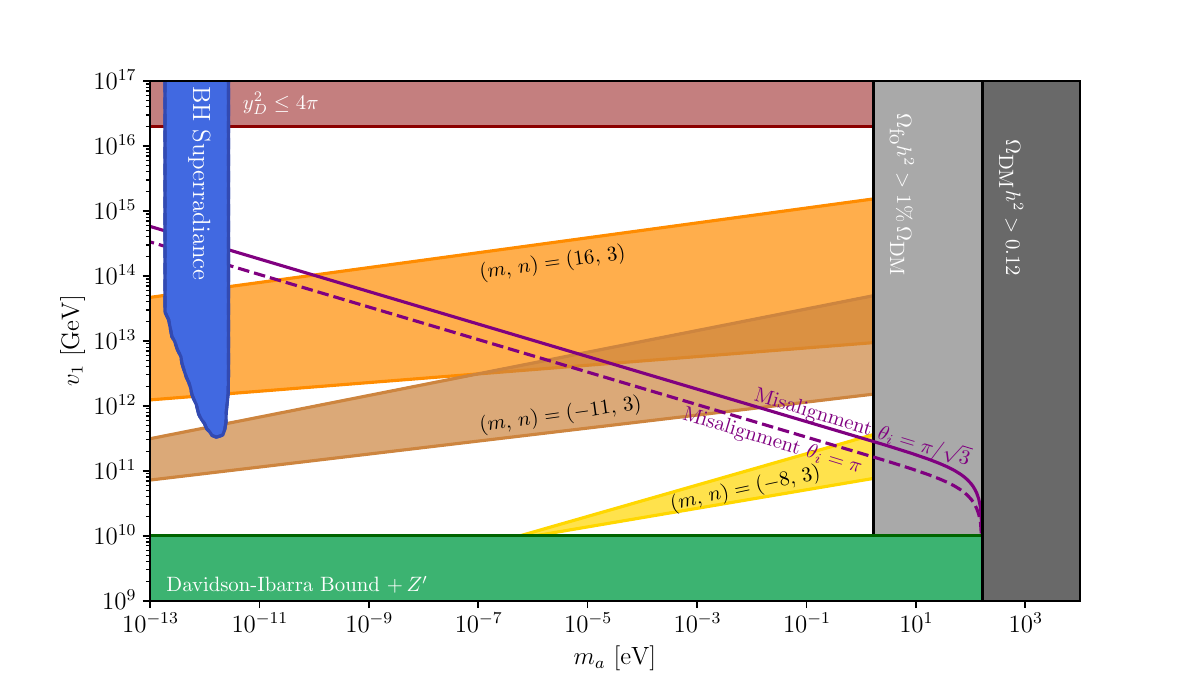}
    \caption{Parameter space allowed by perturbativity (red) and leptogenesis constraints (green). The dark-colored regions indicate where the relic abundance from relativistic freeze-out would exceed observational limits. Black hole superradiance constraints are shown in blue. Three representative models are highlighted in different parameter space regions, with their corresponding viable ranges. The region where misalignment saturates the DM relic abundance for different $\theta_i$ values is plotted in solid purple.}
    \label{fig:light_majoron}
\end{figure*}

\begin{figure*}[t]
    \centering
    \includegraphics[width=0.8\linewidth]{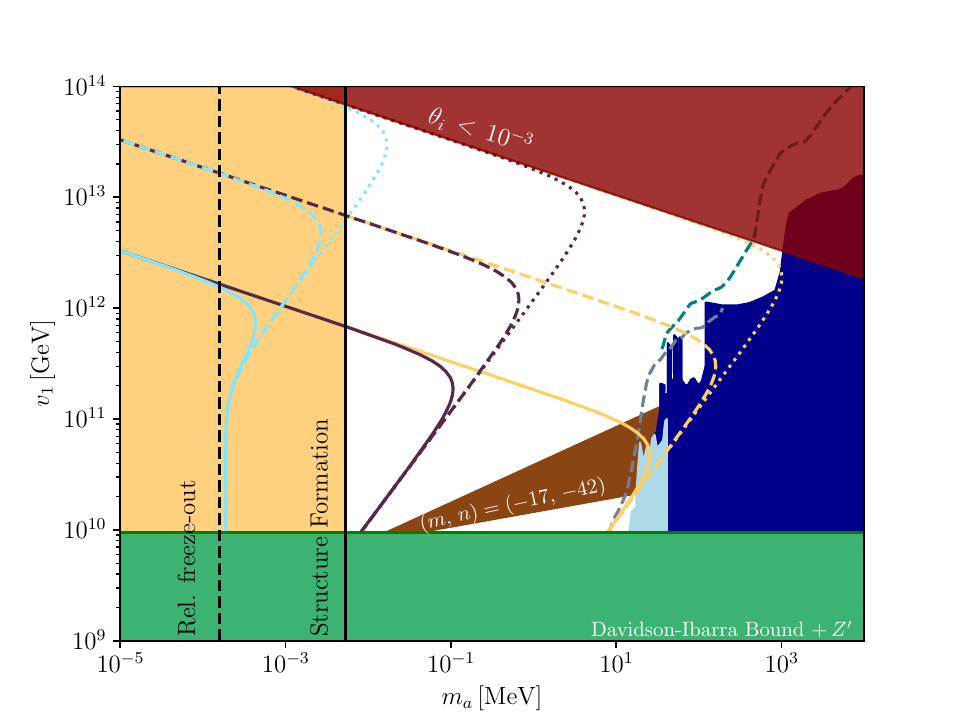}
    \caption{Parameter space of a majoron produced by freeze-in via the $t$-channel $N\, N\to a\, a$ and misalignment. The colored lines correspond to different values of $v_1/v_2=$ \textcolor[rgb]{1, 0.81263, 0.40424}{$0.001$}, \textcolor[rgb]{0.34056,0.16094, 0.28636}{$0.01$} , \textcolor[rgb]{0.50411,0.90708,0.99978}{$0.1$}. The solid, dashed, and dotted lines correspond to the values of the initial misalignment value of $\theta_i=0.1 ,\, 0.01, \, 0.001$ respectively. In yellow, we denote the region where a major thermal component of DM would be excluded by structure formation, and in green, the bound imposed by leptogenesis on $v_1$. Finally, several exclusion regions given by neutrino detectors are colored in blue.}
    \label{fig:heavier_majoron}
\end{figure*}

%%%%%%
\section{Summary: constraints and projections}
\label{sec:summary}
%%%%%
After considering all relevant aspects of the cosmic phenomenology, two distinct scenarios for the majoron emerge: a light and a heavy regime.

\paragraph{Light majoron regime}
The light majoron scenario is recapped in Fig.~\ref{fig:light_majoron}, where three representative models are shown in orange, brown, and yellow. The shaded bands indicate the range of $v_2$ values consistent with thermal leptogenesis. This corresponds to the requirement that $Z'$ interactions decouple before the temperature drops to $T = M_{N_1}$, and that $M_{N_1}$ satisfies the lower bound $M_{N_1}^\textrm{min} \gtrsim 4.9 \times 10^8\,\text{GeV}$—the minimum mass required for successful leptogenesis when the lightest RHN is in thermal equilibrium~\cite{Giudice:2003jh}. Imposing both conditions leads to a lower bound on the symmetry-breaking scale, $v_1 \gtrsim 10^{10}\,$GeV, shown in green. In the majoron mass-region $ m_a \lesssim \SI{5}{keV}$, only the misalignment mechanism discussed in Sect.~\ref{sec:nontherm} can set the correct DM relic abundance of majorons. This is plotted in purple in Fig.~\ref{fig:light_majoron}, where $\theta_i=\pi/\sqrt{3}$. 
In the pre-inflationary scenario (scenario \textit{iii} in Section~\ref{sec:cosmic_history}) where symmetry was never restored after inflation,  there is no prediction for $ \theta_i $, which takes a single value throughout the universe. In that event, $ \theta_i $ could be as large as $ \pi $ (or much lower), giving a wide region that can accommodate majoron DM.  

The gray region represents the parameter space where a significant thermal population of majorons would arise, and hence a bound of the parameter space if $T_\textrm{RH}>v_2$. However, in scenarios where the symmetry is never restored after inflation, the majoron does not thermalize, and these thermal relic bounds become irrelevant, allowing for masses beyond these bounds. Nevertheless, such a regime lies in a part of parameter space with poor detectability and would not give rise to observable topological defects, as they would be inflated away.\footnote{One may argue that requiring a reheating temperature $T_\sscript{RH}$ high enough to enable thermal leptogenesis but low enough to avoid majoron thermalization involves a degree of fine-tuning.}

As an upper limit on the VEV, the red region corresponds to the breakdown of perturbativity in the Dirac Yukawa couplings. Meanwhile, the blue area denotes constraints from black hole superradiance~\cite{Arvanitaki:2010sy, Arvanitaki:2014wva, Arvanitaki:2016qwi}, which place significant bounds on ultralight bosonic fields. In particular, this mechanism excludes most of the parameter space around $m_a \sim 10^{\eminus 12}\,$eV~\cite{Hoof:2024quk}.

Another test of this scenario comes from the measurement of relativistic degrees of freedom $N_\textrm{eff}$, which is modified from the SM value \cite{Cielo:2023bqp} if majorons thermalise in the early universe and $m_a \lesssim 1$\,eV. Future measurements of CMB \cite{CMB-S4:2016ple} will be able to see hints of this type of contribution at the 1$\sigma$ level \cite{DiLuzio:2020wdo}.

\paragraph{Heavy majoron regime}
As discussed in Section~\ref{sec:therm}, a heavier majoron can exist in the universe only if it was out of thermal equilibrium with the SM bath. In this scenario, a population of majorons can be produced by a combination of freeze-in and misalignment, as shown in Fig.~\ref{fig:heavier_majoron}. The different colored lines indicate different sizes of the effective Yukawa $M_{N_1}/v_1\simeq v_1/v_2$, while the different line styles indicate the value of the initial misalignment. One can see that it is easy to populate the full parameter space in $m_a-v_1$ by combining these two free parameters. While smaller values, such as $\theta_i < 10^{-3}$ (shown in red in Fig.~\ref{fig:heavier_majoron}), are possible, they may be considered fine-tuned in the absence of a deeper mechanism explaining their smallness~\cite{Wilczek:2004cr, Graham:2018jyp}.

Once again, in Fig.~\ref{fig:heavier_majoron}, the lower value of $v_1$ is set by the requirements that $Z'$ interactions decouple before $T=M_{N_1}$. The band in brown corresponds to a particular model, whose width is also determined by varying $v_2$. Finally, the parameter space of this scenario is bounded for large $m_a$ by neutrino telescopes~\cite{Garcia-Cely:2017oco, Akita:2023qiz}. In particular, the current bounds (from left to right) are set by Borexino~\cite{Borexino:2019wln}, Kamland~\cite{KamLAND:2021gvi}, and Super-Kamiokande~\cite{Palomares-Ruiz:2007egs, Olivares-DelCampo:2017feq, Super-Kamiokande:2011lwo, Super-Kamiokande:2013ufi, Super-Kamiokande:2021jaq} experiments. In the future, these bounds will be improved by JUNO~\cite{Akita:2022lit} and Hyper-Kamiokande~\cite{Bell:2020rkw}, shown in dashed gray and green lines, respectively. The yellow band to the left shows the lower bound $5.3\,$keV imposed by structure formation~\cite{Irsic:2017ixq}. However, the bound applies only if the fraction of DM produced via freeze-in is larger than a few percent; that is, if the dominant mechanism comes from misalignment, the bound is lifted. 

\begin{figure*}[t]
    \centering \includegraphics[width=0.9\linewidth]{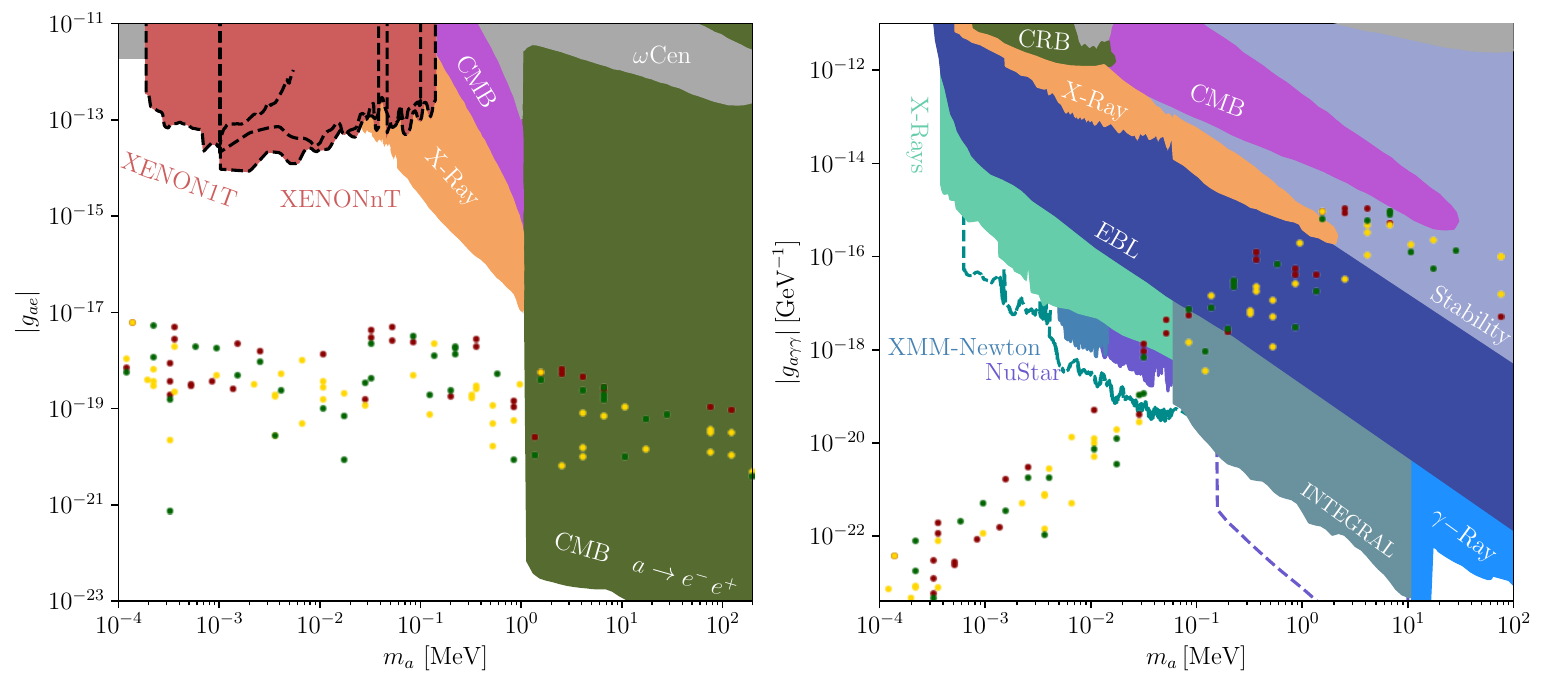}
    \caption{Bounds on majoron models to electron (\textit{left}) and photon (\textit{right}) couplings. The photon coupling is strongly bounded by satellite searches of $X-$ and $\gamma-$ray telescopes. At the same time, for electrons, the dominant bound arises from the CMB when the majoron can decay into a pair of electrons. The model predictions come from the parameter scan, which is compatible with leptogenesis, see Fig.~\ref{fig:leptogenesis_scan}. We aknowledge the use of Ref.~\cite{AxionLimits} for these figures.}
    \label{fig:photon_electron}
\end{figure*}

The $m_a \sim \mathcal{O}(\textrm{MeV})$ region is also widely explored for DM models with electron and photon couplings. In our case, these are generated at one- and two-loop level~\cite{Heeck:2019guh} and typically depend on $\sim Y_N Y_N^\dagger /f_a$. This quantity is related to the neutrino masses in a non-trivial way. Here, we use the points from the scan performed in Fig.~\ref{fig:leptogenesis_scan}, which are compatible with leptogenesis. The points for the different types of textures are plotted in Fig.~\ref{fig:photon_electron}, using the same color reference as in Fig.~\ref{fig:leptogenesis_scan}: we find no major differences between the different textures.

For majorons with $m_a\geq 2 m_e$, the CMB bounds decay into two electrons~\cite{Slatyer:2016qyl}, as can be seen in the excluded green region in Fig.~\ref{fig:photon_electron} (left). Below this threshold, one searches either in photons~\cite{Langhoff:2022bij} or direct detection experiments like the XENON~\cite{XENON:2021qze, XENON:2022ltv}. However, these bounds are many orders of magnitude away from testing the model. Instead, for masses below the MeV, $X-$ray telescopes set very stringent bounds on decaying DM. As shown on the right in Fig.~\ref{fig:photon_electron}, INTEGRAL~\cite{Calore:2022pks} and NuStar~\cite{Perez:2016tcq, Ng:2019gch, Roach:2022lgo} set the strongest limit on the models in the region of interest for our model. In the future, THESEUS~\cite{Thorpe-Morgan:2020rwc} (green dashed line) and GammaTPC $\gamma-$ray telescopes~\cite{Shutt:2025xvc} (purple dashed line) would improve these bounds. All in all, although small differences exist for different models, we can impose a rough bound on the majoron mass at $m_a\lesssim0.1\,$MeV. 

%-----------------------------------------------------------------------------
\section{Conclusions} 
\label{sec:conc}
%-----------------------------------------------------------------------------

In this work, we have explored the phenomenology of a class of models~\cite{Davighi:2022qgb} in which the proton is predicted to be \textit{exactly} stable, independent of the details of the UV completion. This includes protection against quantum gravity effects, which are generally expected to violate all global symmetries. The central feature of the model is a novel mechanism that links baryon number to lepton flavor numbers, effectively embedding proton stability into the structure of a gauged non-universal symmetry.

As detailed in Section~\ref{sec:model}, the model is built on a gauged $\U(1)_{3m(B - L) - n (3 L_p-  L)}$ symmetry that is spontaneously broken to a discrete subgroup containing a $\mathbb{Z}_9$, under which only quarks carry a unit charge while all other SM fields remain neutral. The gauge-invariant operators that carry non-zero baryon number must involve a multiple of nine quark fields. This leads to an exact selection rule: baryon number can be violated only in multiples of three. As a result, proton decay is strictly forbidden, whereas processes with $\Delta B = 3$, such as sphaleron transitions, remain allowed. The minimal anomaly-free matter content requires three right-handed neutrinos, and the structure needed to reproduce the observed neutrino oscillations calls for two scalar fields, $\phi_1$ and $\phi_2$.

Despite its minimal field content, the model gives rise to a rich and compelling phenomenology, the exploration of which has been the central aim of this work. In its original form, the model naturally addresses several key observations beyond the SM: neutrino masses and mixings via a high-scale seesaw mechanism (Section~\ref{sec:PMNS}), the baryon asymmetry of the universe through minimal thermal leptogenesis (Section~\ref{sec:leptogenesis}), and a viable dark matter candidate in the form of the majoron (Section~\ref{sec:DM}). Notably, all three phenomena rely on the high-scale symmetry breaking, pointing to a coherent and unified picture. Despite its high-scale dynamics, the model produces a range of relevant predictions that can be tested by current and future experiments.

Lepton flavor non-universality, central to the proton stability mechanism, gives rise to a specific neutrino mass texture, as detailed in Section~\ref{sec:PMNS}. The light-neutrino mass matrix exhibits a substructure in which a minor of the matrix vanishes at the tree level. As a result, the neutrino parameters describe a triangle~\eqref{eq:Delta_m_generic_form} in the complex plane. Only three out of six scenarios are compatible with current data: normal ordering with $\mu$- or $\tau$-specific charges, and inverted ordering with $e$-specific charges. The triangle equation constrains the allowed ranges of the lightest neutrino mass and the Dirac CP-violating phase, while determining the Majorana phases in terms of these two parameters up to a two-fold ambiguity (see Fig.~\ref{fig:majorana_phases}). As illustrated in Fig.~\ref{fig:neutrinolesss}, a future measurement of the Dirac phase would yield a sharp prediction for the neutrinoless double beta decay rate as a function of the lightest neutrino mass.

The high-scale symmetry breaking in the model naturally supports minimal thermal leptogenesis (Section~\ref{sec:leptogenesis}), consistent with the Davidson–Ibarra bound. The matter--antimatter asymmetry is generated via CP-violating out-of-equilibrium decays of the lightest right-handed neutrino, with the resulting lepton asymmetry converted into a baryon asymmetry through sphaleron processes, permitted by the selection rule in Eq.~\eqref{eq:selectrule}. Unlike the lepton universal case, where a hierarchical spectrum of right-handed neutrinos is typically unexplained, our model can dynamically realize this condition. The structure of the right-handed neutrino mass matrix, together with a mild hierarchy in the scalar VEVs ($v_2 \gg v_1$), naturally leads to a substantial mass separation, $M_{N_1} \ll M_{N_{2,3}}, Z', \rho_{1,2}$, under the assumption of comparable dimensionless couplings. The hierarchy ensures that $N_1$ decouples from the thermal bath in time for out-of-equilibrium decays, thereby suppressing washout effects and enabling successful leptogenesis as required by the Sakharov conditions. We have identified viable regions in parameter space where this mechanism accounts for the observed matter–antimatter asymmetry (see Fig.~\ref{fig:leptogenesis_scan}).

Finally, the model naturally accounts for the presence of cold dark matter in the form of a light pNGB—the majoron (Section~\ref{sec:DM}). To reproduce the observed neutrino masses and mixings, two scalar fields are required to break the symmetry; one of the resulting phases is absorbed as the longitudinal mode of the massive gauge boson, while the other manifests as the majoron. Realistic dark matter phenomenology imposes additional constraints, requiring a high enough symmetry-breaking scale and a light majoron to achieve the correct relic abundance, ensure a cosmologically long lifetime, and remain consistent with current observational limits, see Figs.~\ref{fig:light_majoron} and \ref{fig:heavier_majoron}. Notably, the charge assignments required by the proton stability mechanism suppress gravitational symmetry-breaking effects by pushing them to operators of sufficiently high dimension, which is crucial for keeping the majoron light. We have systematically examined all relevant production mechanisms consistent with cosmological observations and identified potential experimental signatures. In particular, $X$- and $\gamma$-ray telescopes constrain the majoron mass to $m_a \lesssim 100\,\text{keV}$, see Fig.~\ref{fig:photon_electron}. This is especially compelling, as a significant relic abundance can be produced via freeze-in, with structure formation setting a lower bound of $m_a \gtrsim 5\,\text{keV}$. These results strongly motivate continued searches for decaying dark matter in this mass window.

The distinct symmetry-breaking pattern gives rise to topological defects—local and global strings if the breaking occurs after inflation, followed by domain walls at later stages—as discussed in Section~\ref{sec:topological}. In such a scenario, the need to avoid stable domain walls that would overclose the universe imposes further constraints on the allowed \gx charge assignments, as summarized in Tab.~\ref{tab:models}. In models with $N_{W}=1$, the resulting metastable string-wall network leads to a cosmologically viable scenario and may generate observable signatures in the stochastic gravitational wave background. A detailed calculation of the spectrum is beyond the scope of this work and is left for future study.

In conclusion, the proton’s apparent stability may offer a profound insight into the underlying structure of microscopic physics. Remarkably, a simple model that guarantees exact proton stability also accounts for key empirical signatures of physics beyond the SM observed to date, including neutrino masses and the cosmic origins of dark and visible matter.

%-----------------------------------------------------------------------------
\section*{Acknowledgments}
%-----------------------------------------------------------------------------

We thank Joe Davighi and Alessandro Valenti for valuable discussions. This work has received funding from the Swiss National Science Foundation (SNF) through the Eccellenza Professorial Fellowship ``Flavor Physics at the High Energy Frontier,'' project number 186866. The work of AET is funded by the Swiss National Science Foundation (SNSF) through the Ambizione grant ``Matching and Running: Improved Precision in the Hunt for New Physics,'' project number 209042. We also acknowledge the proton for not decaying (so far).

\appendix 
\renewcommand{\thesection}{\Alph{section}}
\renewcommand{\thesubsection}{\Alph{section}.\arabic{subsection}}
\setcounter{section}{0}

%%%%%%%%%%%%%%%%

%-----------------------------------------------------------------------------

\section{Flavor symmetries and parameter counting}
\label{app:flavorcounting}
%-----------------------------------------------------------------------------

The global flavor symmetry of the leptonic kinetic terms is
\begin{equation}
   \U(2)_{\ell_q} \times    \U(2)_{e_q} \times    \U(2)_{N_q} \times   \U(1)_{\ell_p} \times    \U(1)_{e_p} \times    \U(1)_{N_p}.  
\end{equation}
This gives us the freedom to perform field redefinitions and define a suitable gauge basis by eliminating redundant parameters. The chosen basis then defines the UV parameters used in our numerical scan.

By the singular value decomposition theorem, the three $\mathrm{U}(2)$ factors (plus $\U(1)_{e_p}$) can be used to render $\mathcal{P}_p Y_e \mathcal{P}_p$ and $\mathcal{P}_p Y_{\phi_2} \mathcal{P}_p$ diagonal with real, non-negative entries, in Eq.~\eqref{eq:e_yuk}. There is a $ \U(1)^2 $ subgroup of $\mathrm{U}(2)_{\ell_q} \times \mathrm{U}(2)_{e_q}$ which is a symmetry of the $\mathcal{P}_p Y_e \mathcal{P}_p$ term and, together with $\mathrm{U}(1)_{\ell_p}$, it can be used to set the diagonal entries of $\mathcal{P}_p Y_N \mathcal{P}_p$ real. Finally, $\U(1)_{N_p}$ is used to remove one of the phases from $\mathcal{P}_p Y_{\phi,1} \mathcal{P}_p$.

To summarise, we define the flavor basis in Eq.~\eqref{eq:e_yuk} to be
\begin{align}\label{eq:UVparams}
    Y_{e} &\sim 
    \mathcal{P}_p \begin{pmatrix}
    y^e_1 & 0 & 0 \\
    0 & y^e_2 & 0 \\
    0  & 0 & y^e_3 
    \end{pmatrix} \mathcal{P}_p\,,  \nonumber \\ 
    Y_{N} &\sim 
    \mathcal{P}_p \begin{pmatrix}
    y^N_1 & 0 & 0 \\
    0 & y^N_2 & y^N_{23} e^{\delta_2} \\
    0  & y^N_{32} e^{\delta_3} & y^N_3 
    \end{pmatrix} \mathcal{P}_p\,, \nonumber \\
    Y_{\phi,1} &\sim 
    \mathcal{P}_p \begin{pmatrix}
    0 & y^{\phi}_{12} & y^{\phi}_{13} e^{\delta_1}  \\
   y^{\phi}_{12} & 0 & 0 \\
    y^{\phi}_{13} e^{\delta_1}  & 0 & 0 
    \end{pmatrix}\mathcal{P}_p\,, \nonumber \\
    Y_{\phi,2} &\sim 
    \mathcal{P}_p \begin{pmatrix}
    0& 0 &0 \\
    0 & y^{\phi}_2 & 0 \\
    0 & 0 & y^{\phi}_3 \\
    \end{pmatrix}\mathcal{P}_p\,.
    \end{align}
In total, the UV Lagrangian in Eq.~\eqref{eq:L_UV} relevant to the leptonic sector contains nine real parameters ($y$'s) and three complex phases ($\delta$'s).

%-----------------------------------------------------------------------------
\section{Neutrino mass matrix}
\label{app:neutrino_mass_matrix}
%-----------------------------------------------------------------------------
In this appendix, we explore the conditions on the low-energy neutrino data imposed by the structure of the Yukawa matrices and right-handed mass matrix, which is a consequence of the lepton-flavored \gx symmetry. 

\subsection{Majorana mass matrix}
With SSB the right-handed neutrinos receive a Majorana mass matrix $ M_N $. The left-handed neutrinos obtain masses after decoupling of the right-handed neutrinos as per the type-I seesaw mechanism:
    \begin{equation}
    m_\nu = -\vew^2 (Y_N M_N^{\eminus 1} Y_N\transpose)^\ast,  
    \end{equation}
where 
    \begin{equation}
    M_N = \mathcal{P}_p \begin{pmatrix}
        0 & \boldsymbol{\mu}\transpose \\ \boldsymbol{\mu} & M
    \end{pmatrix}\mathcal{P}_p , \qquad 
    Y_N = \mathcal{P}_p \begin{pmatrix}
        y_1 & 0 \\ 0 & Y
    \end{pmatrix} \mathcal{P}_p.
    \end{equation}
The inverse of the $ M_N $ is explicitly given as
    \begin{equation}
    M^{\eminus 1}_N = \dfrac{1}{\boldsymbol{\mu}\transpose M^{\eminus 1} \boldsymbol{\mu}} \mathcal{P}_p \begin{pmatrix}
        \eminus 1 & \boldsymbol{\mu}\transpose M^{\eminus 1} \\ M^{\eminus 1} \boldsymbol{\mu} & (\boldsymbol{\mu}\transpose M^{\eminus 1} \boldsymbol{\mu}) A
    \end{pmatrix} \mathcal{P}_p.
    \end{equation}
One may now observe that the lower right $ 2 \times 2 $ block,
    \begin{equation} \label{eq:mass_matrix_2x2}
    A \equiv M^{\eminus 1} - \dfrac{M^{\eminus 1} \boldsymbol{\mu} \boldsymbol{\mu}\transpose M^{\eminus 1}}{ \boldsymbol{\mu}\transpose M^{\eminus 1} \boldsymbol{\mu}} ,
    \end{equation}
is rank 1, which follows from the observation $ A \boldsymbol{\mu} = 0 $.\footnote{In fact, no explicit formula for $ M_N^{\eminus 1}$ is required to arrive at this conclusion. It is sufficient to consider the first column of the trivial identity $ M_N^{\eminus 1} M_N = \mathds{1} $.}
This observation extends to the light neutrino mass matrix $ m_\nu^\ast $, whose lower right $ 2 \times 2 $ block, $ \vew^2Y A Y\transpose $, is also rank-1: $ (Y^{\eminus 1})\transpose \boldsymbol{\mu} $ is a null vector.

\subsection{Constraints on neutrino masses and mixing parameters}
\label{app:triangle_equations}
The neutrino Majorana mass matrix is given by (for simplicity, we define $m\equiv m_\nu$ of Eq.~\eqref{eq:seesaw_formula} in this section)
    \begin{equation}
    m_\nu = U^* \hat{m}_\nu U^\dagger, \qquad U = U_{\sscript{PMNS}} \diag \big(e^{i\alpha/2},\, e^{i\beta/2},\, 1\big),
    \end{equation}
with $ \alpha, \beta $ being the Majorana phases (an overall unphysical phase has been factored out). We denote the matrix minors
    \begin{equation}
    \begin{split}
    [m]_{aa} &\equiv m_{ii} m_{jj} - m_{ij}^2, \qquad \text{for}\; i<j \; \text{and}\; i,j\neq a \\
    &= \sum_{k<\ell} m_k m_\ell \big(U_{ik}^* U_{j\ell}^* - U_{i\ell}^* U_{jk}^* \big)^2,
    \end{split}
    \end{equation}
where $ m_i $ are the diagonal entries of the diagonal matrix $ \hat{m}_\nu $.
We have shown that the lower $2 \times 2$ block (before applying the scenario-specific permutation matrices $ \mathcal{P}_p $) of the neutrino mass matrix has rank one. A direct consequence is that the corresponding minor vanishes
    \begin{equation}
    \label{eq:minor_equation}
    [m]_{pp} =0,
    \end{equation}
where when indexing with $ p $ we use $ p = 1,2,3 $ for the $e$-,$\mu$-,$\tau$-specific scenarios.

\begin{figure*}[t]
    \centering
    \includegraphics[width=1\linewidth]{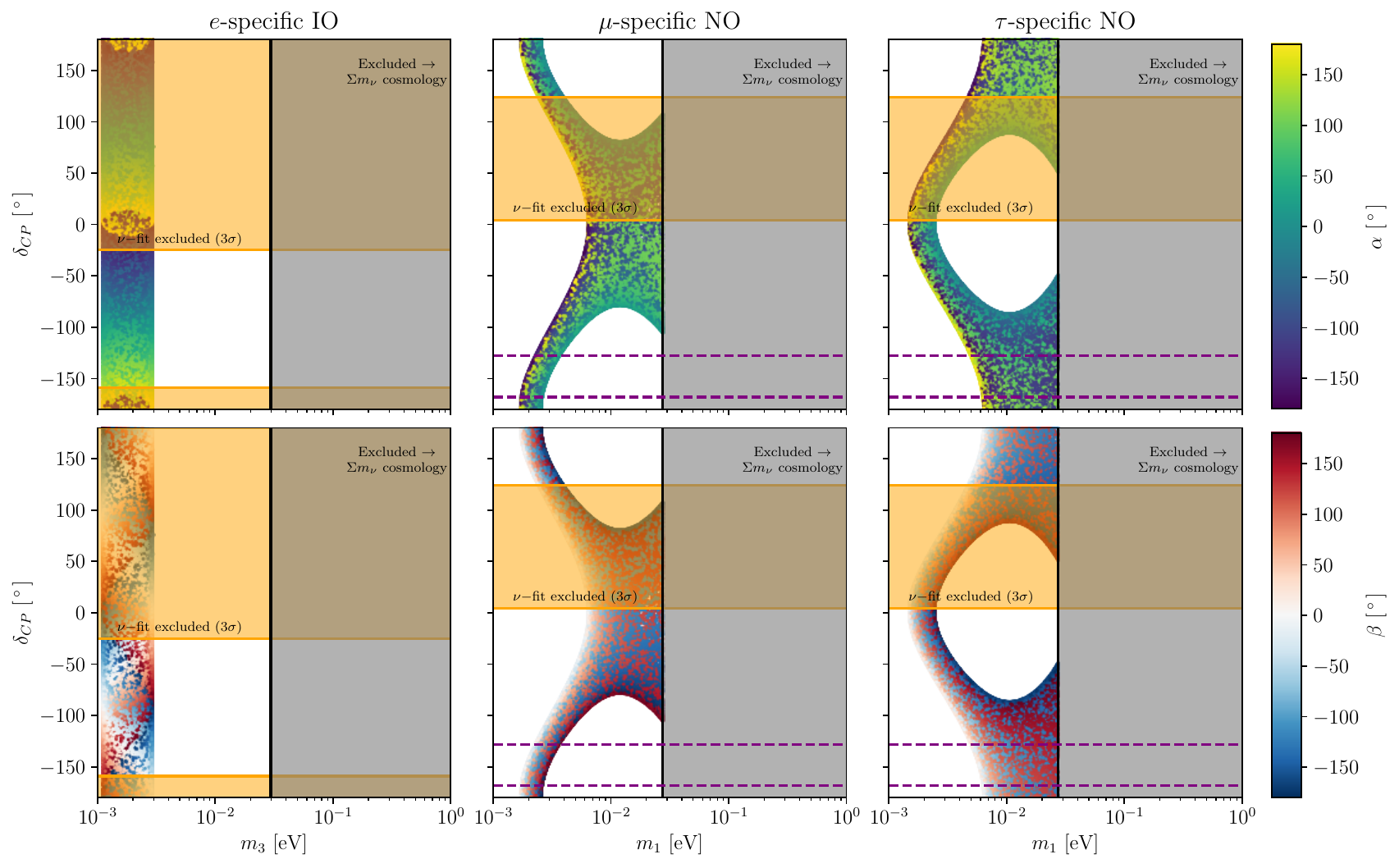}
    \caption{Same as in Fig.~\ref{fig:majorana_phases} but performing a scan of the UV parameter which reproduces the low-energy neutrino data. Here, both solutions to the triangle equations are simultaneously seen.}
    \label{fig:scan_majoranaphases}
\end{figure*}

Each of the three minor conditions imposed by the $p $-specific scenarios can be cast in terms of the neutrino mixing angles, the Dirac phase, the two Majorana phases, and the three neutrino masses. We have 
\begin{widetext}
    \begin{align}
    \dfrac{e^{ i(\alpha + \beta +2\delta_\sscript{CP})}}{m_{1} m_{2} m_{3}} [m]_{11} 
    &= \dfrac{1}{m_3} s_{13}^2 + \dfrac{e^{ i (\beta + 2 \delta_\sscript{CP}) }}{m_2} s_{12}^2 c_{13}^2 + \dfrac{e^{ i (\alpha + 2 \delta_\sscript{CP})}}{m_1} c_{12}^2 c_{13}^2 \label{eq:electron_fav_Delta}\\
    \dfrac{e^{ i(\alpha + \beta)}}{m_{1} m_{2} m_{3}} [m]_{22} 
    &= \dfrac{1}{m_3} c_{13}^2 s_{23}^2 + \dfrac{e^{ i \beta}}{m_2} \big(c_{12} c_{23}- s_{23} s_{12} s_{13} e^{i \delta_\sscript{CP}} \big)^2 + \dfrac{e^{ i \alpha}}{m_1} \big(s_{12} c_{23}+ c_{12} s_{23} s_{13} e^{i \delta_\sscript{CP}} \big)^2\\
    \dfrac{e^{ i(\alpha + \beta)}}{m_{1} m_{2} m_{3}} [m]_{33}
    &= \dfrac{1}{m_3} c_{13}^2 c_{23}^2 + \dfrac{e^{ i \beta}}{m_2} \big(c_{12} s_{23}+ c_{23} s_{12} s_{13} e^{i \delta_\sscript{CP}} \big)^2 + \dfrac{e^{ i \alpha}}{m_1} \big(s_{12} s_{23} - c_{12} c_{23} s_{13} e^{i \delta_\sscript{CP}} \big)^2
    \end{align}
\end{widetext}
where $ c_{ij}= \cos \theta_{ij} $ and $ s_{ij}= \sin \theta_{ij} $. $ \theta_{ij} $ are the mixing angles and $ \delta_\sscript{CP} $ the phase of the PMNS matrix.
Generally, we observe that 
    \begin{equation} \label{eq:Delta_m_generic_form}
    [m]_{pp} \propto \dfrac{f_{p,3}}{m_3} + e^{i \beta} \dfrac{f_{p,2}}{m_2} + e^{ i \alpha} \dfrac{f_{p,1}}{m_1}, \qquad \begin{cases}
    f_{p,3} \in \mathbb{R} \\ f_{p,1(2)} \in \mathbb{C}
    \end{cases}
    \end{equation}
and $ f_{p,i}(\theta_{12},\theta_{23},\theta_{13}, \delta_\sscript{CP}) $ are functions of the PMNS parameters. The requirement $ [m]_{pp} = 0 $ is equivalent to the three terms in~\eqref{eq:Delta_m_generic_form} describing a triangle in the complex plane. The independent Majorana phases $ \alpha, \beta $ ensure that the triangle exists if and only if the side lengths, $ |f_{p,k}|/m_k $, satisfy all the triangle equalities:
    \begin{equation} \label{eq:triangle}
    \dfrac{|f_{p,k}|}{m_k} \leq \dfrac{|f_{p,\ell}|}{m_\ell} + \dfrac{|f_{p,m}|}{m_m}, \qquad \mathrm{for}\; \ell \neq m.
    \end{equation} 
For a valid set of side lengths, there are two solutions for the Majorana phases related by 
    \begin{equation}
    \label{eq:conjugate_solutions}
    \begin{split}
    \alpha &\longrightarrow - \alpha  - 2 \arg f_{p,1}, \\
    \beta &\longrightarrow - \beta - 2 \arg f_{p,2}.
    \end{split}
    \end{equation}
This can be seen from the observation that if a triangle is a solution, then so is the conjugate triangle. 

In the NO neutrino fit, it is particularly instructive to consider the inequality~\eqref{eq:triangle} with $ (k,\ell,m) = (1,2,3) $, which is then equivalent to
    \begin{equation}
    |f_{p,1}| \leq \dfrac{|f_{p,2}|}{\sqrt{1+ \tfrac{\Delta m^2_{21}}{m_1^2} }} + \dfrac{|f_{p,3}|}{ \sqrt{1+ \tfrac{\Delta m^2_{31}}{m_1^2} }}
    \end{equation}
where the r.h.s. approaches $ |f_{p,2}| + |f_{p,3}| $ from below as $ m_1 \to \infty $. On the other hand, the r.h.s. vanishes as $ m_1 \to 0 $. Hence, there will always be a finite $ m_1 > 0 $ below which no solution exists. This is a sharp prediction from the theory.

In the electron-specific scenario the neutrino oscillation data constrains the coefficients~\eqref{eq:electron_fav_Delta} to $ |f_{p,3}| + |f_{p,2}| < |f_{p,1}| $. This implies that there is no solution consistent with current experimental data, and the scenario is ruled out. For the tau-specific scenario, the size of the coefficients $ |f_{p,k}| $ satisfy the triangle equality, and solutions are guaranteed in the degeneracy limit for the neutrino masses.  

One of the two explicit solutions for the Majorana phases is plotted in Fig.~\ref{fig:majorana_phases}, where only three models have solutions consistent with current neutrino data. Depending on the value of $\delta_\sscript{CP}$, there exists a limit on the lightest neutrino mass. Moreover, the three models can be distinguished clearly by measuring $\delta_\sscript{CP}$ and the lightest neutrino mass. The determination of the $\theta_{23}$ octant could have played an important role, as it determines whether an upper bound on the mass exists (as seen in $\tau$-specific models) or not (as in the $\mu$-specific model). However, this possibility is strongly ruled out by cosmological constraints.

The same solutions can be obtained by scanning over the UV parameters and fitting to low-energy neutrino oscillation data, as shown in Fig.~\ref{fig:scan_majoranaphases}. Here, the two Majorana phase solutions overlap in the same parameter space. This demonstrates that the Majorana phases can be predicted simply by solving the minor equations~\eqref{eq:minor_equation}, without relying on UV parameters and using only low-energy data.

\subsection{Higher order corrections to the Majorana mass matrix} \label{app:radiative}
Radiative corrections to the tree-level seesaw formula for the neutrino mass matrix are expected to break the rank-1 condition for the submatrix. An exact determination would require computing both matching and running corrections in a spontaneously broken gauge theory, which is beyond the scope of this project. We can, nevertheless, estimate the impact from EW corrections, in the scenario $ M_{N_1} \ll M_{N_2}, M_{N_3} $ ($ v_1 \ll v_2$). 

First, we go to a block-diagonal not-quite-mass basis for the heavy neutrinos:
    \begin{equation}
    M_N \to M'_N = V\transpose M_N V \andeq Y_N \to Y_N' = Y_N V^\ast,
    \end{equation}
where 
    \begin{align}
    M'_N &= \begin{pmatrix}
        - \boldsymbol{\mu}\transpose M^{\eminus 1} \boldsymbol{\mu} & \\ & M + M^{\eminus 1\, \ast} \boldsymbol{\mu}^\ast \boldsymbol{\mu}\transpose + \boldsymbol{\mu} \boldsymbol{\mu}^\dagger M^{\eminus 1\, \ast} 
    \end{pmatrix}\\
    V &= \begin{pmatrix}
       1 - \tfrac{1}{2} \boldsymbol{\mu}\transpose M^{\eminus 1} M^{\eminus 1\, \ast} \boldsymbol{\mu}^\ast & \boldsymbol{\mu}\transpose M^{\eminus 1} \\ \eminus M^{\eminus 1\, \ast} \boldsymbol{\mu}^\ast & \mathds{1} - \tfrac{1}{2} M^{\eminus 1\, \ast} \boldsymbol{\mu}^\ast \boldsymbol{\mu}\transpose M^{\eminus 1} 
    \end{pmatrix}
    \end{align}
neglecting terms of $ \mathcal{O}(\boldsymbol{\mu}/M)^3 $. For this subsection, we leave out the permutation matrices $ \mathcal{P}_p $, effectively focusing on the $ e $-specific scenario, to keep things simple. In particular, we can parametrize $ M'_N = \diag(M_{N_1},\, M_{23}) $, where $ M_{23} $ is the mass matrix of the two heavy neutrino mass eigenstates, with singular values $ M_{N_2} \sim M_{N_3} $. 

Upon integrating out the two heavy neutrinos, the tree-level contribution to the Weinberg operator is
    \begin{equation}
    C_{\nu\nu}(\mu = \sqrt{M_{N_2} M_{N_3}}) = Y_N' \begin{pmatrix}
    0 & \\ & M_{23}^{\eminus 1}
    \end{pmatrix} Y_N^{\prime \intercal}.
    \end{equation}
EW interactions introduce a running of the Weinberg operator given by $ \dot{C}_{\nu\nu} = \gamma(\lambda,\, g_2,\, g_1) C_{\nu\nu} $. Accounting for this down to the scale $ M_1 $, where the last heavy neutrino is integrated out, giving a second contribution to the Weinberg operator, we find that 
    \begin{equation}
    C_{\nu\nu}(M_{N_1}) = Y_N' \begin{pmatrix}
    M_{N_1}^{\eminus 1} & \\ & (1 - \gamma \,t) M_{23}^{\eminus 1}
    \end{pmatrix} Y_N^{\prime \intercal}.
    \end{equation}
in the leading log approximation, where $ t = \log(\sqrt{M_{N_2} M_{N_3}}/ M_{N_1}) $. In principle, there is also a running of $ Y_N' $; however, as long as the renormalization of $ Y_N' $ does not involve the $ N_1 $ leg of the vertex (expected to be a reasonable assumption), this running is also part of $ \gamma $. We simply take $ \gamma $ to be the effective running of $ C_{\nu\nu} $ subtracting off the running of the Yukawas. Returning to the gauge basis, we have 
    \begin{equation} \label{eq:C_nn_at_M1}
    C_{\nu\nu}(M_{N_1}) = Y_N M_\mathrm{eff}^{\eminus 1} Y_N\transpose.
    \end{equation}
The inverse effective mass matrix is 
    \begin{equation}
    M_\mathrm{eff}^{\eminus 1} = M_N^{\eminus 1} - \gamma\, t \, V^\ast \diag\! \big( 0,\, M_{23}^{\eminus 1}\big) V^\dagger\, . 
    \end{equation}
The effective modification of the mass matrix $2 \times 2 $ block~\eqref{eq:mass_matrix_2x2} is 
    \begin{equation}
    A_\mathrm{eff} = A - \gamma\, t \, M^{\eminus 1}
    \end{equation}
at leading order in $ \boldsymbol{\mu}/ M$. The deviation lifts the rank of $ A $:\footnote{We liberally use that for $ 2\times 2$ matrices it holds that $ \mathds{1} \det X = X \tr X - X^2 $ and, thus, also $ \det X = \tfrac{1}{2} \tr^2 X - \tfrac{1}{2} \tr X^2$.}
\begin{widetext}
    \begin{equation}
    \begin{split}
    \det A_\mathrm{eff} \simeq\,& - \gamma \, t \big( \tr A \tr M^{\eminus 1} - \mathrm{tr} [A M^{\eminus 1}] \big) \\
    =&- \gamma \, t \left(\tr^2 M^{\eminus 1} - \tr M^{\eminus 2} + \dfrac{\boldsymbol{\mu}\transpose M^{\eminus 1} \big(M^{\eminus 1} - \tr M^{\eminus 1} \big) M^{\eminus 1} \boldsymbol{\mu}}{\boldsymbol{\mu}\transpose M^{\eminus 1} \boldsymbol{\mu}} \right) 
    =- \gamma \, t \det M^{\eminus 1}
    \end{split}
    \end{equation}
\end{widetext}
to leading log. 
Everything in this derivation generalizes with permutation matrices to the generic $ p $-specific scenarios. In summary, a good approximation for the $ pp $-minor of the light neutrino mass matrix is 
    \begin{equation}
    \begin{split}
    [m_\nu^\ast]_{pp} &= \det\big( \eminus \vew^2 Y A_\mathrm{eff} Y\transpose \big)  \\
    &\simeq - \gamma \, t \det\big(\vew^2 Y M^{\eminus 1} Y\transpose \big)  
    \end{split}
    \end{equation}
when including EW corrections. 

A reasonable approximation for the EW contribution to $ \gamma $ is to use the contribution to the anomalous dimension of $ C_{\nu\nu} $ that come from vertex correction to the counterterm~\cite{Ibarra:2024tpt}: $ \gamma = \tfrac{1}{16\pi^2} (2\lambda + \tfrac{1}{2} g_1^2- \tfrac{3}{2} g_2^2) $. All contributions from wave-function renormalization is reproduced in the running of $ (Y'_N)_{i1} $ and are expected to become an overall (irrelevant) multiplication factor in~\eqref{eq:C_nn_at_M1}.\footnote{This simplified analysis ignores vertex corrections to the running of $ Y'_N$ involving the right-handed neutrino, not to mention the running of $ M_1 $.} It has proven difficult to obtain a strict upper bound on $ |\det Y| $ even knowing how it enters the tree-level neutrino mass formula. As a best guess, we observe that $ Y M^{\eminus 1} Y\transpose $ is part of the contribution to the $ 2\times 2$ block of $ m_\nu^\ast $. In the absence of large cancellations, its singular values are expected not to exceed the largest singular values in $ m_\nu $. A rough numerical estimate for the $ pp $ minor from radiative corrections is therefore   
    \begin{equation}
    |[m_\nu]_{pp}| \overset{?}{\lesssim} 0.01\cdot  \begin{cases}
    m_2 m_3 & (\mathrm{NO}) \\
    m_1 m_2 & (\mathrm{IO}) 
    \end{cases}.
    \end{equation}
We have numerically verified that even using the upper bound does not qualitatively change our predictions for the lightest neutrino mass and Majorana phases.

\section{Interactions of the majoron}
\label{app:Majoron}
Here we derive some of the basic properties of the majoron starting from the unbroken theory~\eqref{eq:L_UV}.  

\subsection{Angular eigenstates}
The VEVs of the scalar fields $ \phi_i $ break the \gx symmetry with its associated current 
    \begin{equation}
    J_\mu = \sum_i -i \mathcal{X}_i \phi_i^\ast \overset{\leftrightarrow}{\partial}_\mu \phi_i + \ldots \supset \frac{1}{\sqrt{2}} \sum_i \mathcal{X}_i v_i a_i,
    \end{equation}
obtained by parameterizing $\phi_i \sim  \tfrac{1}{\sqrt{2}} v_i \exp(i a_i/v_i) $ with $a_i$ the angular component of each scalar field and $v_i$ its VEV. The breaking of the gauge symmetry causes one linear combination $ \varphi_X $  of the radial component to be identified as the longitudinal component of the massive gauge boson $ X_\mu $, whereas the orthogonal component $ a $ is our pNGB majoron. We write 
    \begin{equation} \label{eq:rotation}
    \begin{pmatrix}
        a \\ \varphi_X 
    \end{pmatrix}
    =\begin{pmatrix}
        c_\varphi & -s_\varphi \\
        s_\varphi & c_\varphi
    \end{pmatrix} \begin{pmatrix}
        a_1 \\ a_2
    \end{pmatrix} \, ,
    \end{equation}
with the mixing angle $ \varphi $. The majoron $ a $ has no kinetic mixing with the gauge field, meaning that 
    \begin{equation} 
    \label{eq:Goldstone}
    0 = \bra{0} J_\mu \ket{a} = \frac{i p_\mu}{\sqrt{2}} \big( c_\varphi v_1 \mathcal{X}_1 - s_\varphi v_1 \mathcal{X}_1 \big)\, .
    \end{equation}
It follows that the mixing angle satisfy 
    \begin{equation} \label{eq:angles}
    \binom{s_\varphi}{c_\varphi} = \dfrac{1}{v_X} \binom{\mathcal{X}_1 v_1}{\mathcal{X}_2 v_2}, \qquad v_{X} = \sqrt{\mathcal{X}_1^2 v_1^2 + \mathcal{X}_2^2 v_2^2} \,,
    \end{equation}
where $ v_X $ is the decay constant of $ \varphi_X $. Furthermore, we let 
    \begin{equation}
    f_a \equiv  \dfrac{s_{2\varphi}}{2} v_X = \mathcal{X}_1 \mathcal{X}_2 \dfrac{v_1 v_2}{v_X} 
    \end{equation}
be the effective decay constant of the majoron such that 
    \begin{equation} \label{eq:a_in_phi}
    \phi_1 \supset \dfrac{v_1}{\sqrt{2}} \exp\!\left[ \mathcal{X}_1 c_\varphi^2 \dfrac{a}{f_a} \right]\,, \quad 
    \phi_2 \supset \dfrac{v_2}{\sqrt{2}} \exp\!\left[ - \mathcal{X}_2 s_\varphi^2 \dfrac{a}{f_a} \right]\,.
    \end{equation}
In the limit $ \mathcal{X}_2 v_2 \gg v_1 \mathcal{X}_1 $, the mixing between the angular modes is suppressed, and we obtain
    \begin{equation}
    v_X \simeq \mathcal{X}_2 v_2, \qquad f_a \simeq \mathcal{X}_1 v_1.
    \end{equation}

\subsection{Majoron interactions with heavy neutrinos}
\label{sec:C2}

The Yukawa interactions between the $ \phi_i $ fields and the right-handed Neutrinos are given by~\eqref{eq:L_UV}. Expanding the interactions up to linear order in the majoron, we find that  
\begin{equation}
\label{eq:RHNLag1}
     \mathcal{L} \supset  -\frac{1}{2} \overline{N}^c \! \left[M_N +i \frac{a}{\sqrt{2}}  \left(Y_{\phi,1} c_\varphi- Y_{\phi,2} s_\varphi \right) \right]\! N \, + \, \textrm{H.c.}
\end{equation}
where the mass of the RHNs is defined as usual: $M_N=(Y_{\phi,1} v_1+Y_{\phi,2} v_2)/\sqrt{2}$. Gauge invariance of the $ Y_{\phi, i} $ interactions ultimately constrain the charge matrix $ \mathcal{X}_N $ of the \gx group acting on the right-handed neutrinos to satisfy\footnote{This relation is somewhat analogous to the non-universal DFSZ axion models, see Ref.~\cite{DiLuzio:2020wdo}.}
    \begin{equation}
    \label{eq:YukawaRelation1}
    \left\{\mathcal{X}_N, M_N\right\} = -\frac{v_X}{\sqrt{2}} \left( Y_{\phi,1} s_\varphi +Y_{\phi,2} c_\varphi\right) \, .
    \end{equation}
Thus, we can cast~\eqref{eq:RHNLag1} as 
\begin{align}\label{eq:RHNLag2}
     \mathcal{L} \supset  &-\frac{1}{2} \overline{N}^c \! \Bigg[ M_N -i \frac{a }{f_a} \!  \left( \dfrac{1}{\mathcal{X}_2} - \dfrac{1}{\mathcal{X}_1} \right)^{\!\! \eminus 1}\\ &\times \! \left( \! M_N + \left[ \left( \dfrac{c_\varphi^2 }{\mathcal{X}_2} + \dfrac{s_\varphi^2}{\mathcal{X}_1} \right) \left\{\mathcal{X}_N, M_N\right\}  \right] \right) \Bigg]\! N \, + \, \textrm{H.c.} \nonumber    
\end{align}

In our model, the charges are not universal, which means that the anti-commutator is not trivial. To be explicit about it, the charge matrix for the right-handed neutrinos is 
    \begin{equation}
    \begin{split}
    \mathcal{X}_N &= (n-3m) \mathds{1} -3n\, \xi_p \\
    &= - \tfrac{1}{2} \mathcal{X}_2 \mathds{1} + (\mathcal{X}_2 - \mathcal{X}_1) \xi_p \, , 
    \end{split}
    \end{equation}
where 
    \begin{equation}
    \xi_p \equiv \mathcal{P}_p \diag(1,0,0) \mathcal{P}_p.
    \end{equation}
The interactions between the majoron and the right-handed neutrinos is then given by 
    \begin{multline} \label{eq:RHNLaginitial}
     \mathcal{L}_{aNN} =  -\frac{i}{2} \frac{a }{f_a} \overline{N}^c \! \Bigg[ \! \mathcal{X}_2 s_\varphi^2 M_N \\
     - \left( \mathcal{X}_1 c_\varphi^2 +\mathcal{X}_2 s_\varphi^2 \right) \! \left\{ \xi_p, M_N \right\} \!\Bigg]\! N \, + \, \textrm{H.c.}
    \end{multline}
The first term is proportional to the mass matrix of the right-handed neutrinos and is, therefore, diagonal in the mass eigenbasis; however, the second term is expected to dominate when, as in our baseline scenario, $ c_\varphi \gg s_\varphi $.

\subsection{Majoron interactions with light neutrinos}
Integrating out the heavy neutrinos from the theory gives rise to the Weinberg operator and an operator with an extra majoron field: 
    \begin{equation}\label{eq:alightnu}
    \mathcal{L}\supset \frac{1}{2} \left(C_{ij}-i\frac{a}{f_a}D_{ij}\right) \bar{\ell}^c_i \tilde{H}^c H^\dagger \ell_j  +\textrm{H.c.} \, ,
    \end{equation}
where 
    \begin{align}
    \label{eq:Ccoupling}
    C^\ast  &=Y_N M_N^{\eminus 1} Y_N\transpose \, , \\ 
    \label{eq:Dcoupling}
    D^\ast &= - \mathcal{X}_2 s_\varphi^2 C^\ast + \left( \mathcal{X}_1 c_\varphi^2 +\mathcal{X}_2 s_\varphi^2 \right) \! Y_N \big\{ \xi_p, M_N^{\eminus 1} \big\}  Y_N\transpose\, .
    \end{align}
We can express this in a simpler way by utilizing gauge invariance of the $ Y_N $ coupling. As the Higgs field is uncharged, this manifests as the requirement $ 0 = -\mathcal{X}_\ell Y_N + Y_N \mathcal{X}_N $. All lepton species are charged the same, so $ \mathcal{X}_N = \mathcal{X}_\ell $, and with $ \xi_p $ being the non-trivial part of the lepton charge matrix, we find that $ \big[\xi_p,\, Y_N \big] = 0 $. It follows that the coupling matrix for the majoron to light neutrinos is given by 
    \begin{equation}
    \vew^2 D^\ast = - \mathcal{X}_2 s_\varphi^2 m_\nu^\ast + \left( \mathcal{X}_1 c_\varphi^2 +\mathcal{X}_2 s_\varphi^2 \right) \! \big\{ \xi_p, m_\nu^\ast \big\}  .
    \end{equation}
The first term is again proportional to the neutrino mass matrix and conserves neutrino flavor. All non-diagonal couplings in the mass basis come from the second term.

\section{The formation of local strings} \label{sec:senario_simultaneous_breaking}
In this appendix, we recap the topological vacuum solutions related to the breaking of a gauge symmetry, which may clarify the discussion in Section~\ref{sec:topological}. As a topical example,e we use our model, but consider the scenario where both $ \phi_a $ fields develop VEVS $ v_1 \sim v_2 $, such that the transition from the broken to unbroken phases changes the symmetry  
	\begin{equation}
	\U(1)_X \longrightarrow \Gamma
	\end{equation}
in a single step. Since both scalars are charged under the \gx gauge symmetry, any string formed during the phase transition will be \emph{local}, meaning that the gauge field will end up curling around the string to minimize the energy of the classical field solution~\cite{Hindmarsh:1994re}. 

A string (or vortex line) is a non-trivial static field configuration that cannot continuously be deformed into a vacuum solution. To illustrate the prototypical string solution to the field equations, we work in cylindrical coordinates $ (r, \theta) $. When we are far removed from the core of the string, the solution should reduce to the (static) ground state, implying that 
    \begin{equation}
    \phi_a(r,\theta) \xrightarrow[r\to \infty]{} \bar{\phi}_a(\theta), \qquad \big| \bar{\phi}_a(\theta) \big| = \tfrac{1}{\sqrt{2}} v_a.
    \end{equation}
While the gauge field need not vanish at long distances, it should reduce to a pure gauge configuration (minimizing energy):
    \begin{equation}
    X_i(r,\theta) \xrightarrow[r\to \infty]{} i g^{\eminus 1}(\theta) \partial_i g(\theta),
    \end{equation}
where $ X_i $ denotes the spatial components of the gauge field.

The string energy contribution from the kinetic energy of the scalar fields, $ \mathcal{E}_\mathrm{kin} =  \int \! \dd^2 x |D_i \phi_a|^2 $, is minimized with the requirement that 
	\begin{equation}
	D_i \phi_a \xrightarrow[r\to \infty]{} 0,
	\end{equation}
which is possible for \emph{local} strings because of the presence of a non-trivial gauge field. At infinity, we now observe that 
    \begin{equation}
    \begin{split}
    0 = \rho_a[g^{\eminus1}(\theta)] D_i \bar{\phi}_a(\theta)
    = \partial_i \Big( \rho_a[ g^{\eminus1}(\theta)] \bar{\phi}_a(\theta) \Big),
    \end{split}
    \end{equation}
where $ \rho_a[g] $ is the representation of $ g\in \U(1)_{X_p} $ acting on $ \phi_a $, i.e., $ \rho_a[e^{i\alpha}] = e^{i \mathcal{X}_a \alpha} $. With the combination above being constant, it follows that 
	\begin{equation}
	\bar{\phi}_a(\theta) = \tfrac{1}{\sqrt{2}} v_a \rho_a[ g(\theta)] ,
	\end{equation}
and, thus, also the long distance behavior of $ \phi_a $ is governed by the gauge transformation $ g(\theta) $.

Finally, a consistent field solution must respect the periodicity of the cylindrical coordinates:
    \begin{equation}
    \bar{\phi}_a(2\pi) = \bar{\phi}_a(0) \implies \rho_a[ g(2\pi)] = 1 \implies g(2\pi) \in \Gamma,
    \end{equation}  
taking $ g(0) = 1 $ for simplicity. A generating element of the stability group $ \Gamma $ is $ e^{2\pi i/ |\Gamma|} $, and so we may parameterize 
    \begin{equation}
    g(2\pi) = e^{2\pi i\, w/ |\Gamma|}, \qquad w \in \mathbb{Z}.
    \end{equation} 
What will make our solution a string solution is when $ g $ possesses a non-trivial topology, that is, when we may smoothly deform $ g(\theta) \to e^{i\,\theta w/ |\Gamma|} $ for $ w\neq 0 $. In this case, the winding numbers for the two scalar fields are 
    \begin{equation} \label{eq:col_break_windings}
    \binom{w_1}{w_2} = \dfrac{w}{|\Gamma|} \binom{\mathcal{X}_1 }{\mathcal{X}_2 } = w\,\binom{t}{-s}, 
    \end{equation} 
applying~\eqref{eq:combined} to rewrite the charges in terms of the field powers in the gravitational potential~\eqref{eq:explicit_breaking_term}. The winding numbers for the fields determine how many full turns their phases make through a closed circuit around the string.

\bibliographystyle{JHEP}
\bibliography{refs.bib}

@article{Davighi:2022qgb,
    author = "Davighi, Joe and Greljo, Admir and Thomsen, Anders Eller",
    title = "{Leptoquarks with exactly stable protons}",
    eprint = "2202.05275",
    archivePrefix = "arXiv",
    primaryClass = "hep-ph",
    doi = "10.1016/j.physletb.2022.137310",
    journal = "Phys. Lett. B",
    volume = "833",
    pages = "137310",
    year = "2022"
}

@article{Graham:2018jyp,
    author = "Graham, Peter W. and Scherlis, Adam",
    title = "{Stochastic axion scenario}",
    eprint = "1805.07362",
    archivePrefix = "arXiv",
    primaryClass = "hep-ph",
    doi = "10.1103/PhysRevD.98.035017",
    journal = "Phys. Rev. D",
    volume = "98",
    number = "3",
    pages = "035017",
    year = "2018"
}

@article{Wilczek:2004cr,
    author = "Wilczek, Frank",
    editor = "Carr, Bernard J.",
    title = "{A Model of anthropic reasoning, addressing the dark to ordinary matter coincidence}",
    eprint = "hep-ph/0408167",
    archivePrefix = "arXiv",
    pages = "151--162",
    month = "8",
    year = "2004"
}

@article{Petrossian-Byrne:2025mto,
    author = "Petrossian-Byrne, Rudin and Villadoro, Giovanni",
    title = "{Open String Axiverse}",
    eprint = "2503.16387",
    archivePrefix = "arXiv",
    primaryClass = "hep-ph",
    month = "3",
    year = "2025"
}

@article{Esteban:2024eli,
    author = "Esteban, Ivan and Gonzalez-Garcia, M. C. and Maltoni, Michele and Martinez-Soler, Ivan and Pinheiro, Jo\~ao Paulo and Schwetz, Thomas",
    title = "{NuFit-6.0: updated global analysis of three-flavor neutrino oscillations}",
    eprint = "2410.05380",
    archivePrefix = "arXiv",
    primaryClass = "hep-ph",
    reportNumber = "IFT-UAM/CSIC-24-140, YITP-SB-2024-24, IPPP/24/64, IPPP/24/64, IFT-UAM/CSIC-24-140, YITP-SB-2024-24",
    doi = "10.1007/JHEP12(2024)216",
    journal = "JHEP",
    volume = "12",
    pages = "216",
    year = "2024"
}

@article{Ohlsson:2023ddi,
    author = "Ohlsson, Tommy",
    title = "{Proton decay}",
    eprint = "2306.02401",
    archivePrefix = "arXiv",
    primaryClass = "hep-ph",
    doi = "10.1016/j.nuclphysb.2023.116268",
    journal = "Nucl. Phys. B",
    volume = "993",
    pages = "116268",
    year = "2023"
}

@article{Auclair:2019wcv,
    author = "Auclair, Pierre and others",
    title = "{Probing the gravitational wave background from cosmic strings with LISA}",
    eprint = "1909.00819",
    archivePrefix = "arXiv",
    primaryClass = "astro-ph.CO",
    doi = "10.1088/1475-7516/2020/04/034",
    journal = "JCAP",
    volume = "04",
    pages = "034",
    year = "2020"
}

@article{NANOGrav:2023hvm,
    author = "Afzal, Adeela and others",
    collaboration = "NANOGrav",
    title = "{The NANOGrav 15 yr Data Set: Search for Signals from New Physics}",
    eprint = "2306.16219",
    archivePrefix = "arXiv",
    primaryClass = "astro-ph.HE",
    reportNumber = "FERMILAB-PUB-23-589-T",
    doi = "10.3847/2041-8213/acdc91",
    journal = "Astrophys. J. Lett.",
    volume = "951",
    number = "1",
    pages = "L11",
    year = "2023",
    note = "[Erratum: Astrophys.J.Lett. 971, L27 (2024), Erratum: Astrophys.J. 971, L27 (2024)]"
}

@article{JUNO:2015sjr,
    author = "Djurcic, Zelimir and others",
    collaboration = "JUNO",
    title = "{JUNO Conceptual Design Report}",
    eprint = "1508.07166",
    archivePrefix = "arXiv",
    primaryClass = "physics.ins-det",
    month = "8",
    year = "2015"
}

@article{DUNE:2020lwj,
    author = "Abi, Babak and others",
    collaboration = "DUNE",
    title = "{Deep Underground Neutrino Experiment (DUNE), Far Detector Technical Design Report, Volume I Introduction to DUNE}",
    eprint = "2002.02967",
    archivePrefix = "arXiv",
    primaryClass = "physics.ins-det",
    reportNumber = "FERMILAB-PUB-20-024-ND, FERMILAB-DESIGN-2020-01",
    doi = "10.1088/1748-0221/15/08/T08008",
    journal = "JINST",
    volume = "15",
    number = "08",
    pages = "T08008",
    year = "2020"
}

@article{Hyper-Kamiokande:2018ofw,
    author = "Abe, K. and others",
    collaboration = "Hyper-Kamiokande",
    title = "{Hyper-Kamiokande Design Report}",
    eprint = "1805.04163",
    archivePrefix = "arXiv",
    primaryClass = "physics.ins-det",
    month = "5",
    year = "2018"
}

@article{Boos:2022jvc,
    author = "Boos, Jens and Carone, Christopher D. and Donald, Noah L. and Musser, Mikkie R.",
    title = "{Asymptotic safety and gauged baryon number}",
    eprint = "2206.02686",
    archivePrefix = "arXiv",
    primaryClass = "hep-ph",
    doi = "10.1103/PhysRevD.106.035015",
    journal = "Phys. Rev. D",
    volume = "106",
    number = "3",
    pages = "035015",
    year = "2022"
}

@article{Giddings:1988cx,
    author = "Giddings, Steven B. and Strominger, Andrew",
    title = "{Loss of incoherence and determination of coupling constants in quantum gravity}",
    reportNumber = "HUTP-88/A006",
    doi = "10.1016/0550-3213(88)90109-5",
    journal = "Nucl. Phys. B",
    volume = "307",
    pages = "854--866",
    year = "1988"
}

@article{Ma:2020quj,
    author = "Ma, Ernest",
    title = "{Gauged baryon number and dibaryonic dark matter}",
    eprint = "2011.13887",
    archivePrefix = "arXiv",
    primaryClass = "hep-ph",
    reportNumber = "UCRHEP-T604 (Nov 2020)",
    doi = "10.1016/j.physletb.2021.136066",
    journal = "Phys. Lett. B",
    volume = "813",
    pages = "136066",
    year = "2021"
}

@article{FileviezPerez:2010gw,
    author = "Fileviez Perez, Pavel and Wise, Mark B.",
    title = "{Baryon and lepton number as local gauge symmetries}",
    eprint = "1002.1754",
    archivePrefix = "arXiv",
    primaryClass = "hep-ph",
    doi = "10.1103/PhysRevD.82.079901",
    journal = "Phys. Rev. D",
    volume = "82",
    pages = "011901",
    year = "2010",
    note = "[Erratum: Phys.Rev.D 82, 079901 (2010)]"
}

@article{Banks:1988yz,
    author = "Banks, Tom and Dixon, Lance J.",
    title = "{Constraints on String Vacua with Space-Time Supersymmetry}",
    reportNumber = "PUPT-1086, SCIPP-8805",
    doi = "10.1016/0550-3213(88)90523-8",
    journal = "Nucl. Phys. B",
    volume = "307",
    pages = "93--108",
    year = "1988"
}

@article{Banks:2010zn,
    author = "Banks, Tom and Seiberg, Nathan",
    title = "{Symmetries and Strings in Field Theory and Gravity}",
    eprint = "1011.5120",
    archivePrefix = "arXiv",
    primaryClass = "hep-th",
    doi = "10.1103/PhysRevD.83.084019",
    journal = "Phys. Rev. D",
    volume = "83",
    pages = "084019",
    year = "2011"
}

@article{Babu:2003qh,
    author = "Babu, K. S. and Gogoladze, Ilia and Wang, Kai",
    title = "{Gauged baryon parity and nucleon stability}",
    eprint = "hep-ph/0306003",
    archivePrefix = "arXiv",
    reportNumber = "OSU-HEP-03-8",
    doi = "10.1016/j.physletb.2003.07.036",
    journal = "Phys. Lett. B",
    volume = "570",
    pages = "32--38",
    year = "2003"
}

@article{Fukugita:1986hr,
    author = "Fukugita, M. and Yanagida, T.",
    title = "{Baryogenesis Without Grand Unification}",
    reportNumber = "RIFP-641",
    doi = "10.1016/0370-2693(86)91126-3",
    journal = "Phys. Lett. B",
    volume = "174",
    pages = "45--47",
    year = "1986"
}

@article{Davidson:2002qv,
    author = "Davidson, Sacha and Ibarra, Alejandro",
    title = "{A Lower bound on the right-handed neutrino mass from leptogenesis}",
    eprint = "hep-ph/0202239",
    archivePrefix = "arXiv",
    reportNumber = "OUTP-02-10P, IPPP-02-16, DCPT-02-32",
    doi = "10.1016/S0370-2693(02)01735-5",
    journal = "Phys. Lett. B",
    volume = "535",
    pages = "25--32",
    year = "2002"
}

@article{Super-Kamiokande:2020wjk,
    author = "Takenaka, A. and others",
    collaboration = "Super-Kamiokande",
    title = "{Search for proton decay via $p\to e^+\pi^0$ and $p\to \mu^+\pi^0$ with an enlarged fiducial volume in Super-Kamiokande I-IV}",
    eprint = "2010.16098",
    archivePrefix = "arXiv",
    primaryClass = "hep-ex",
    doi = "10.1103/PhysRevD.102.112011",
    journal = "Phys. Rev. D",
    volume = "102",
    number = "11",
    pages = "112011",
    year = "2020"
}

@article{CentellesChulia:2024uzv,
    author = "Centelles Chuli\'a, Salvador and Herrero-Brocal, Antonio and Vicente, Avelino",
    title = "{The Type-I Seesaw family}",
    eprint = "2404.15415",
    archivePrefix = "arXiv",
    primaryClass = "hep-ph",
    month = "4",
    year = "2024"
}

@article{Mohapatra:1979ia,
    author = "Mohapatra, Rabindra N. and Senjanovic, Goran",
    title = "{Neutrino Mass and Spontaneous Parity Nonconservation}",
    reportNumber = "MDDP-TR-80-060, MDDP-PP-80-105, CCNY-HEP-79-10",
    doi = "10.1103/PhysRevLett.44.912",
    journal = "Phys. Rev. Lett.",
    volume = "44",
    pages = "912",
    year = "1980"
}

@article{Gell-Mann:1979vob,
    author = "Gell-Mann, Murray and Ramond, Pierre and Slansky, Richard",
    title = "{Complex Spinors and Unified Theories}",
    eprint = "1306.4669",
    archivePrefix = "arXiv",
    primaryClass = "hep-th",
    reportNumber = "PRINT-80-0576",
    journal = "Conf. Proc. C",
    volume = "790927",
    pages = "315--321",
    year = "1979"
}

@article{Yanagida:1980xy,
    author = "Yanagida, Tsutomu",
    title = "{Horizontal Symmetry and Masses of Neutrinos}",
    reportNumber = "TU-80-208",
    doi = "10.1143/PTP.64.1103",
    journal = "Prog. Theor. Phys.",
    volume = "64",
    pages = "1103",
    year = "1980"
}

@article{Minkowski:1977sc,
    author = "Minkowski, Peter",
    title = "{$\mu \to e\gamma$ at a Rate of One Out of $10^{9}$ Muon Decays?}",
    reportNumber = "Print-77-0182 (BERN)",
    doi = "10.1016/0370-2693(77)90435-X",
    journal = "Phys. Lett. B",
    volume = "67",
    pages = "421--428",
    year = "1977"
}

@article{DESI:2024mwx,
    author = "Adame, A. G. and others",
    collaboration = "DESI",
    title = "{DESI 2024 VI: cosmological constraints from the measurements of baryon acoustic oscillations}",
    eprint = "2404.03002",
    archivePrefix = "arXiv",
    primaryClass = "astro-ph.CO",
    reportNumber = "FERMILAB-PUB-24-0154-PPD",
    doi = "10.1088/1475-7516/2025/02/021",
    journal = "JCAP",
    volume = "02",
    pages = "021",
    year = "2025"
}

@article{Schechter:1980gr,
    author = "Schechter, J. and Valle, J. W. F.",
    title = "{Neutrino Masses in SU(2) x U(1) Theories}",
    reportNumber = "SU-4217-167, COO-3533-167",
    doi = "10.1103/PhysRevD.22.2227",
    journal = "Phys. Rev. D",
    volume = "22",
    pages = "2227",
    year = "1980"
}

@article{Maki:1962mu,
    author = "Maki, Ziro and Nakagawa, Masami and Sakata, Shoichi",
    title = "{Remarks on the unified model of elementary particles}",
    doi = "10.1143/PTP.28.870",
    journal = "Prog. Theor. Phys.",
    volume = "28",
    pages = "870--880",
    year = "1962"
}

@article{Ibarra:2024tpt,
    author = "Ibarra, Alejandro and Leister, Nicholas and Zhang, Di",
    title = "{Complete two-loop renormalization group equation of the Weinberg operator}",
    eprint = "2411.08011",
    archivePrefix = "arXiv",
    primaryClass = "hep-ph",
    reportNumber = "MITP-24-081",
    doi = "10.1007/JHEP03(2025)214",
    journal = "JHEP",
    volume = "03",
    pages = "214",
    year = "2025"
}

@article{Frigerio:2011in,
    author = "Frigerio, Michele and Hambye, Thomas and Masso, Eduard",
    title = "{Sub-GeV dark matter as pseudo-Goldstone from the seesaw scale}",
    eprint = "1107.4564",
    archivePrefix = "arXiv",
    primaryClass = "hep-ph",
    reportNumber = "ULB-TH-11-17",
    doi = "10.1103/PhysRevX.1.021026",
    journal = "Phys. Rev. X",
    volume = "1",
    pages = "021026",
    year = "2011"
}

@article{Rothstein:1992rh,
    author = "Rothstein, I. Z. and Babu, K. S. and Seckel, D.",
    title = "{Planck scale symmetry breaking and majoron physics}",
    eprint = "hep-ph/9301213",
    archivePrefix = "arXiv",
    reportNumber = "UM-TH-92-31, BA-92-77",
    doi = "10.1016/0550-3213(93)90368-Y",
    journal = "Nucl. Phys. B",
    volume = "403",
    pages = "725--748",
    year = "1993"
}

@article{deGiorgi:2023tvn,
    author = "de Giorgi, Arturo and Merlo, Luca and Ponce D\'\i{}az, Xavier and Rigolin, Stefano",
    title = "{The minimal massive Majoron Seesaw Model}",
    eprint = "2312.13417",
    archivePrefix = "arXiv",
    primaryClass = "hep-ph",
    doi = "10.1007/JHEP03(2024)094",
    journal = "JHEP",
    volume = "03",
    pages = "094",
    year = "2024"
}

@article{Barr:1992qq,
    author = "Barr, Stephen M. and Seckel, D.",
    title = "{Planck scale corrections to axion models}",
    reportNumber = "BA-92-11",
    doi = "10.1103/PhysRevD.46.539",
    journal = "Phys. Rev. D",
    volume = "46",
    pages = "539--549",
    year = "1992"
}

@article{Garcia-Cely:2017oco,
    author = "Garcia-Cely, Camilo and Heeck, Julian",
    title = "{Neutrino Lines from Majoron Dark Matter}",
    eprint = "1701.07209",
    archivePrefix = "arXiv",
    primaryClass = "hep-ph",
    reportNumber = "ULB-TH-17-01",
    doi = "10.1007/JHEP05(2017)102",
    journal = "JHEP",
    volume = "05",
    pages = "102",
    year = "2017"
}

@article{Heeck:2019guh,
    author = "Heeck, Julian and Patel, Hiren H.",
    title = "{Majoron at two loops}",
    eprint = "1909.02029",
    archivePrefix = "arXiv",
    primaryClass = "hep-ph",
    reportNumber = "UCI-TR-2019-23",
    doi = "10.1103/PhysRevD.100.095015",
    journal = "Phys. Rev. D",
    volume = "100",
    number = "9",
    pages = "095015",
    year = "2019"
}

@article{Chikashige:1980ui,
    author = "Chikashige, Y. and Mohapatra, Rabindra N. and Peccei, R. D.",
    title = "{Are There Real Goldstone Bosons Associated with Broken Lepton Number?}",
    reportNumber = "MPI-PAE-PTH-36-80",
    doi = "10.1016/0370-2693(81)90011-3",
    journal = "Phys. Lett. B",
    volume = "98",
    pages = "265--268",
    year = "1981"
}

@article{Chikashige:1980qk,
    author = "Chikashige, Y. and Mohapatra, Rabindra N. and Peccei, R. D.",
    title = "{Spontaneously Broken Lepton Number and Cosmological Constraints on the Neutrino Mass Spectrum}",
    reportNumber = "MPI-PAE/PTh 40/80",
    doi = "10.1103/PhysRevLett.45.1926",
    journal = "Phys. Rev. Lett.",
    volume = "45",
    pages = "1926",
    year = "1980"
}

@article{Gu:2009hn,
	archiveprefix = {arXiv},
	author = {Gu, Pei-Hong and Sarkar, Utpal},
	doi = {10.1140/epjc/s10052-011-1560-2},
	eprint = {0909.5468},
	journal = {Eur. Phys. J. C},
	pages = {1560},
	primaryclass = {hep-ph},
	title = {{Leptogenesis Bound on Spontaneous Symmetry Breaking of Global Lepton Number}},
	volume = {71},
	year = {2011}}

@article{Heeck:2017xbu,
	archiveprefix = {arXiv},
	author = {Heeck, Julian and Teresi, Daniele},
	doi = {10.1103/PhysRevD.96.035018},
	eprint = {1706.09909},
	journal = {Phys. Rev. D},
	number = {3},
	pages = {035018},
	primaryclass = {hep-ph},
	reportnumber = {ULB-TH-17-09},
	title = {{Cold keV dark matter from decays and scatterings}},
	volume = {96},
	year = {2017}}

@article{Reig:2019sok,
	archiveprefix = {arXiv},
	author = {Reig, Mario and Valle, Jos\'e W. F. and Yamada, Masaki},
	doi = {10.1088/1475-7516/2019/09/029},
	eprint = {1905.01287},
	journal = {JCAP},
	pages = {029},
	primaryclass = {hep-ph},
	title = {{Light majoron cold dark matter from topological defects and the formation of boson stars}},
	volume = {09},
	year = {2019}}

@article{Akhmedov:1992hi,
    author = "Akhmedov, Evgeny K. and Berezhiani, Z. G. and Mohapatra, R. N. and Senjanovic, G.",
    title = "{Planck scale effects on the majoron}",
    eprint = "hep-ph/9209285",
    archivePrefix = "arXiv",
    reportNumber = "UMDHEP-93-020, LMU-12-92, SISSA-149-92-EP, IC-92-328",
    doi = "10.1016/0370-2693(93)90887-N",
    journal = "Phys. Lett. B",
    volume = "299",
    pages = "90--93",
    year = "1993"
}

@article{Boulebnane:2017fxw,
    author = "Boulebnane, Sami and Heeck, Julian and Nguyen, Anne and Teresi, Daniele",
    title = "{Cold light dark matter in extended seesaw models}",
    eprint = "1709.07283",
    archivePrefix = "arXiv",
    primaryClass = "hep-ph",
    reportNumber = "ULB-TH-17-17",
    doi = "10.1088/1475-7516/2018/04/006",
    journal = "JCAP",
    volume = "04",
    pages = "006",
    year = "2018"
}

@article{Biggio:2023gtm,
	archiveprefix = {arXiv},
	author = {Biggio, Carla and Calibbi, Lorenzo and Ota, Toshihiko and Zanchini, Samuele},
	doi = {10.1103/PhysRevD.108.115003},
	eprint = {2304.12527},
	journal = {Phys. Rev. D},
	number = {11},
	pages = {115003},
	primaryclass = {hep-ph},
	reportnumber = {IFT-UAM/CSIC-23-45},
	title = {{Majoron dark matter from a type II seesaw model}},
	volume = {108},
	year = {2023}}

@article{Hindmarsh:1994re,
    author = "Hindmarsh, M. B. and Kibble, T. W. B.",
    title = "{Cosmic strings}",
    eprint = "hep-ph/9411342",
    archivePrefix = "arXiv",
    reportNumber = "SUSX-TP-94-74, IMPERIAL-TP-94-95-5, NI-94025",
    doi = "10.1088/0034-4885/58/5/001",
    journal = "Rept. Prog. Phys.",
    volume = "58",
    pages = "477--562",
    year = "1995"
}

@article{Gelmini:1980re,
    author = "Gelmini, G. B. and Roncadelli, M.",
    title = "{Left-Handed Neutrino Mass Scale and Spontaneously Broken Lepton Number}",
    reportNumber = "MPI-PAE-PTH-50-80",
    doi = "10.1016/0370-2693(81)90559-1",
    journal = "Phys. Lett. B",
    volume = "99",
    pages = "411--415",
    year = "1981"
}

@article{Cline:1993ht,
    author = "Cline, James M. and Kainulainen, Kimmo and Olive, Keith A.",
    title = "{Constraints on majoron models, neutrino masses and baryogenesis}",
    eprint = "hep-ph/9304229",
    archivePrefix = "arXiv",
    reportNumber = "UMN-TH-1113-93, UMN-TH-1129-93, TPI-MINN-93-13-T",
    doi = "10.1016/0927-6505(93)90005-X",
    journal = "Astropart. Phys.",
    volume = "1",
    pages = "387--398",
    year = "1993"
}

@article{Schechter:1981cv,
    author = "Schechter, J. and Valle, J. W. F.",
    title = "{Neutrino Decay and Spontaneous Violation of Lepton Number}",
    reportNumber = "SU-4217-203, COO-3533-203",
    doi = "10.1103/PhysRevD.25.774",
    journal = "Phys. Rev. D",
    volume = "25",
    pages = "774",
    year = "1982"
}

@article{Giudice:2003jh,
    author = "Giudice, G. F. and Notari, A. and Raidal, M. and Riotto, A. and Strumia, A.",
    title = "{Towards a complete theory of thermal leptogenesis in the SM and MSSM}",
    eprint = "hep-ph/0310123",
    archivePrefix = "arXiv",
    reportNumber = "IFUP-TH-2003-37, CERN-TH-2003-240",
    doi = "10.1016/j.nuclphysb.2004.02.019",
    journal = "Nucl. Phys. B",
    volume = "685",
    pages = "89--149",
    year = "2004"
}

@article{Baur:2017stq,
    author = "Baur, Julien and Palanque-Delabrouille, Nathalie and Yeche, Christophe and Boyarsky, Alexey and Ruchayskiy, Oleg and Armengaud, \'Eric and Lesgourgues, Julien",
    title = "{Constraints from Ly-$\alpha$ forests on non-thermal dark matter including resonantly-produced sterile neutrinos}",
    eprint = "1706.03118",
    archivePrefix = "arXiv",
    primaryClass = "astro-ph.CO",
    doi = "10.1088/1475-7516/2017/12/013",
    journal = "JCAP",
    volume = "12",
    pages = "013",
    year = "2017"
}

@article{Yeche:2017upn,
    author = "Y\`eche, Christophe and Palanque-Delabrouille, Nathalie and Baur, Julien and du Mas des Bourboux, H\'elion",
    title = "{Constraints on neutrino masses from Lyman-alpha forest power spectrum with BOSS and XQ-100}",
    eprint = "1702.03314",
    archivePrefix = "arXiv",
    primaryClass = "astro-ph.CO",
    doi = "10.1088/1475-7516/2017/06/047",
    journal = "JCAP",
    volume = "06",
    pages = "047",
    year = "2017"
}

@article{Abbott:1982af,
    author = "Abbott, L. F. and Sikivie, P.",
    editor = "Srednicki, M. A.",
    title = "{A Cosmological Bound on the Invisible Axion}",
    reportNumber = "PRINT-82-0695 (BRANDEIS)",
    doi = "10.1016/0370-2693(83)90638-X",
    journal = "Phys. Lett. B",
    volume = "120",
    pages = "133--136",
    year = "1983"
}

@article{Dine:1982ah,
    author = "Dine, Michael and Fischler, Willy",
    editor = "Srednicki, M. A.",
    title = "{The Not So Harmless Axion}",
    reportNumber = "UPR-0201T",
    doi = "10.1016/0370-2693(83)90639-1",
    journal = "Phys. Lett. B",
    volume = "120",
    pages = "137--141",
    year = "1983"
}

@article{Preskill:1982cy,
    author = "Preskill, John and Wise, Mark B. and Wilczek, Frank",
    editor = "Srednicki, M. A.",
    title = "{Cosmology of the Invisible Axion}",
    reportNumber = "HUTP-82-A048, NSF-ITP-82-103",
    doi = "10.1016/0370-2693(83)90637-8",
    journal = "Phys. Lett. B",
    volume = "120",
    pages = "127--132",
    year = "1983"
}

@article{Arias:2012az,
    author = "Arias, Paola and Cadamuro, Davide and Goodsell, Mark and Jaeckel, Joerg and Redondo, Javier and Ringwald, Andreas",
    title = "{WISPy Cold Dark Matter}",
    eprint = "1201.5902",
    archivePrefix = "arXiv",
    primaryClass = "hep-ph",
    reportNumber = "DESY-11-226, MPP-2011-140, CERN-PH-TH-2011-323, IPPP-11-80, DCPT-11-160",
    doi = "10.1088/1475-7516/2012/06/013",
    journal = "JCAP",
    volume = "06",
    pages = "013",
    year = "2012"
}

@article{Blinov:2019rhb,
    author = "Blinov, Nikita and Dolan, Matthew J and Draper, Patrick and Kozaczuk, Jonathan",
    title = "{Dark matter targets for axionlike particle searches}",
    eprint = "1905.06952",
    archivePrefix = "arXiv",
    primaryClass = "hep-ph",
    reportNumber = "FERMILAB-PUB-19-197-A-T",
    doi = "10.1103/PhysRevD.100.015049",
    journal = "Phys. Rev. D",
    volume = "100",
    number = "1",
    pages = "015049",
    year = "2019"
}

@article{Chun:2023eqc,
    author = "Chun, Eung Jin and Jung, Tae Hyun",
    title = "{Leptogenesis driven by a Majoron}",
    eprint = "2311.09005",
    archivePrefix = "arXiv",
    primaryClass = "hep-ph",
    reportNumber = "CTPU-PTC-23-48",
    doi = "10.1103/PhysRevD.109.095004",
    journal = "Phys. Rev. D",
    volume = "109",
    number = "9",
    pages = "095004",
    year = "2024"
}

@article{GrillidiCortona:2015jxo,
    author = "Grilli di Cortona, Giovanni and Hardy, Edward and Pardo Vega, Javier and Villadoro, Giovanni",
    title = "{The QCD axion, precisely}",
    eprint = "1511.02867",
    archivePrefix = "arXiv",
    primaryClass = "hep-ph",
    doi = "10.1007/JHEP01(2016)034",
    journal = "JHEP",
    volume = "01",
    pages = "034",
    year = "2016"
}

@article{Beltran:2005xd,
    author = "Beltran, Maria and Garcia-Bellido, Juan and Lesgourgues, Julien and Liddle, Andrew R and Slosar, Anze",
    title = "{Bayesian model selection and isocurvature perturbations}",
    eprint = "astro-ph/0501477",
    archivePrefix = "arXiv",
    reportNumber = "IFT-UAM-CSIC-05-02, LAPTH-1085-05, DFPD-05-A05",
    doi = "10.1103/PhysRevD.71.063532",
    journal = "Phys. Rev. D",
    volume = "71",
    pages = "063532",
    year = "2005"
}

@article{Beltran:2006sq,
    author = "Beltran, Maria and Garcia-Bellido, Juan and Lesgourgues, Julien",
    title = "{Isocurvature bounds on axions revisited}",
    eprint = "hep-ph/0606107",
    archivePrefix = "arXiv",
    reportNumber = "LAPTH-1149-06",
    doi = "10.1103/PhysRevD.75.103507",
    journal = "Phys. Rev. D",
    volume = "75",
    pages = "103507",
    year = "2007"
}

@article{Crotty:2003rz,
    author = "Crotty, Patrick and Garcia-Bellido, Juan and Lesgourgues, Julien and Riazuelo, Alain",
    title = "{Bounds on isocurvature perturbations from CMB and LSS data}",
    eprint = "astro-ph/0306286",
    archivePrefix = "arXiv",
    reportNumber = "LAPTH-985-03, CERN-TH-2003-124, SACLAY-SPHT-T03-076, IFT-UNAM-CSIC-03-19",
    doi = "10.1103/PhysRevLett.91.171301",
    journal = "Phys. Rev. Lett.",
    volume = "91",
    pages = "171301",
    year = "2003"
}

@article{Hoof:2024quk,
    author = "Hoof, Sebastian and Marsh, David J. E. and Sisk-Reyn\'es, J\'ulia and Matthews, James H. and Reynolds, Christopher",
    title = "{Getting More Out of Black Hole Superradiance: a Statistically Rigorous Approach to Ultralight Boson Constraints}",
    eprint = "2406.10337",
    archivePrefix = "arXiv",
    primaryClass = "hep-ph",
    month = "6",
    year = "2024"
}

@article{Arvanitaki:2016qwi,
    author = "Arvanitaki, Asimina and Baryakhtar, Masha and Dimopoulos, Savas and Dubovsky, Sergei and Lasenby, Robert",
    title = "{Black Hole Mergers and the QCD Axion at Advanced LIGO}",
    eprint = "1604.03958",
    archivePrefix = "arXiv",
    primaryClass = "hep-ph",
    doi = "10.1103/PhysRevD.95.043001",
    journal = "Phys. Rev. D",
    volume = "95",
    number = "4",
    pages = "043001",
    year = "2017"
}

@article{Arvanitaki:2014wva,
    author = "Arvanitaki, Asimina and Baryakhtar, Masha and Huang, Xinlu",
    title = "{Discovering the QCD Axion with Black Holes and Gravitational Waves}",
    eprint = "1411.2263",
    archivePrefix = "arXiv",
    primaryClass = "hep-ph",
    doi = "10.1103/PhysRevD.91.084011",
    journal = "Phys. Rev. D",
    volume = "91",
    number = "8",
    pages = "084011",
    year = "2015"
}

@article{Arvanitaki:2010sy,
    author = "Arvanitaki, Asimina and Dubovsky, Sergei",
    title = "{Exploring the String Axiverse with Precision Black Hole Physics}",
    eprint = "1004.3558",
    archivePrefix = "arXiv",
    primaryClass = "hep-th",
    doi = "10.1103/PhysRevD.83.044026",
    journal = "Phys. Rev. D",
    volume = "83",
    pages = "044026",
    year = "2011"
}

@article{CMB-S4:2016ple,
    author = "Abazajian, Kevork N. and others",
    collaboration = "CMB-S4",
    title = "{CMB-S4 Science Book, First Edition}",
    eprint = "1610.02743",
    archivePrefix = "arXiv",
    primaryClass = "astro-ph.CO",
    reportNumber = "FERMILAB-FN-1024-A-AE",
    month = "10",
    year = "2016"
}

@article{Cielo:2023bqp,
    author = "Cielo, Mattia and Escudero, Miguel and Mangano, Gianpiero and Pisanti, Ofelia",
    title = "{Neff in the Standard Model at NLO is 3.043}",
    eprint = "2306.05460",
    archivePrefix = "arXiv",
    primaryClass = "hep-ph",
    reportNumber = "CERN-TH-2023-103",
    doi = "10.1103/PhysRevD.108.L121301",
    journal = "Phys. Rev. D",
    volume = "108",
    number = "12",
    pages = "L121301",
    year = "2023"
}

@article{Akita:2023qiz,
    author = "Akita, Kensuke and Niibo, Michiru",
    title = "{Updated constraints and future prospects on majoron dark matter}",
    eprint = "2304.04430",
    archivePrefix = "arXiv",
    primaryClass = "hep-ph",
    reportNumber = "CTPU-PTC-23-10",
    doi = "10.1007/JHEP07(2023)132",
    journal = "JHEP",
    volume = "07",
    pages = "132",
    year = "2023"
}

@article{Borexino:2019wln,
    author = "Agostini, M. and others",
    collaboration = "Borexino",
    title = "{Search for low-energy neutrinos from astrophysical sources with Borexino}",
    eprint = "1909.02422",
    archivePrefix = "arXiv",
    primaryClass = "hep-ex",
    reportNumber = "FERMILAB-PUB-21-152-AE",
    doi = "10.1016/j.astropartphys.2020.102509",
    journal = "Astropart. Phys.",
    volume = "125",
    pages = "102509",
    year = "2021"
}

@article{KamLAND:2021gvi,
    author = "Abe, S. and others",
    collaboration = "KamLAND",
    title = "{Limits on Astrophysical Antineutrinos with the KamLAND Experiment}",
    eprint = "2108.08527",
    archivePrefix = "arXiv",
    primaryClass = "astro-ph.HE",
    doi = "10.3847/1538-4357/ac32c1",
    journal = "Astrophys. J.",
    volume = "925",
    number = "1",
    pages = "14",
    year = "2022"
}

@article{Palomares-Ruiz:2007egs,
    author = "Palomares-Ruiz, Sergio",
    title = "{Model-independent bound on the dark matter lifetime}",
    eprint = "0712.1937",
    archivePrefix = "arXiv",
    primaryClass = "astro-ph",
    reportNumber = "IPPP-07-96, DCPT-07-192",
    doi = "10.1016/j.physletb.2008.05.040",
    journal = "Phys. Lett. B",
    volume = "665",
    pages = "50--53",
    year = "2008"
}

@article{Olivares-DelCampo:2017feq,
    author = "Olivares-Del Campo, Andr\'es and B\oe{}hm, C\'eline and Palomares-Ruiz, Sergio and Pascoli, Silvia",
    title = "{Dark matter-neutrino interactions through the lens of their cosmological implications}",
    eprint = "1711.05283",
    archivePrefix = "arXiv",
    primaryClass = "hep-ph",
    reportNumber = "IFIC-17-54, IPPP-17-84",
    doi = "10.1103/PhysRevD.97.075039",
    journal = "Phys. Rev. D",
    volume = "97",
    number = "7",
    pages = "075039",
    year = "2018"
}

@article{Super-Kamiokande:2011lwo,
    author = "Bays, K. and others",
    collaboration = "Super-Kamiokande",
    title = "{Supernova Relic Neutrino Search at Super-Kamiokande}",
    eprint = "1111.5031",
    archivePrefix = "arXiv",
    primaryClass = "hep-ex",
    doi = "10.1103/PhysRevD.85.052007",
    journal = "Phys. Rev. D",
    volume = "85",
    pages = "052007",
    year = "2012"
}

@article{Super-Kamiokande:2013ufi,
    author = "Zhang, H. and others",
    collaboration = "Super-Kamiokande",
    title = "{Supernova Relic Neutrino Search with Neutron Tagging at Super-Kamiokande-IV}",
    eprint = "1311.3738",
    archivePrefix = "arXiv",
    primaryClass = "hep-ex",
    doi = "10.1016/j.astropartphys.2014.05.004",
    journal = "Astropart. Phys.",
    volume = "60",
    pages = "41--46",
    year = "2015"
}

@article{Super-Kamiokande:2021jaq,
    author = "Abe, K. and others",
    collaboration = "Super-Kamiokande",
    title = "{Diffuse supernova neutrino background search at Super-Kamiokande}",
    eprint = "2109.11174",
    archivePrefix = "arXiv",
    primaryClass = "astro-ph.HE",
    doi = "10.1103/PhysRevD.104.122002",
    journal = "Phys. Rev. D",
    volume = "104",
    number = "12",
    pages = "122002",
    year = "2021"
}

@article{Liang:2024vnd,
    author = "Liang, Qiuyue and Ponce D\'\i{}az, Xavier and Yanagida, Tsutomu T.",
    title = "{Axion Detection Experiments Can Probe Majoron Models}",
    eprint = "2406.19083",
    archivePrefix = "arXiv",
    primaryClass = "hep-ph",
    doi = "10.1103/PhysRevLett.134.151803",
    journal = "Phys. Rev. Lett.",
    volume = "134",
    number = "15",
    pages = "151803",
    year = "2025"
}

@article{Bell:2020rkw,
    author = "Bell, Nicole F. and Dolan, Matthew J. and Robles, Sandra",
    title = "{Searching for Sub-GeV Dark Matter in the Galactic Centre using Hyper-Kamiokande}",
    eprint = "2005.01950",
    archivePrefix = "arXiv",
    primaryClass = "hep-ph",
    doi = "10.1088/1475-7516/2020/09/019",
    journal = "JCAP",
    volume = "09",
    pages = "019",
    year = "2020"
}

@article{Akita:2022lit,
    author = "Akita, Kensuke and Lambiase, Gaetano and Niibo, Michiru and Yamaguchi, Masahide",
    title = "{Neutrino lines from MeV dark matter annihilation and decay in JUNO}",
    eprint = "2206.06755",
    archivePrefix = "arXiv",
    primaryClass = "hep-ph",
    reportNumber = "CTPU-PTC-22-14",
    doi = "10.1088/1475-7516/2022/10/097",
    journal = "JCAP",
    volume = "10",
    pages = "097",
    year = "2022"
}

@article{Irsic:2017ixq,
    author = "Ir\v{s}i\v{c}, Vid and others",
    title = "{New Constraints on the free-streaming of warm dark matter from intermediate and small scale Lyman-$\alpha$ forest data}",
    eprint = "1702.01764",
    archivePrefix = "arXiv",
    primaryClass = "astro-ph.CO",
    doi = "10.1103/PhysRevD.96.023522",
    journal = "Phys. Rev. D",
    volume = "96",
    number = "2",
    pages = "023522",
    year = "2017"
}

@article{Slatyer:2016qyl,
    author = "Slatyer, Tracy R. and Wu, Chih-Liang",
    title = "{General Constraints on Dark Matter Decay from the Cosmic Microwave Background}",
    eprint = "1610.06933",
    archivePrefix = "arXiv",
    primaryClass = "astro-ph.CO",
    reportNumber = "MIT-CTP-4842",
    doi = "10.1103/PhysRevD.95.023010",
    journal = "Phys. Rev. D",
    volume = "95",
    number = "2",
    pages = "023010",
    year = "2017"
}

@article{XENON:2021qze,
	archiveprefix = {arXiv},
	author = {Aprile, E. and others},
	collaboration = {XENON},
	doi = {10.1103/PhysRevD.106.022001},
	eprint = {2112.12116},
	journal = {Phys. Rev. D},
	number = {2},
	pages = {022001},
	primaryclass = {hep-ex},
	title = {{Emission of single and few electrons in XENON1T and limits on light dark matter}},
	volume = {106},
	year = {2022}}

@article{XENON:2022ltv,
	archiveprefix = {arXiv},
	author = {Aprile, E. and others},
	collaboration = {XENON},
	doi = {10.1103/PhysRevLett.129.161805},
	eprint = {2207.11330},
	journal = {Phys. Rev. Lett.},
	number = {16},
	pages = {161805},
	primaryclass = {hep-ex},
	title = {{Search for New Physics in Electronic Recoil Data from XENONnT}},
	volume = {129},
	year = {2022}}

@article{Langhoff:2022bij,
    author = "Langhoff, Kevin and Outmezguine, Nadav Joseph and Rodd, Nicholas L.",
    title = "{Irreducible Axion Background}",
    eprint = "2209.06216",
    archivePrefix = "arXiv",
    primaryClass = "hep-ph",
    reportNumber = "CERN-TH-2022-148",
    doi = "10.1103/PhysRevLett.129.241101",
    journal = "Phys. Rev. Lett.",
    volume = "129",
    number = "24",
    pages = "241101",
    year = "2022"
}

@article{Thorpe-Morgan:2020rwc,
    author = {Thorpe-Morgan, Charles and Malyshev, Denys and Santangelo, Andrea and Jochum, Josef and J\"ager, Barbara and Sasaki, Manami and Saeedi, Sara},
    title = "{THESEUS insights into axionlike particles, dark photon, and sterile neutrino dark matter}",
    eprint = "2008.08306",
    archivePrefix = "arXiv",
    primaryClass = "astro-ph.HE",
    doi = "10.1103/PhysRevD.102.123003",
    journal = "Phys. Rev. D",
    volume = "102",
    number = "12",
    pages = "123003",
    year = "2020"
}

@article{Roach:2022lgo,
    author = "Roach, Brandon M. and Rossland, Steven and Ng, Kenny C. Y. and Perez, Kerstin and Beacom, John F. and Grefenstette, Brian W. and Horiuchi, Shunsaku and Krivonos, Roman and Wik, Daniel R.",
    title = "{Long-exposure NuSTAR constraints on decaying dark matter in the Galactic halo}",
    eprint = "2207.04572",
    archivePrefix = "arXiv",
    primaryClass = "astro-ph.HE",
    doi = "10.1103/PhysRevD.107.023009",
    journal = "Phys. Rev. D",
    volume = "107",
    number = "2",
    pages = "023009",
    year = "2023"
}

@article{Ng:2019gch,
    author = "Ng, Kenny C. Y. and Roach, Brandon M. and Perez, Kerstin and Beacom, John F. and Horiuchi, Shunsaku and Krivonos, Roman and Wik, Daniel R.",
    title = "{New Constraints on Sterile Neutrino Dark Matter from $NuSTAR$ M31 Observations}",
    eprint = "1901.01262",
    archivePrefix = "arXiv",
    primaryClass = "astro-ph.HE",
    doi = "10.1103/PhysRevD.99.083005",
    journal = "Phys. Rev. D",
    volume = "99",
    pages = "083005",
    year = "2019"
}

@article{Perez:2016tcq,
    author = "Perez, Kerstin and Ng, Kenny C. Y. and Beacom, John F. and Hersh, Cora and Horiuchi, Shunsaku and Krivonos, Roman",
    title = "{Almost closing the \ensuremath{\nu}MSM sterile neutrino dark matter window with NuSTAR}",
    eprint = "1609.00667",
    archivePrefix = "arXiv",
    primaryClass = "astro-ph.HE",
    doi = "10.1103/PhysRevD.95.123002",
    journal = "Phys. Rev. D",
    volume = "95",
    number = "12",
    pages = "123002",
    year = "2017"
}

@article{Calore:2022pks,
    author = "Calore, Francesca and Dekker, Ariane and Serpico, Pasquale Dario and Siegert, Thomas",
    title = "{Constraints on light decaying dark matter candidates from 16~yr of INTEGRAL/SPI observations}",
    eprint = "2209.06299",
    archivePrefix = "arXiv",
    primaryClass = "hep-ph",
    doi = "10.1093/mnras/stad457",
    journal = "Mon. Not. Roy. Astron. Soc.",
    volume = "520",
    number = "3",
    pages = "4167--4172",
    year = "2023",
    note = "[Erratum: Mon.Not.Roy.Astron.Soc. 538, 132 (2025)]"
}

@article{Shutt:2025xvc,
    author = "Shutt, Tom and others",
    title = "{The GammaTPC Gamma-Ray Telescope Concept}",
    eprint = "2502.14841",
    archivePrefix = "arXiv",
    primaryClass = "astro-ph.IM",
    reportNumber = "FERMILAB-PUB-25-0108-PPD",
    month = "2",
    year = "2025"
}

@misc{AxionLimits,
  author       = {Ciaran O'Hare},
  title        = {cajohare/AxionLimits: AxionLimits},
  month        = jul,
  year         = 2020,
  publisher    = {Zenodo},
  version      = {v1.0},
  doi          = {10.5281/zenodo.3932430},
  howpublished = {\url{https://cajohare.github.io/AxionLimits/}}
}

@article{Weinberg:1977ma,
    author = "Weinberg, Steven",
    title = "{A New Light Boson?}",
    reportNumber = "HUTP-77/A074",
    doi = "10.1103/PhysRevLett.40.223",
    journal = "Phys. Rev. Lett.",
    volume = "40",
    pages = "223--226",
    year = "1978"
}

@article{Wilczek:1977pj,
    author = "Wilczek, Frank",
    title = "{Problem of Strong  $P$  and  $T$  Invariance in the Presence of Instantons}",
    reportNumber = "Print-77-0939 (COLUMBIA)",
    doi = "10.1103/PhysRevLett.40.279",
    journal = "Phys. Rev. Lett.",
    volume = "40",
    pages = "279--282",
    year = "1978"
}

@article{Pontecorvo:1967fh,
    author = "Pontecorvo, B.",
    title = "{Neutrino Experiments and the Problem of Conservation of Leptonic Charge}",
    journal = "Zh. Eksp. Teor. Fiz.",
    volume = "53",
    pages = "1717--1725",
    year = "1967"
}

@article{DUNE:2015lol,
    author = "Acciarri, R. and others",
    collaboration = "DUNE",
    title = "{Long-Baseline Neutrino Facility (LBNF) and Deep Underground Neutrino Experiment (DUNE)}: {Conceptual Design Report, Volume 2: The Physics Program for DUNE at LBNF}",
    eprint = "1512.06148",
    archivePrefix = "arXiv",
    primaryClass = "physics.ins-det",
    reportNumber = "FERMILAB-DESIGN-2016-02",
    month = "12",
    year = "2015"
}

@article{Marsh:2015xka,
    author = "Marsh, David J. E.",
    title = "{Axion Cosmology}",
    eprint = "1510.07633",
    archivePrefix = "arXiv",
    primaryClass = "astro-ph.CO",
    reportNumber = "KCL-PH-TH-2015-50",
    doi = "10.1016/j.physrep.2016.06.005",
    journal = "Phys. Rept.",
    volume = "643",
    pages = "1--79",
    year = "2016"
}

@article{Cirelli:2024ssz,
    author = "Cirelli, Marco and Strumia, Alessandro and Zupan, Jure",
    title = "{Dark Matter}",
    eprint = "2406.01705",
    archivePrefix = "arXiv",
    primaryClass = "hep-ph",
    month = "6",
    year = "2024"
}

@article{Sikivie:2006ni,
    author = "Sikivie, Pierre",
    editor = "Kuster, Markus and Raffelt, Georg and Beltran, Berta",
    title = "{Axion Cosmology}",
    eprint = "astro-ph/0610440",
    archivePrefix = "arXiv",
    reportNumber = "UFIFT-HEP-06-16",
    doi = "10.1007/978-3-540-73518-2_2",
    journal = "Lect. Notes Phys.",
    volume = "741",
    pages = "19--50",
    year = "2008"
}

@article{Sikivie:1982qv,
    author = "Sikivie, P.",
    title = "{Of Axions, Domain Walls and the Early Universe}",
    reportNumber = "UFTP-82-3",
    doi = "10.1103/PhysRevLett.48.1156",
    journal = "Phys. Rev. Lett.",
    volume = "48",
    pages = "1156--1159",
    year = "1982"
}

@article{DiLuzio:2020wdo,
    author = "Di Luzio, Luca and Giannotti, Maurizio and Nardi, Enrico and Visinelli, Luca",
    title = "{The landscape of QCD axion models}",
    eprint = "2003.01100",
    archivePrefix = "arXiv",
    primaryClass = "hep-ph",
    reportNumber = "DESY 20-036, DESY-20-036",
    doi = "10.1016/j.physrep.2020.06.002",
    journal = "Phys. Rept.",
    volume = "870",
    pages = "1--117",
    year = "2020"
}

@article{DUNE:2022aul,
    author = "Abed Abud, A. and others",
    collaboration = "DUNE",
    title = "{Snowmass Neutrino Frontier: DUNE Physics Summary}",
    eprint = "2203.06100",
    archivePrefix = "arXiv",
    primaryClass = "hep-ex",
    reportNumber = "FERMILAB-FN-1168-LBNF",
    month = "3",
    year = "2022"
}

@article{DiValentino:2024xsv,
    author = "Di Valentino, Eleonora and Gariazzo, Stefano and Mena, Olga",
    title = "{Neutrinos in Cosmology}",
    eprint = "2404.19322",
    archivePrefix = "arXiv",
    primaryClass = "astro-ph.CO",
    month = "4",
    year = "2024"
}

@article{Baur:2015jsy,
    author = "Baur, Julien and Palanque-Delabrouille, Nathalie and Y\`eche, Christophe and Magneville, Christophe and Viel, Matteo",
    title = "{Lyman-alpha Forests cool Warm Dark Matter}",
    eprint = "1512.01981",
    archivePrefix = "arXiv",
    primaryClass = "astro-ph.CO",
    doi = "10.1088/1475-7516/2016/08/012",
    journal = "JCAP",
    volume = "08",
    pages = "012",
    year = "2016"
}

@article{CUPID:2022jlk,
    author = "Alfonso, K. and others",
    collaboration = "CUPID",
    title = "{CUPID: The Next-Generation Neutrinoless Double Beta Decay Experiment}",
    doi = "10.1007/s10909-022-02909-3",
    journal = "J. Low Temp. Phys.",
    volume = "211",
    number = "5-6",
    pages = "375--383",
    year = "2023"
}

@article{nEXO:2021ujk,
    author = "Adhikari, G. and others",
    collaboration = "nEXO",
    title = "{nEXO: neutrinoless double beta decay search beyond 10$^{28}$ year half-life sensitivity}",
    eprint = "2106.16243",
    archivePrefix = "arXiv",
    primaryClass = "nucl-ex",
    doi = "10.1088/1361-6471/ac3631",
    journal = "J. Phys. G",
    volume = "49",
    number = "1",
    pages = "015104",
    year = "2022"
}

@article{LEGEND:2021bnm,
    author = "Abgrall, N. and others",
    collaboration = "LEGEND",
    title = "{The Large Enriched Germanium Experiment for Neutrinoless $\beta\beta$ Decay}: {LEGEND-1000 Preconceptual Design Report}",
    eprint = "2107.11462",
    archivePrefix = "arXiv",
    primaryClass = "physics.ins-det",
    month = "7",
    year = "2021"
}

@article{Adams:2022jwx,
    author = "Adams, C. and others",
    title = "{Neutrinoless Double Beta Decay}",
    eprint = "2212.11099",
    archivePrefix = "arXiv",
    primaryClass = "nucl-ex",
    reportNumber = "FERMILAB-CONF-22-962-ND",
    month = "12",
    year = "2022"
}

@article{Granelli:2025cho,
    author = "Granelli, Alessandro and Hamaguchi, Koichi and Ramirez-Quezada, Maura E. and Shimada, Kengo and Wada, Juntaro and Yokoyama, Tatsuya",
    title = "{Insights on the Scale of Leptogenesis from Neutrino Masses and Neutrinoless Double-Beta Decay}",
    eprint = "2502.10093",
    archivePrefix = "arXiv",
    primaryClass = "hep-ph",
    month = "2",
    year = "2025"
}

@article{Vilenkin:1982ks,
    author = "Vilenkin, A. and Everett, A. E.",
    title = "{Cosmic Strings and Domain Walls in Models with Goldstone and PseudoGoldstone Bosons}",
    doi = "10.1103/PhysRevLett.48.1867",
    journal = "Phys. Rev. Lett.",
    volume = "48",
    pages = "1867--1870",
    year = "1982"
}

@article{CUORE:2024ikf,
    author = "Adams, D. Q. and others",
    collaboration = "CUORE",
    title = "{With or without $\nu$? Hunting for the seed of the matter-antimatter asymmetry}",
    eprint = "2404.04453",
    archivePrefix = "arXiv",
    primaryClass = "nucl-ex",
    month = "4",
    year = "2024"
}

@article{GERDA:2020xhi,
    author = "Agostini, M. and others",
    collaboration = "GERDA",
    title = "{Final Results of GERDA on the Search for Neutrinoless Double-$\beta$ Decay}",
    eprint = "2009.06079",
    archivePrefix = "arXiv",
    primaryClass = "nucl-ex",
    doi = "10.1103/PhysRevLett.125.252502",
    journal = "Phys. Rev. Lett.",
    volume = "125",
    number = "25",
    pages = "252502",
    year = "2020"
}

@article{KamLAND-Zen:2024eml,
    author = "Abe, S. and others",
    collaboration = "KamLAND-Zen",
    title = "{Search for Majorana Neutrinos with the Complete KamLAND-Zen Dataset}",
    eprint = "2406.11438",
    archivePrefix = "arXiv",
    primaryClass = "hep-ex",
    month = "6",
    year = "2024"
}

@article{Sakharov:1967dj,
    author = "Sakharov, A. D.",
    title = "{Violation of CP Invariance, C asymmetry, and baryon asymmetry of the universe}",
    doi = "10.1070/PU1991v034n05ABEH002497",
    journal = "Pisma Zh. Eksp. Teor. Fiz.",
    volume = "5",
    pages = "32--35",
    year = "1967"
}

@article{Plumacher:1996kc,
    author = "Plumacher, Michael",
    title = "{Baryogenesis and lepton number violation}",
    eprint = "hep-ph/9604229",
    archivePrefix = "arXiv",
    reportNumber = "DESY-96-052",
    doi = "10.1007/s002880050418",
    journal = "Z. Phys. C",
    volume = "74",
    pages = "549--559",
    year = "1997"
}

@article{Heeck:2016oda,
    author = "Heeck, Julian and Teresi, Daniele",
    title = "{Leptogenesis and neutral gauge bosons}",
    eprint = "1609.03594",
    archivePrefix = "arXiv",
    primaryClass = "hep-ph",
    reportNumber = "ULB-TH-16-15",
    doi = "10.1103/PhysRevD.94.095024",
    journal = "Phys. Rev. D",
    volume = "94",
    number = "9",
    pages = "095024",
    year = "2016"
}

@article{Martucci:2024trp,
    author = "Martucci, Luca and Risso, Nicol\`o and Valenti, Alessandro and Vecchi, Luca",
    title = "{Wormholes in the axiverse, and the species scale}",
    eprint = "2404.14489",
    archivePrefix = "arXiv",
    primaryClass = "hep-th",
    doi = "10.1007/JHEP07(2024)240",
    journal = "JHEP",
    volume = "07",
    pages = "240",
    year = "2024"
}

@article{Cohen:1997rt,
    author = "Cohen, Andrew G. and Kaplan, David B. and Nelson, Ann E.",
    title = "{Counting 4 pis in strongly coupled supersymmetry}",
    eprint = "hep-ph/9706275",
    archivePrefix = "arXiv",
    reportNumber = "BUHEP-97-16, DOE-ER-40561-328:-R-INT-97-00-172, UW-PT-97-13",
    doi = "10.1016/S0370-2693(97)00995-7",
    journal = "Phys. Lett. B",
    volume = "412",
    pages = "301--308",
    year = "1997"
}

@article{DiValentino:2014eea,
    author = "Di Valentino, Eleonora and Melchiorri, Alessandro",
    title = "{Planck constraints on neutrino isocurvature density perturbations}",
    eprint = "1405.5418",
    archivePrefix = "arXiv",
    primaryClass = "astro-ph.CO",
    doi = "10.1103/PhysRevD.90.083531",
    journal = "Phys. Rev. D",
    volume = "90",
    number = "8",
    pages = "083531",
    year = "2014"
}

\end{document}